\documentclass{aa}

\usepackage{color}

\usepackage{soul}
\usepackage{graphicx}

\usepackage{mathtools}
\usepackage{siunitx}
\usepackage[bottom]{footmisc}
\usepackage{tablefootnote}
\usepackage{threeparttable}
\usepackage{adjustbox}
\usepackage{subcaption}
\usepackage{float}
\usepackage{booktabs}
\usepackage{varwidth}
\usepackage{hyperref}
\usepackage{afterpage}
\usepackage{lineno}
\hypersetup{
colorlinks = true,
citecolor = blue
}

\usepackage{txfonts}
\begin{document} 

   \title{Lopsided distribution of MATLAS and ELVES dwarf satellite systems around isolated host galaxies} 
   \author{Nick Heesters
          \inst{1}
          \and
          Helmut Jerjen
          \inst{2} 
          \and  
          Oliver Müller
          \inst{1}
          \and
          Marcel S. Pawlowski 
          \inst{3}
          \and
          Kosuke Jamie Kanehisa
          \inst{3,4}
          }

   \institute{Institute of Physics, Laboratory of Astrophysics, Ecole Polytechnique F\'ed\'erale de Lausanne (EPFL), 1290 Sauverny, Switzerland\\
   \email{nick.heesters@epfl.ch}
   \and
    Research School of Astronomy and Astrophysics, Australian National University, Canberra, ACT 2611, Australia
    \and
    Leibniz-Institut für Astrophysik Potsdam (AIP), An der Sternwarte 16, D-14482 Potsdam, Germany
    \and
    Institut für Physik und Astronomie, Universität Potsdam, Karl-Liebknecht-Straße 24/25, 14476 Potsdam, Germany
             }
    
   \date{Received XX, XXXX; accepted YY, YYYY}

 
  \abstract
   {
   The properties of satellite dwarf galaxies pose important empirical constraints to verify cosmological models on galaxy scales. Their phase-space correlations, in particular, offer interesting insights into a broad range of models, since they are dominated by gravity and are largely independent of the specific dark matter flavor or baryonic processes that are considered. Next to the much-debated planes-of-satellites phenomenon, the lopsided distribution of satellites relative to their host galaxy has been studied observationally and in cosmological simulations. The degree to which observed lopsidedness is consistent with expectations from simulations is still unclear. We quantify the level of lopsidedness in isolated observed satellite systems under six different metrics. We study 47 systems from the MATLAS survey beyond the Local Volume (LV) as well as 21 LV satellite systems from the ELVES survey. The satellite systems are complete to an estimated absolute magnitude $M\sim$ -9. We find that the so-called wedge metric, counting the number of dwarfs in wedges with varying opening angles, is best suited to capture a system's overall lopsidedness. Under this metric, our analysis reveals that $\sim$16 percent of the tested satellite systems exhibit a statistically significant degree of lopsidedness when compared to systems with randomly generated satellite position angles. This presents a notable excess over the expected 5\% (2$\sigma$ level) of significantly lopsided systems in a sample with no overall inherent lopsidedness. To gain the most well-rounded picture, however, a combination of metrics that are sensitive to different features of lopsidedness should be used. Combining all tested metrics, the number of significantly lopsided systems increases to $\sim$21 percent. Contrary to recent results from the literature, we find more lopsided systems among the red early-type galaxies in the MATLAS survey compared to the mostly blue late-type hosts in ELVES. We further find that satellite galaxies at larger distances from the host, potentially recently accreted, are likely the primary contributors to the reported excess of lopsidedness. Our results set the groundwork that allows a comparison with similar systems in cosmological simulations to assess the consistency with the standard model.
   }


   \maketitle
%

\section{Introduction}
The phase-space distribution of dwarf galaxy satellite systems has proven to be a critical test for the standard model of cosmology \citep{2010A&A...523A..32K,2013MNRAS.435.1928P,2014Natur.511..563I,2015MNRAS.452.3838C,2017MNRAS.468L..41C,2019MNRAS.490.3786L,2021MNRAS.503.6170B,2022NatAs...6..897S}. The most discussed of these tests is the plane-of-satellite problem \citep{Pawlowski2018,2021Galax...9...66P}, which describes a tension between the predicted and observed motion and distribution of dwarf satellite galaxies around their hosts \citep[e.g.,][]{2013Natur.493...62I,2018Sci...359..534M,2024MNRAS.tmp..191K}.
Another more recent observation of the properties of satellite systems is the lopsidedness of dwarf galaxies toward one side of their host galaxies. Most prominently, this phenomenon was discovered for the Andromeda galaxy, where 20 out of 27 satellites (74\%) were shown to reside in one hemisphere \citep{mcconnachie2006satellite,2013ApJ...766..120C}. Recent distance measurements based on RR Lyrae presented in \citet{2022ApJ...938..101S} introduced a larger sample size and improved distance errors. The authors note an enhancement of the degree of anisotropy based on these new measurements such that 80\% of the Andromeda satellites are located in one hemisphere. Intriguingly, this overdensity of satellites seems to be aligned with the connecting line between the Milky Way (MW) and the Andromeda galaxy.

Motivated by these findings in the Local Group LG) system, \citet{2016ApJ...830..121L} studied thousands of galaxies in the Sloan Digital Sky Survey (SDSS, \citealt{2000AJ....120.1579Y}) and again found an excess of satellites with a 5$\sigma$ significance,  preferably occupying the space between pairs of host galaxies. An excess of up to 10\% of the satellites occupy the space between the pairs when compared to the expectations from a uniform distribution. Taken into account in the analysis was the signal arising from two overlapping halos, which will naturally boost the number of dwarf galaxies found between the pair of hosts. \citet{2016ApJ...830..121L}  argued that these findings highlight the unrelaxed and interacting nature of galaxies in pairs. 

Is such a lopsidedness expected for satellite distributions in standard cosmology simulations? \citet{2017ApJ...850..132P} studied pairs of host galaxies in different suites of cosmological simulations -- both dark matter-only and hydrodynamical simulations. They find a signal up to twice as strong as that reported for the SDSS systems. Assuming that the SDSS signal is a lower limit due to contamination from back- and foreground objects, the cosmological simulations appear to be consistent with the observed lopsidedness of satellite galaxies around galaxy pairs.

The origin of the lopsidedness in cosmological simulations has been studied by \citet{2019MNRAS.488.3100G}. Tracing back the orbital trajectories of satellites which at redshift $z = 0$ display the effect of lopsidedness, it was uncovered that a) lopsidedness was stronger in the past, and b) the signal is driven by satellites that are on their first approach along filaments connecting the two members of the pair and have yet to experience a `flyby'.
These findings are in line with previous arguments from \citet{2016ApJ...830..121L} that lopsidedness is a signal of an unrelaxed/young system. 

Once a galaxy environment has accreted the satellites and is fully virialized, the lopsidedness signal should in principle vanish. This begs the question if lopsidedness should be observed around spatially isolated luminous galaxies. \citet{2020ApJ...898L..15B} studied the spatial distribution of satellites around bright isolated host galaxies in the NASA-Sloan Atlas (NSA) catalog that were required to have at least two satellites. They report a statistically significant excess of satellite pairs on the same side of their host when compared to isotropic satellite distributions. This excess is most conspicuous for blue hosts and is mostly driven by satellites at large distances ($\gtrsim$300\,kpc) from their host, that were likely recently accreted. In \citet{2023ApJ...947...56S} the authors compared these observational findings directly with results from mock redshift surveys based on the Millennium simulation \citep{2005Natur.435..629S}. They report a significant excess of satellite pairs on the same side of the host, particularly pronounced among blue hosts, consistent with prior observational results. Moreover, they rule out the incompleteness of the observed catalog, interlopers or recently infalling satellites as the sole explanations for the observed lopsidedness. \citet{2021ApJ...914...78W} studied the lopsidedness of satellite systems around isolated host galaxies in Illustris-TNG \citep{2018MNRAS.473.4077P}, a hydrodynamical cosmological simulation. They found a statistically significant lopsided signal in projected positions. The strength of the lopsidedness depends on galaxy mass, color, and the large-scale environment. Satellite galaxies that inhabit low-mass blue hosts, or which are located further from the hosts, exhibit a stronger lopsidedness, as well as galaxy systems with massive neighbors. These findings are consistent with the results from \citet{2020ApJ...898L..15B} and \citet{2023ApJ...947...56S}. The origin of such lopsidedness for isolated host galaxies is not yet understood but may come from the merger history and/or large-scale environment.

Empirical results of lopsidedness either relied on stacking thousands of galaxies in blind surveys such as SDSS, where the foreground/background contamination may be high and the survey depth in terms of satellite luminosity or mass shallow, or on individual nearby galaxy systems such as our own Local Group, the M\,101 galaxy \citep{2017ApJ...850..109B}, or the M\,83 group \citep{2015A&A...583A..79M}. However, deep systematic studies of a larger sample of galaxy groups, where contamination is kept to a minimum, have not yet been conducted. Here, we present such a study based on galaxy environments selected from two recent dwarf galaxy surveys, the Mass Assembly of early-Type GaLAxies with their fine Structures (MATLAS, \citealt{2013IAUS..295..358D,2020MNRAS.491.1901H,2021MNRAS.506.5494P}) survey and the Exploration of Local VolumE Satellites (ELVES, \citealt{2022ApJ...933...47C}) survey. The MATLAS survey provides a large number of satellite systems beyond the Local Volume (LV) where satellite membership is, however, not yet confirmed via distance/velocity measurements for the majority of the systems. The ELVES survey contains fewer but complete systems with confirmed membership information within the LV. For our study, we use 47 early-type galaxies (ETGs) from the MATLAS survey and 21 MW-like galaxies from ELVES as host galaxies, which are residing in medium to low mass environments (see Tables\, \ref{tab:matlas_prop} and \ref{tab:elves_prop}).

This paper is structured as follows. In Section \ref{sec:data} we discuss the data we use for this study, in Section \ref{sec:sample_selection} we describe the isolation criteria we apply in order to obtain our galaxy sample. We then describe the different metrics of lopsidedness we tested on our sample and discuss details on how statistical significance is assessed in Section \ref{sec:metrics}. The results and discussions on this analysis follow in Section \ref{sec:results} and we summarize and conclude our findings in Section \ref{sec:summary}.

\begin{table*}[!htb]
\caption{Parameters for five of the six tested lopsidedness metrics for satellite distributions around 47 isolated ETGs from the dwarf galaxy survey MATLAS. The "wedge" metric is missing because it is a graph for each system rather than a number. 
\label{tab:matlas_prop}}
\begin{threeparttable}
\begin{tabular}{lllllllllll}
\toprule
      &  &  &  &  &  & Centroid & Max angle & Hemisphere & Pairwise & MRL \\
     Host & R.A. & DEC & D & N & $r_{cover}$ & $d_{norm}$ & $\theta_{max}$ & $(2-N/N_{lop})_{max}$ & $\theta_{lop}$ &  $\overline{R}$ \\
     name & (deg) & (deg) & (Mpc) &  & (kpc) &  & (deg) &  & (deg) &  \\
     (1) & (2) & (3) & (4) & (5) & (6) & (7) & (8) & (9) & (10) & (11) \\
\midrule
IC0598 & 153.2024 & 43.1455 & 35.3 & 4 & 320 & 0.07 & 140.6 & 0.667 & 112.6 & 0.16 \\
NGC0448 & 18.8189 & -1.6261 & 29.5 & 13 & 268 & 0.44 & 113.1 & 0.917 & 64.2 & 0.61 \\
NGC0661 & 26.0610 & 28.7060 & 30.6 & 12 & 278 & 0.28 & 92.2 & 0.500 & 91.8 & 0.26 \\
NGC0936 & 36.9061 & -1.1563 & 22.4 & 18 & 203 & 0.07 & 75.2 & 0.364 & 93.7 & 0.10 \\
NGC1121 & 42.6634 & -1.7340 & 35.3 & 15 & 321 & 0.16 & 85.1 & 0.636 & 92.1 & 0.19 \\
NGC1222 & 47.2364 & -2.9552 & 33.3 & 19 & 303 & 0.37 & 51.6 & 0.643 & 89.3 & 0.25 \\
NGC2679 & 132.8872 & 30.8654 & 31.1 & 3 & 283 & 0.26 & 172.9 & 0.500 & 120.0 & 0.30 \\
NGC2685 & 133.8948 & 58.7344 & 16.7 & 9 & 123 & 0.44 & 120.6 & 0.714 & 94.8 & 0.23 \\
NGC2764 & 137.0730 & 21.4434 & 39.6 & 4 & 360 & 0.40 & 207.8 & 1.000 & 79.7 & 0.58 \\
NGC2824 & 139.7593 & 26.2700 & 40.7 & 4 & 370 & 0.16 & 124.2 & 0.667 & 114.8 & 0.11 \\
NGC2859 & 141.0773 & 34.5134 & 27.0 & 16 & 246 & 0.28 & 48.9 & 0.545 & 88.9 & 0.29 \\
NGC2880 & 142.3942 & 62.4906 & 21.3 & 9 & 194 & 0.20 & 102.0 & 0.500 & 97.8 & 0.08 \\
NGC2950 & 145.6463 & 58.8512 & 14.5 & 7 & 132 & 0.34 & 113.8 & 0.600 & 100.6 & 0.17 \\
NGC2974 & 145.6386 & -3.6991 & 20.9 & 12 & 190 & 0.06 & 107.3 & 0.500 & 96.3 & 0.08 \\
NGC3032 & 148.0341 & 29.2363 & 21.4 & 3 & 195 & 0.98 & 337.5 & 1.000 & 15.0 & 0.99 \\
NGC3098 & 150.5695 & 24.7111 & 23.0 & 6 & 209 & 0.12 & 120.8 & 0.500 & 102.3 & 0.20 \\
NGC3182 & 154.8876 & 58.2058 & 34.0 & 7 & 309 & 0.55 & 159.7 & 0.833 & 74.4 & 0.51 \\
NGC3230 & 155.9331 & 12.5679 & 40.8 & 16 & 371 & 0.58 & 114.3 & 0.857 & 70.4 & 0.56 \\
NGC3248 & 156.9393 & 22.8472 & 24.6 & 5 & 223 & 0.59 & 184.0 & 1.000 & 82.2 & 0.56 \\
NGC3457 & 163.7026 & 17.6212 & 20.1 & 7 & 142 & 0.65 & 177.2 & 0.833 & 69.1 & 0.62 \\
NGC3522 & 166.6685 & 20.0856 & 25.5 & 2 & 231 & 0.79 & 287.4 & 1.000 & 72.6 & 0.81 \\
NGC3648 & 170.6312 & 39.8770 & 31.9 & 11 & 289 & 0.09 & 134.0 & 0.625 & 96.7 & 0.14 \\
NGC3694 & 172.2256 & 35.4139 & 35.2 & 9 & 320 & 0.70 & 180.4 & 1.000 & 62.6 & 0.64 \\
NGC3941 & 178.2307 & 36.9864 & 11.9 & 5 & 108 & 0.24 & 171.9 & 0.750 & 103.9 & 0.25 \\
NGC4119 & 182.0402 & 10.3787 & 16.5 & 1 & 150 & 1.00 & - & 1.000 & - & 1.00 \\
NGC4150 & 182.6402 & 30.4015 & 13.4 & 8 & 122 & 0.50 & 109.0 & 0.667 & 91.6 & 0.34 \\
NGC4643 & 190.8339 & 1.9784 & 16.5 & 18 & 150 & 0.40 & 93.9 & 0.800 & 79.3 & 0.44 \\
NGC4697 & 192.1496 & -5.8008 & 11.4 & 12 & 104 & 0.38 & 124.4 & 0.800 & 81.9 & 0.43 \\
NGC4753 & 193.0921 & -1.1997 & 22.9 & 8 & 208 & 0.38 & 152.1 & 0.667 & 89.1 & 0.36 \\
NGC5493 & 212.8724 & -5.0437 & 38.8 & 8 & 352 & 0.29 & 80.0 & 0.400 & 98.7 & 0.16 \\
NGC5582 & 215.1797 & 39.6936 & 27.7 & 9 & 204 & 0.62 & 148.1 & 0.875 & 70.0 & 0.58 \\
NGC5611 & 216.0199 & 33.0475 & 24.5 & 8 & 223 & 0.38 & 162.0 & 0.857 & 77.7 & 0.51 \\
NGC5631 & 216.6387 & 56.5827 & 27.0 & 9 & 216 & 0.38 & 160.9 & 0.875 & 87.6 & 0.38 \\
NGC5866 & 226.6232 & 55.7633 & 14.9 & 10 & 136 & 0.17 & 130.4 & 0.571 & 95.1 & 0.23 \\
NGC6547 & 271.2917 & 25.2326 & 40.8 & 6 & 371 & 0.39 & 116.8 & 0.800 & 99.0 & 0.23 \\
NGC7710 & 353.9423 & -2.8809 & 34.0 & 7 & 309 & 0.59 & 191.4 & 1.000 & 62.4 & 0.68 \\
PGC028887 & 149.9313 & 11.6608 & 41.0 & 11 & 371 & 0.39 & 103.8 & 0.625 & 91.8 & 0.11 \\
PGC056772 & 240.5483 & 7.0860 & 39.5 & 6 & 358 & 0.63 & 159.9 & 0.800 & 85.4 & 0.48 \\
PGC058114 & 246.5178 & 2.9066 & 23.8 & 3 & 216 & 0.78 & 261.1 & 1.000 & 65.9 & 0.75 \\
PGC061468 & 272.3607 & 19.1177 & 36.2 & 3 & 329 & 0.23 & 173.6 & 0.500 & 120.0 & 0.31 \\
UGC03960 & 115.0949 & 23.2751 & 33.2 & 3 & 281 & 0.40 & 149.0 & 0.500 & 120.0 & 0.22 \\
UGC04551 & 131.0246 & 49.7940 & 28.0 & 11 & 255 & 0.72 & 178.9 & 0.900 & 60.5 & 0.68 \\
UGC05408 & 150.9661 & 59.4361 & 45.8 & 3 & 417 & 0.43 & 203.9 & 1.000 & 104.1 & 0.47 \\
UGC06062 & 164.6567 & 9.0505 & 38.7 & 9 & 351 & 0.05 & 106.3 & 0.500 & 99.5 & 0.08 \\
UGC06176 & 166.8528 & 21.6572 & 40.1 & 3 & 365 & 0.73 & 203.6 & 1.000 & 104.3 & 0.40 \\
UGC08876 & 209.2419 & 45.9732 & 33.9 & 1 & 308 & 1.00 & - & 1.000 & - & 1.00 \\
UGC09519 & 221.5880 & 34.3707 & 27.6 & 4 & 251 & 0.63 & 210.1 & 1.000 & 79.6 & 0.61 \\
\bottomrule
\end{tabular}
Columns: (1) name of ETG, (2,3) coordinates in the J2000 frame, (4) distance to ETG, (5) number of satellites within maximum circle covered by FoV, (6) radius of maximum circle covered by FoV in kpc at the distance of the host, (7) normalized offset between ETG and centroid of 2D satellite distribution, (8) maximum opening angle without a satellite, (9) maximum contrast in satellite distribution from a 180-degree separation line where $N_{lop}$ is the maximum number of satellites found on one side, (10) mean pairwise angle between the satellites, (11) mean resultant length. Note: empty entries (-) mark systems for which a given angular metric is not applicable because they feature only a single satellite.
\end{threeparttable}
\end{table*}

\begin{table*}[!htb]
\caption{Lopsidedness $p$-values for the metrics applied to the MATLAS systems.\label{tab:matlas_pvals}}
\centering
\begin{threeparttable}
\begin{tabular}{lllllll}
\toprule
Host & $p_{centroid}$ & $p_{\theta_{max}}$ & $p_{hemisphere}$  & $p_{pairwise}$ & $p_{MRL}$ & $p_{wedge}$ \\
(1) & (2) & (3) & (4) & (5) & (6) & (7) \\
\midrule
IC0598 & 0.987 & 0.843 & 1.000 & 0.921 & 0.907 & 0.902 \\ 
NGC0448 & 0.072 & 0.139 & \textbf{0.034} & \textbf{0.003} & \textbf{0.006} & \textbf{0.002} \\ 
NGC0661 & 0.413 & 0.438 & 0.965 & 0.496 & 0.452 & 0.320 \\ 
NGC0936 & 0.921 & 0.319 & 0.995 & 0.863 & 0.826 & 0.748 \\  
NGC1121 & 0.698 & 0.330 & 0.574 & 0.565 & 0.573 & 0.788 \\ 
NGC1222 & 0.075 & 0.818 & 0.392 & 0.316 & 0.300 & 0.218 \\ 
NGC2679 & 0.918 & 0.805 & 1.000 & 1.000 & 0.818 & 0.829 \\ 
NGC2685 & 0.172 & 0.340 & 0.695 & 0.637 & 0.624 & 0.724 \\ 
NGC2764 & 0.600 & 0.301 & 0.495 & 0.258 & 0.267 & 0.252 \\ 
NGC2824 & 0.927 & 0.945 & 1.000 & 0.961 & 0.950 & 0.986 \\ 
NGC2859 & 0.291 & 0.963 & 0.794 & 0.301 & 0.271 & 0.634 \\ 
NGC2880 & 0.724 & 0.582 & 0.995 & 0.870 & 0.949 & 0.715 \\ 
NGC2950 & 0.461 & 0.661 & 0.983 & 0.873 & 0.828 & 0.920 \\ 
NGC2974 & 0.956 & 0.240 & 0.966 & 0.928 & 0.927 & 0.541 \\ 
NGC3032 & \textbf{0.013} & \textbf{0.012} & 0.746 & \textbf{0.012} & \textbf{0.012} & \textbf{0.019} \\ 
NGC3098 & 0.924 & 0.717 & 1.000 & 0.868 & 0.807 & 0.929 \\ 
NGC3182 & 0.117 & 0.207 & 0.541 & 0.100 & 0.167 & \textbf{0.001} \\ 
NGC3230 & \textbf{0.002} & 0.052 & \textbf{0.042} & \textbf{0.005} & \textbf{0.005} & \textbf{0.017} \\ 
NGC3248 & 0.181 & 0.283 & 0.306 & 0.246 & 0.217 & 0.542 \\ 
NGC3457 & \textbf{0.025} & 0.119 & 0.540 & 0.061 & 0.065 & 0.078 \\ 
NGC3522 & 0.402 & 0.402 & 0.997 & 0.402 & 0.402 & 0.407 \\ 
NGC3648 & 0.925 & 0.106 & 0.802 & 0.911 & 0.827 & 0.333 \\ 
NGC3694 & \textbf{0.007} & \textbf{0.035} & \textbf{0.034} & \textbf{0.014} & \textbf{0.021} & \textbf{0.004} \\ 
NGC3941 & 0.779 & 0.375 & 0.935 & 0.800 & 0.747 & 0.579 \\ 
NGC4119 & 1.000 & - & 1.000 & - & 1.000 & 1.000 \\ 
NGC4150 & 0.143 & 0.599 & 0.871 & 0.448 & 0.416 & 0.761 \\ 
NGC4643 & 0.057 & 0.105 & 0.074 & \textbf{0.029} & \textbf{0.031} & 0.130 \\ 
NGC4697 & 0.181 & 0.113 & 0.255 & 0.108 & 0.114 & 0.110 \\ 
NGC4753 & 0.327 & 0.173 & 0.870 & 0.351 & 0.374 & 0.253 \\ 
NGC5493 & 0.524 & 0.947 & 1.000 & 0.852 & 0.816 & 0.612 \\ 
NGC5582 & \textbf{0.021} & 0.128 & 0.240 & \textbf{0.036} & \textbf{0.041} & \textbf{0.009} \\ 
NGC5611 & 0.342 & 0.123 & 0.371 & 0.113 & 0.124 & \textbf{0.030} \\ 
NGC5631 & 0.292 & 0.079 & 0.241 & 0.292 & 0.282 & 0.241 \\ 
NGC5866 & 0.773 & 0.175 & 0.934 & 0.704 & 0.613 & 0.494 \\ 
NGC6547 & 0.424 & 0.766 & 0.746 & 0.697 & 0.753 & 0.701 \\ 
NGC7710 & 0.084 & 0.075 & 0.108 & \textbf{0.032} & \textbf{0.033} & \textbf{0.019} \\ 
PGC028887 & 0.202 & 0.357 & 0.799 & 0.484 & 0.869 & 0.601 \\ 
PGC056772 & 0.072 & 0.318 & 0.746 & 0.285 & 0.264 & 0.578 \\ 
PGC058114 & 0.184 & 0.228 & 0.748 & 0.228 & 0.224 & 0.342 \\ 
PGC061468 & 0.878 & 0.801 & 1.000 & 1.000 & 0.806 & 0.757 \\ 
UGC03960 & 0.711 & 0.942 & 1.000 & 1.000 & 0.917 & 0.887 \\ 
UGC04551 & \textbf{0.001} & \textbf{0.011} & 0.095 & \textbf{0.004} & \textbf{0.004} & \textbf{0.046} \\ 
UGC05408 & 0.581 & 0.564 & 0.745 & 0.564 & 0.527 & 0.694 \\ 
UGC06062 & 0.977 & 0.519 & 0.996 & 0.975 & 0.953 & 0.857 \\ 
UGC06176 & 0.246 & 0.568 & 0.748 & 0.568 & 0.625 & 0.215 \\ 
UGC08876 & 1.000 & - & 1.000 & - & 1.000 & 1.000 \\ 
UGC09519 & 0.213 & 0.290 & 0.498 & 0.261 & 0.232 & 0.499 \\
 \addlinespace
 \hline
 \addlinespace
\multicolumn{7}{c}{Number of statistically significant systems per metric} \\
\addlinespace
\hline
\addlinespace
 & 6 ($\sim$12.7\%) & 3 ($\sim$6.4\%) & 3 & 8 ($\sim$17.0\%) & 8 & 9 ($\sim$19.1\%) \\
 \bottomrule
\end{tabular}
Columns: (1) name of the host galaxy, (2-7) $p$-values for the different metrics: centroid, maximum opening angle, hemisphere, mean pairwise angle, mean resultant length (MRL), and wedge. A $p$-value of 0.05 or lower is considered statistically significant at the 2$\sigma$ level and highlighted in bold.
\end{threeparttable}
\end{table*}

\begin{table*}[t]
\caption{List of 21 ELVES host galaxies that are considered to be spatially isolated}.\label{tab:elves_prop}
\centering
\begin{threeparttable}  
\begin{tabular}{lllllllllll}
\toprule
   &  &  &  &  &  & Centroid & Max angle & Hemisphere & Pairwise & MRL \\
     Host & R.A. & DEC & D & N & $r_{cover}$ &$d_{norm}$ & $\theta_{max}$ & $(2-N/N_{lop})_{max}$ & $\theta_{lop}$ &  $\overline{R}$ \\
     name & (deg) & (deg) & (Mpc) &  & (kpc) &  & (deg) &  & (deg) &  \\
     (1) & (2) & (3) & (4) & (5) & (6) & (7) & (8) & (9) & (10) & (11) \\
\midrule
M104 & 189.9976 & -11.6231 & 9.6 & 15 & 150 & 0.21 & 55.4 & 0.500 & 93.9 & 0.12 \\
NGC0253 & 11.8881 & -25.2888 & 3.6 & 6 & 300 & 0.36 & 142.1 & 0.500 & 98.6 & 0.31 \\
NGC0628 & 24.1739 & 15.7836 & 9.8 & 14 & 300 & 0.31 & 86.3 & 0.727 & 89.6 & 0.29 \\
NGC0891 & 35.6371 & 42.3483 & 9.1 & 7 & 200 & 0.35 & 136.4 & 0.833 & 94.9 & 0.25 \\
NGC1023 & 40.1001 & 39.0632 & 10.4 & 17 & 200 & 0.37 & 64.8 & 0.692 & 88.3 & 0.29 \\
NGC1291 & 49.3274 & -41.1080 & 9.1 & 18 & 300 & 0.10 & 65.9 & 0.500 & 92.9 & 0.14 \\
NGC1808 & 76.9264 & -37.5130 & 9.3 & 14 & 300 & 0.17 & 111.3 & 0.727 & 83.7 & 0.40 \\ 
NGC2683 & 133.1721 & 33.4218 & 9.4 & 10 & 300 & 0.29 & 76.0 & 0.571 & 93.5 & 0.26 \\
NGC2903 & 143.0421 & 21.5008 & 9.0 & 7 & 300 & 0.41 & 119.8 & 0.600 & 94.4 & 0.25 \\
NGC3115 & 151.3080 & -7.7186 & 10.2 & 19 & 300 & 0.22 & 76.1 & 0.417 & 94.3 & 0.02 \\
NGC3344 & 160.8798 & 24.9222 & 9.8 & 7 & 300 & 0.38 & 118.3 & 0.833 & 93.9 & 0.30 \\
NGC3521 & 166.4524 & -0.0359 & 11.2 & 12 & 330 & 0.31 & 111.5 & 0.500 & 96.2 & 0.09 \\
NGC3556 & 167.8790 & 55.6741 & 9.6 & 14 & 300 & 0.35 & 97.4 & 0.600 & 84.8 & 0.37 \\
NGC4258 & 184.7401 & 47.3037 & 7.2 & 8 & 150 & 0.44 & 100.2 & 0.667 & 94.1 & 0.29 \\
NGC4517 & 188.1899 & 0.1150 & 8.3 & 9 & 300 & 0.49 & 191.6 & 1.000 & 76.6 & 0.47 \\
NGC4631 & 190.5334 & 32.5415 & 7.4 & 13 & 200 & 0.10 & 63.2 & 0.556 & 93.5 & 0.19 \\
NGC4736 & 192.7211 & 41.1203 & 4.2 & 14 & 300 & 0.16 & 90.1 & 0.600 & 92.6 & 0.21 \\
NGC4826 & 194.1821 & 21.6827 & 5.3 & 9 & 300 & 0.31 & 86.3 & 0.714 & 89.1 & 0.34 \\
NGC5236 & 204.2538 & -29.8658 & 4.7 & 11 & 300 & 0.11 & 66.2 & 0.429 & 94.8 & 0.17 \\
NGC5457 & 210.8024 & 54.3488 & 6.5 & 9 & 300 & 0.37 & 148.9 & 0.714 & 85.5 & 0.41 \\ 
NGC6744 & 287.4422 & -63.8575 & 8.9 & 11 & 200 & 0.53 & 176.6 & 0.900 & 74.2 & 0.54 \\ 
\bottomrule
\end{tabular}
Columns: (1) name of ELVES host, (2,3) coordinates in degrees in the J2000 frame, (4) distance to ELVES host in Mpc, (5) number of satellites, (6) radius of the covered area around the host in kpc at the distance of the host, (7) normalized offset between host and center of satellite distribution, (7) maximum wedge angle without a satellite, (8) maximum contrast in satellite distribution from a 180-degree separation line where $a$ is the maximum number of satellites found on one side (note: by definition $N/2 \leq a \leq N $), (9) mean pairwise angle between the satellites in degrees, (10) mean resultant length (MRL).
\end{threeparttable}
\end{table*}

\begin{table*}
\caption{Lopsidedness $p$-values for the metrics applied to the ELVES systems.\label{tab:elves_pvals}}
\centering
\begin{threeparttable}
\begin{tabular}{lllllll}
\toprule
Host & $p_{centroid}$ & $p_{\theta_{max}}$ & $p_{hemisphere}$ & $p_{pairwise}$ & $p_{MRL}$ & $p_{wedge}$ \\
     (1) & (2) & (3) & (4) & (5) & (6) & (7) \\
\midrule
M104 & 0.519 & 0.912 & 0.911 & 0.790 & 0.805 & 0.761 \\
NGC0253 & 0.498 & 0.480 & 1.000 & 0.680 & 0.586 & 0.261 \\
NGC0628 & 0.272 & 0.379 & 0.350 & 0.349 & 0.314 & 0.199 \\
NGC0891 & 0.441 & 0.397 & 0.543 & 0.573 & 0.665 & 0.427 \\
NGC1023 & 0.099 & 0.608 & 0.322 & 0.262 & 0.240 & 0.101 \\
NGC1291 & 0.853 & 0.516 & 0.842 & 0.725 & 0.702 & 0.685 \\
NGC1808 & 0.690 & 0.113 & 0.348 & 0.122 & 0.112 & \textbf{0.020} \\
NGC2683 & 0.458 & 0.874 & 0.933 & 0.584 & 0.514 & 0.891 \\
NGC2903 & 0.328 & 0.589 & 0.983 & 0.550 & 0.655 & 0.732 \\
NGC3115 & 0.403 & 0.254 & 0.960 & 0.986 & 0.991 & 0.628 \\
NGC3344 & 0.377 & 0.607 & 0.543 & 0.528 & 0.560 & 0.117 \\
NGC3521 & 0.335 & 0.202 & 0.966 & 0.921 & 0.911 & 0.588 \\
NGC3556 & 0.177 & 0.228 & 0.726 & 0.148 & 0.144 & 0.134 \\
NGC4258 & 0.222 & 0.722 & 0.870 & 0.562 & 0.530 & 0.833 \\
NGC4517 & 0.119 & \textbf{0.021} & \textbf{0.035} & 0.082 & 0.137 & \textbf{0.009} \\
NGC4631 & 0.895 & 0.883 & 0.865 & 0.664 & 0.621 & 0.367 \\
NGC4736 & 0.703 & 0.321 & 0.727 & 0.599 & 0.556 & 0.652 \\
NGC4826 & 0.438 & 0.813 & 0.698 & 0.347 & 0.366 & 0.512 \\
NGC5236 & 0.885 & 0.934 & 0.999 & 0.721 & 0.725 & 0.798 \\
NGC5457 & 0.315 & 0.124 & 0.695 & 0.230 & 0.217 & 0.381 \\
NGC6744 & \textbf{0.040} & \textbf{0.013} & 0.095 & \textbf{0.038} & \textbf{0.035} & 0.057 \\

\addlinespace
\hline
\addlinespace
\multicolumn{7}{c}{Number of statistically significant systems per metric} \\
\addlinespace
\hline
\addlinespace
 & 1 ($\sim$4.7\%) & 2 ($\sim$9.5\%) & 1 & 1 & 1 & 2 \\
\bottomrule
\end{tabular}
Columns: (1) name of the host galaxy, (2-7) $p$-values for the different metrics: centroid, maximum opening angle, hemisphere, mean pairwise angle, mean resultant length (MRL), and wedge. A $p$-value of 0.05 or lower is considered statistically significant at the 2$\sigma$ level and highlighted in bold. \\
\end{threeparttable}
\end{table*}

\section{Data}
\label{sec:data}
We use the MATLAS dwarf galaxy catalog \citep{2021MNRAS.506.5494P}, which consists of 2210 dwarfs \citep{2020MNRAS.491.1901H} discovered in targeted CFHT+MegaCam observations of 142 fields of $\sim$1deg$^{2}$ (63\,arcmin x 69\,arcmin) each. Each field is centered on an ETG and may contain several more ETGs and late-type galaxies (LTGs). The line-of-sight distances to these hosts are in the range 10--45\,Mpc and are for the most part based on redshift and surface brightness fluctuation (SBF) measurements \citep[see][]{2011MNRAS.413..813C}. The absolute magnitude limit of the MATLAS dwarfs is estimated at $M_g\approx-9$\,mag, however, the exact magnitudes of the dwarfs depend on their distances, which itself depends on their association with the central host ETG. This association is not yet determined for most of the dwarfs by means of a distance or velocity measurement. However, an initial MUSE study of 56 MATLAS dwarf galaxies revealed that $\approx$57\% are indeed associated with their putative hosts based on similar recession velocities \citep{2023A&A...676A..33H}. Approximately 75\% of the dwarfs in this spectroscopic sample are located in the distance range probed by the MATLAS survey. 

We further use the dwarf catalog from ELVES \citep{2022ApJ...933...47C}, which is a compilation of dwarf galaxies from the literature \citep[e.g., ][]{2009AJ....137.3009C,2014MNRAS.445..881C,2015A&A...583A..79M,2016ApJ...830L..21T,2017ApJ...850..109B,2018ApJ...863..152S}, as well as detections of their own with publicly available data. Excluding the Local Group satellites, they provide a list of 414 dwarf galaxies around 29 hosts in the Local Volume, a sphere of approximately 10\,Mpc around our point of view. The majority (308; $\sim$74\%) of these dwarfs are confirmed satellites through different distance estimates such as tip of the red giant branch (TRGB, e.g., \citealt{2002A&A...385...21K,2013AJ....146..126C,2019A&A...629L...2M,2020ApJ...893L...9B}) or surface brightness fluctuation (SBF, e.g., \citealt{2000AJ....119..166J,2001A&A...380...90J,2019ApJ...878L..16C,2022ApJ...933...47C}) measurements. The catalog is complete to a magnitude of $M_V\approx-9$, similar to the limiting magnitude of MATLAS.

\section{Sample selection} 
\label{sec:sample_selection}
Our lopsidedness analysis shall be focused on satellite systems around isolated hosts. This decision is motivated by the properties of the MATLAS survey: MATLAS consists of individual 1 deg$^{2}$ pointings on a selection of ETGs and does not allow for the study of a large continuous region around them. This would be required to investigate e.g., pairs of galaxies similar to the LG, where a strong degree of lopsidedness in the Andromeda satellite system toward the MW has been reported (Kanehisa, Pawlowski \& Libeskind, in prep.). The limited field of view (FoV) of the MATLAS pointings does not cover the full extent of such paired satellite systems and would bias us to find more dwarf galaxies between the two hosts rather than on opposite sides. Such a study would be more suitable for an all-sky survey. Furthermore, with isolated hosts, we decrease the probability of dwarf contamination from other systems, since membership of all dwarfs is not yet established.

We selected our candidate host\,-\,satellite systems based on the following isolation criterion: no other galaxy within 1\,Mpc of the host with an absolute K-band magnitude $M_{K}$ brighter or up to one magnitude fainter than the host. We choose the magnitude constraint in line with \citet{2020ApJ...898L..15B} and the distance constraint to clearly separate the analysis from grouped hosts such as the MW and Andromeda \citep[separation $\sim$760\,kpc;][]{2021ApJ...920...84L}. To make this cut, we utilize the Heraklion Extragalactic Catalogue \citep[HECATE,][]{2021MNRAS.506.1896K}, which is a value-added galaxy catalog, containing over 200k galaxies at distances $\leq$ 200\,Mpc. We search for the MATLAS and ELVES host galaxies in the catalog and use the positions, distances, and K-band absolute magnitudes to extract a host sample. A total of 48 ETGs meet this isolation criterion in the MATLAS survey (see column 1 of Table\, \ref{tab:matlas_prop}) and 21 MW-like hosts in the ELVES survey (see column 1 of Table\, \ref{tab:elves_prop}). One system (NGC1023) was observed in both surveys. For this case, we choose to use the data from the ELVES survey, due to the available distance estimates.

The MATLAS fields have a rectangular FoV with the ETG slightly offset from the center in some cases. In order to avoid radial incompleteness we discard any satellites found in the corners of the field. We do this by drawing the maximum possible circle within the MATLAS FoV centered on the ETG and only keep dwarfs that fall within this circle. All dwarfs within this circle, although not yet fully confirmed via follow-up measurements, are considered satellites of the ETG for the purpose of this study. Since the MATLAS hosts reside at a range of different distances, these circles have varying physical radii. This could potentially have an impact on the signal for lopsidedness since results from the literature suggest that one of the driving factors are satellites located at large distances ($\gtrsim$300\,kpc) from their host \citep[][]{2020ApJ...898L..15B,2021ApJ...914...78W}. The physical radii of the circles centered on the hosts that fit within the MATLAS FoV are shown in column 6 of Table\, \ref{tab:matlas_prop}. Radial incompleteness is not a concern for the ELVES hosts since the satellite systems are probed in circular footprints. The majority of the selected hosts ($\sim$ 70\%) were surveyed to 300\,kpc (approximately one virial radius $r_{vir}$), and all of them have coverage to at least $r_{vir}$/2. Radial coverage is shown for every ELVES system in column 6 of Table\,\ref{tab:elves_prop}. In this analysis, we treat all dwarfs from the ELVES catalog, which includes confirmed and potential satellites as documented in \citet{2022ApJ...933...47C}, as satellites of their respective putative hosts. The number of satellite galaxies for each host in the MATLAS and ELVES surveys is given in column 5 of Table\, \ref{tab:matlas_prop} and Table\,\ref{tab:elves_prop}, respectively. The median number of satellites that fall within the circle enclosed in the MATLAS FoV is eight. For the ELVES sample, the median number of satellites per system is 11.

\section{Description of metrics of lopsidedness}
\label{sec:metrics}
\citet{2021ApJ...914...78W} used two different 2D metrics to investigate the projected lopsided distribution of satellite galaxies around isolated central galaxies in a hydrodynamic cosmological simulation: the pairwise angle distribution $ \theta^{ij}$ and the lopsided angle $\theta_{lop}=\langle \theta \rangle^{ij}$. In our study we employ six different types of metric, including the mean pairwise angle from \citet{2021ApJ...914...78W}, to quantify the presence or absence of lopsidedness in 2D. They were in part inspired by the 3D metrics introduced and used by Kanehisa, Pawlowski \& Libeskind (in prep.). We assess the statistical significance for all metrics by employing Monte Carlo (MC) simulations. For each observed host environment we create 10$^{5}$ random realizations with the same number of satellites as in the observed system. We achieve this by assigning a new position angle -- drawn from a random uniform distribution -- to every observed satellite in the system. The radial distribution between the observed and the simulated systems thus remains unchanged. We create the random isotropic realizations once for all observed systems and utilize these same sets to study the six metrics. With this, we ensure that the comparison between the metrics is most meaningful. After computing the respective lopsidedness metric for all simulated systems in the same way as for the observed one, we count the number of metric values from the randomized systems that are equal to or more lopsided than the observed one. The number of randomized systems that are equally or more lopsided than the observed one divided by the number of MC realizations gives the $p$-value or the statistical significance of the degree of lopsidedness in a system. We consider $p$-values $\leq$ 0.05 to be statistically significant.

\subsection{Centroid Offset}
We calculate the unweighted centroid of satellites for a given system and measure its distance from the host galaxy. Since this centroid shift depends on the system’s degree of lopsidedness and the characteristic distance of its satellites, we normalize the offset by the root-mean-square distance of the satellites from their host, $d_{rms}$. This yields $d_{norm}$, the normalized centroid offset from the system’s host galaxy. The statistical significance is assessed via metric values equal to or larger than the corresponding observed value.

\subsection{Maximum Opening Angle}
We examine the position angles of all satellites in a system centered on the host galaxy and compute the maximum angle subtended by two satellites where no other dwarf resides in between. Metric values equal to or larger than the observed one determine the $p$-value.

\subsection{Maximum Hemisphere Population}
This metric counts the number of satellites on each side of an arbitrarily oriented straight line with position angle $\alpha$ going through the host galaxy, which divides the distribution into two half-disks. By rotating this line ($0^\circ<\alpha< 180^\circ$) we search for the maximum number of satellites in one half-disk $N_{lop}$. We rotate the dividing line in steps of 1 degree and count the number of satellites on either side. By definition $N/2 \leq N_{lop} \leq N $, where $N$ is the total number of satellites. The maximum contrast metric $(2-N/N_{lop})_{max}$ then takes values from 0 to 1, where 1 means there exists a line going through the host galaxy where all satellites are found on one side. A value of 0 means that the line splits the population with $N/2$ satellites on either side. Thus, higher metric values indicate a higher degree of lopsidedness.

\subsection{Pairwise Angle}
\label{sec:pairwise}
As described in \citet{2021ApJ...914...78W}, the degree of lopsidedness is quantified by inspecting the angles $\theta^{ij}$ $\in$ [0,180]$^\circ$ between all possible pairs of satellites. For perfectly isotropic systems we would expect the mean ($\langle \theta \rangle^{ij}$) of these angles to be 90$^\circ$. A smaller mean value indicates a lopsided system. This metric is, however, sensitive to the number of satellites in a given system. Lopsided systems with a small number of satellites can show mean pairwise angles $>$ 90$^\circ$ since any outliers located at the opposing side of the satellite concentration have a significant influence on the mean at low numbers. Using the median instead of the mean may be beneficial in some instances where a few outliers are located on the opposite side of the bulk of the satellite population. This can, however, also have adverse effects at low numbers. Since some of our satellite systems have such low numbers, we choose to use the mean. This behavior at low numbers should be considered when selecting a metric to describe the lopsidedness of a specific system. Pairwise angles from artificial satellite systems that are equal to or lower than in the observations determine the $p$-value for this metric.

\subsection{Mean Resultant Length (MRL)}
The so-called Mean Resultant Length metric was recently used in \citet{2021RNAAS...5...57B} and \citet{2023ApJ...947...56S} to classify lopsided satellite systems. This metric is a measure of the overall direction of satellites relative to their host. For $N$ satellites with position angles $\phi_{i}$ and the host at the origin of the coordinate system, the MRL is defined as:

\begin{equation}
\label{eq:MRL}
    \overline{R} = \frac{1}{N}\sqrt{\left(\sum_{i=1}^{N} \cos(\phi_i)\right)^2 + \left(\sum_{i=1}^{N} \sin(\phi_i)\right)^2}.
\end{equation}

A perfectly uniform angular distribution would result in $\overline{R}$ = 0, while satellites clustered along a straight radial line from the host would yield $\overline{R}$ = 1. Thus, metric values in the Monte Carlo realizations that are equal to or greater than the observed value signify statistical significance. The MRL metric is similar to the centroid metric in the sense that both describe the mean direction of the satellite system. The MRL only considers the angular distribution as the satellites are placed on a unit circle around the host. The centroid additionally accounts for the projected distance between the satellites and their host.

\subsection{Maximum Wedge Population}
\label{sec:wedge_metric}
The final metric we are testing in this work is the Maximum Wedge Population metric (hereafter the 'wedge metric' for short). The wedge metric is a generalization of the Maximum Hemisphere Population metric. A wedge centered on the host galaxy with an opening angle $\theta$ is defined. For a given opening angle we rotate the wedge by 360 degrees in 1 degree increments and compute the number of satellites enclosed within each wedge. For each $\theta$ we record the most populated wedge with an associated population number $N_{\theta}$.

\begin{figure*}[!htbp]
\centering
\includegraphics[width=0.49\linewidth]{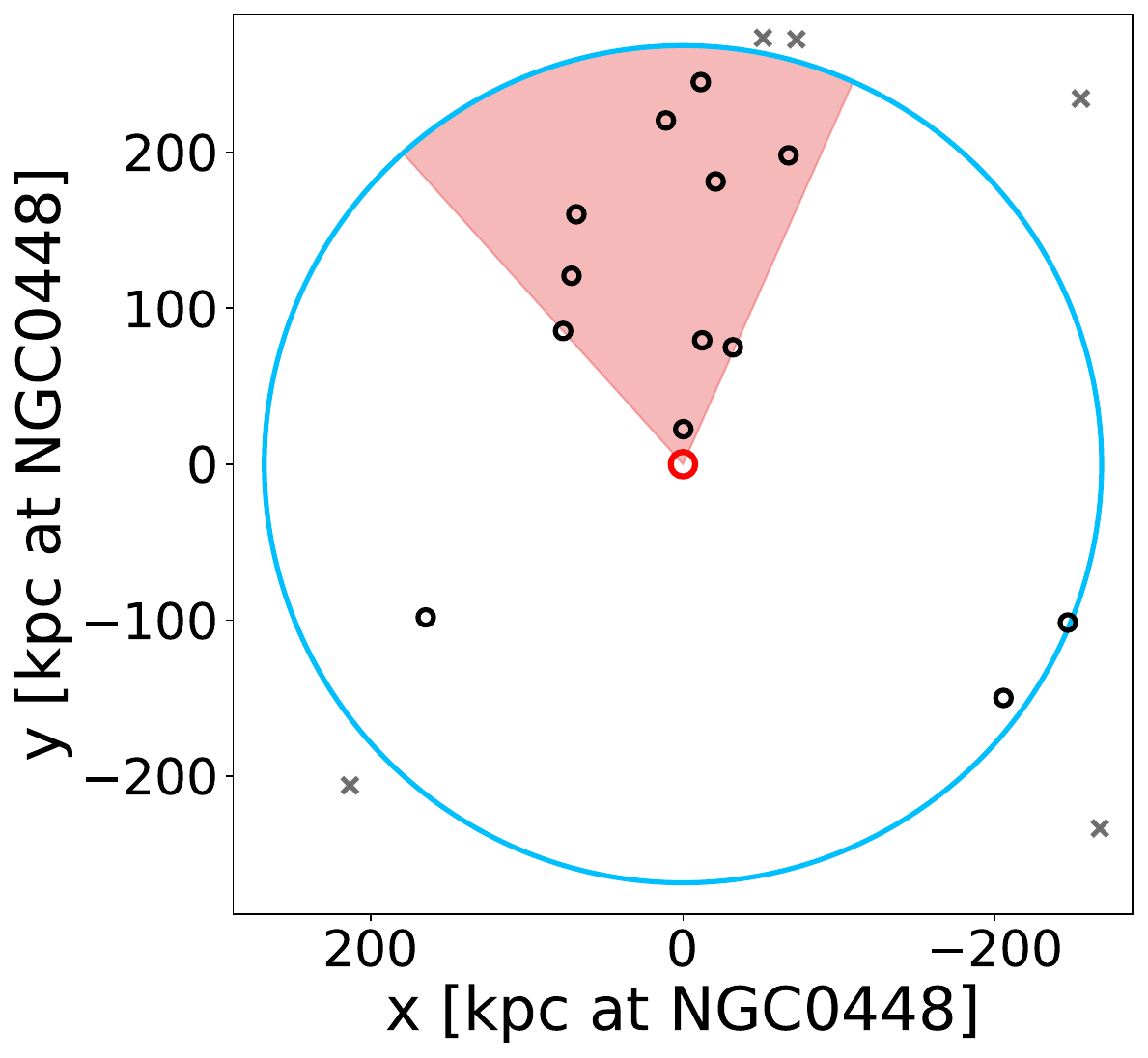}
\includegraphics[width=0.49\linewidth]{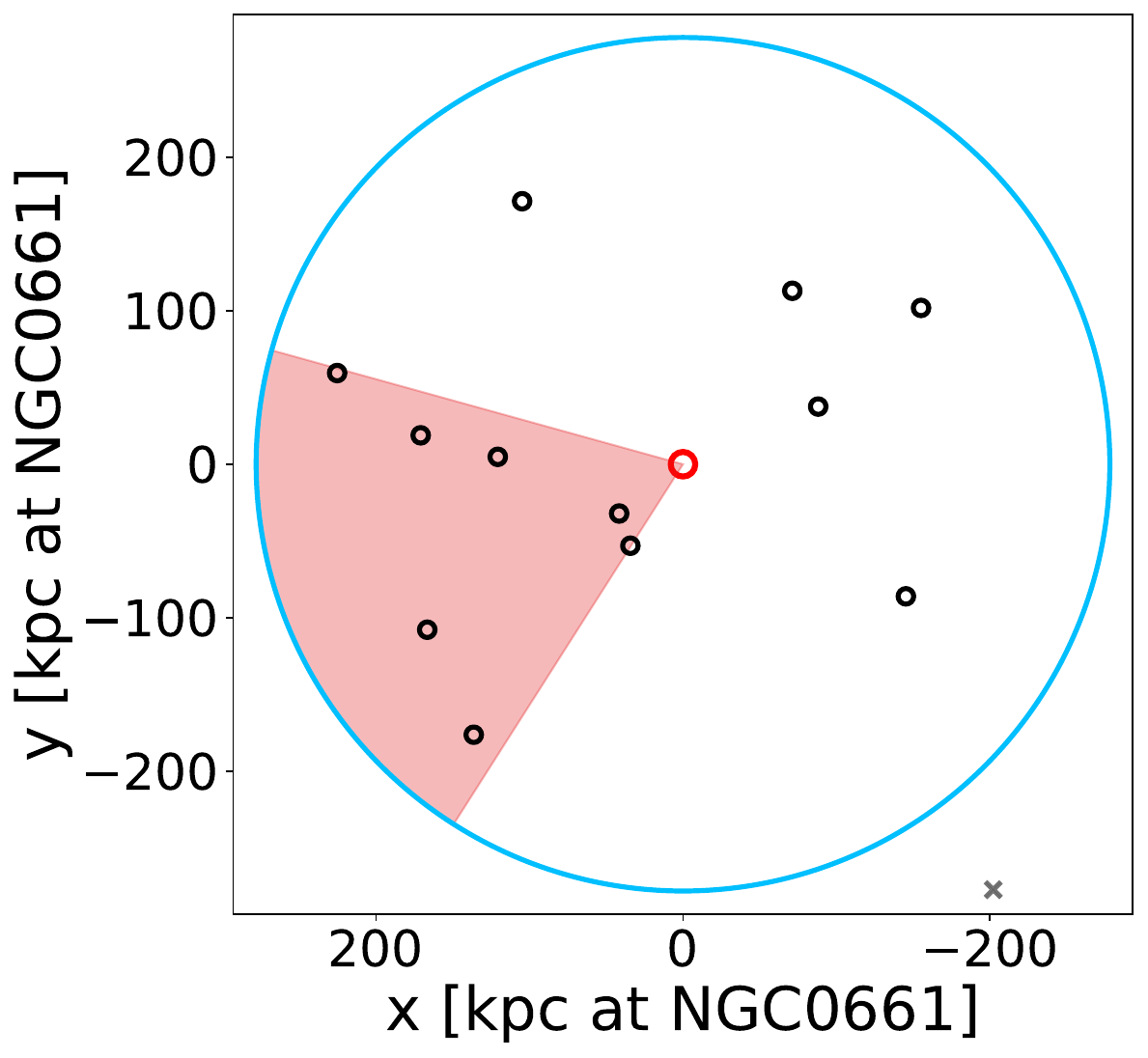}
\caption{The wedge metric applied to two example fields from the MATLAS survey. We plot the systems in a coordinate system with the host (red circle) at the origin. The blue circle illustrates the biggest possible circle around the targeted ETG that fits into the MATLAS FoV. The dwarf satellites (black circles) within this blue region are analyzed in this work, and the ones that fall outside (gray crosses) are discarded to avoid possible radial incompleteness. A wedge, shown as the red-shaded area, is rotated around the host galaxy with increasing opening angles after every rotation. We count the maximum satellite population $N_{\theta}$ and wedge orientation at every opening angle. The corresponding graphs for these systems are shown in Figure \ref{fig:wedge_metric}. The red-shaded areas show the wedges that feature the most unusually high number of satellites when compared to isotropic systems. Left: satellite system NGC0448. Due to the high degree of angular clustering along with few outliers, this configuration is statistically significant under the wedge metric ($p$ = 0.002). Right: system around NGC0661. This system is less unusual compared to isotropic realizations ($p$ = 0.320).}
\label{fig:wedge_examples}
\end{figure*}

This metric differs from the other five because, with the different wedge opening angles, we are dealing with an additional dimension. This warrants extra considerations when quantifying the statistical significance. Initially, we determine the significance of the system's dwarf configuration by applying the metric to all MC realizations, which are generated by assigning random position angles to the satellites. For each wedge opening angle $\theta$, we determine the frequency of isotropic systems that are as or more populated than the most populated wedge in the data at that angle. We do not consider the orientation of these wedges, so the most populated wedge could point in any direction for each realization. Picking the minimum isotropic frequency across all probed wedge opening angles gives the initial $p$-value for the observed degree of lopsidedness. We illustrate this concept in the Appendix, in Figure \ref{fig:wedge_metric}. This procedure extracts the most unusual feature in terms of lopsidedness in the system compared to isotropic realizations. Because some of the isotropic systems might show equal or more extreme wedge populations at different opening angles, we need to quantify the significance of lopsidedness for every isotropic system in the same way. This process works as follows: for every host environment, we treat its isotropic realizations like observed systems. We do this by iteratively picking one of the isotropic random realizations and extracting the most unusual lopsided wedge compared to the other random realizations. The frequency at which this wedge is as or more populated in the other 99,999 isotropic realizations gives a $p$-value for the random realization at hand. In this way, we obtain $p$-values for all isotropic random systems. The overall meta-$p$-value of the observed system is then determined by counting the number of isotropic $p$-values that are smaller or equal to the observed $p$-value and dividing by the number of Monte Carlo realizations. We illustrate the concept of the wedge metric in Figure \ref{fig:wedge_examples} where we show two example systems from the MATLAS survey. The degree of lopsidedness in NGC0448 is statistically significant under the wedge metric, the one in NGC0661 is not.

\section{Results}
\label{sec:results}
In the following, we present the results from our analysis of applying these six different metrics of lopsidedness to isolated galaxy satellite systems in both the Local Volume with the ELVES survey and beyond with the MATLAS survey. We split our results between the two surveys due to their intrinsically different properties such as distance to the systems, confirmed satellite populations, completeness, and FoV. First, we discuss the different metrics on MATLAS and then on the ELVES data. In particular, we will elaborate on the different features of satellite distributions the metrics are more or less sensitive to and what may be the most suitable metric to classify specific distribution properties. 

\begin{figure*}[!htbp]
\centering
\includegraphics[width=0.49\linewidth]{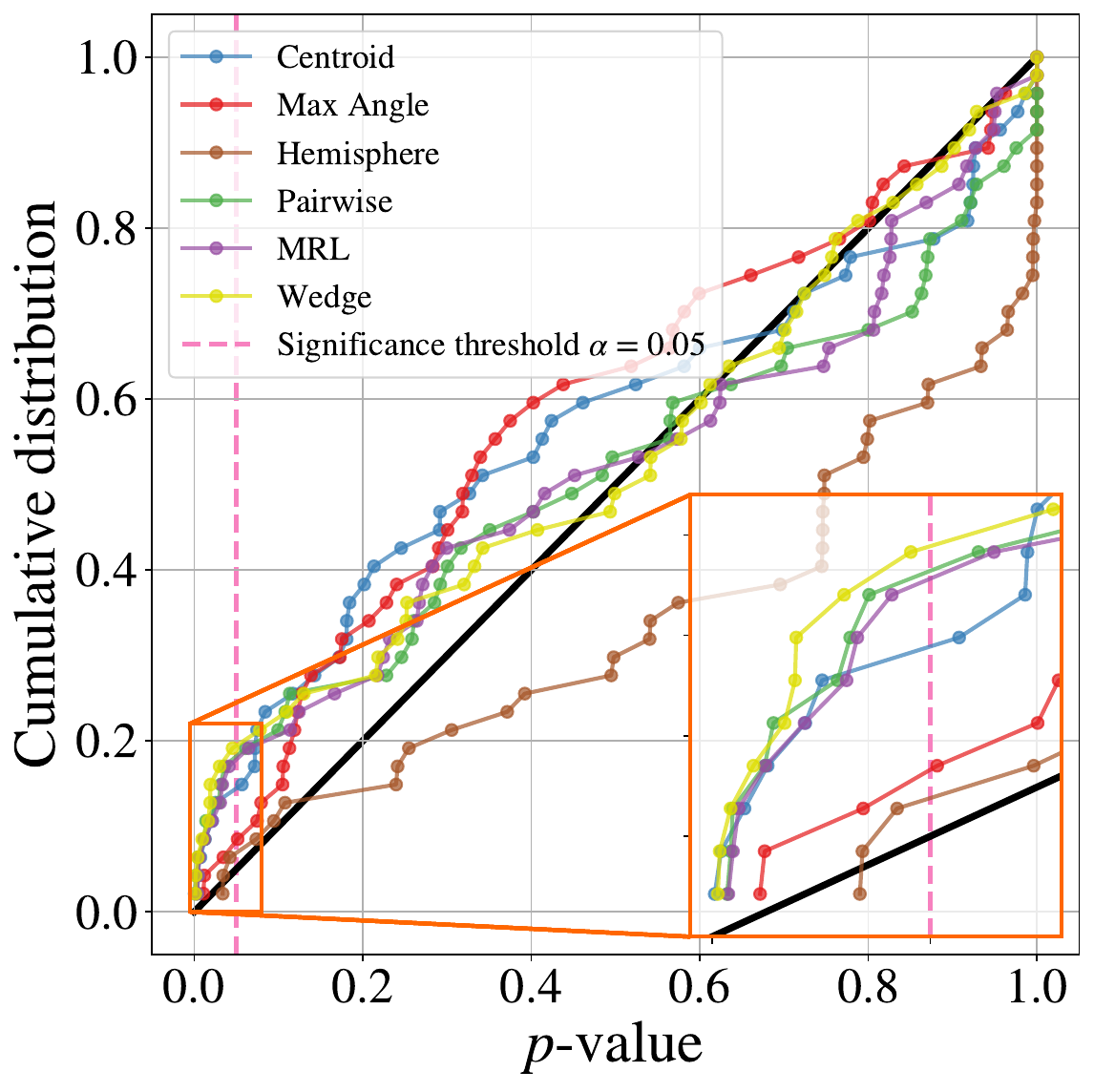}
\includegraphics[width=0.49\linewidth]{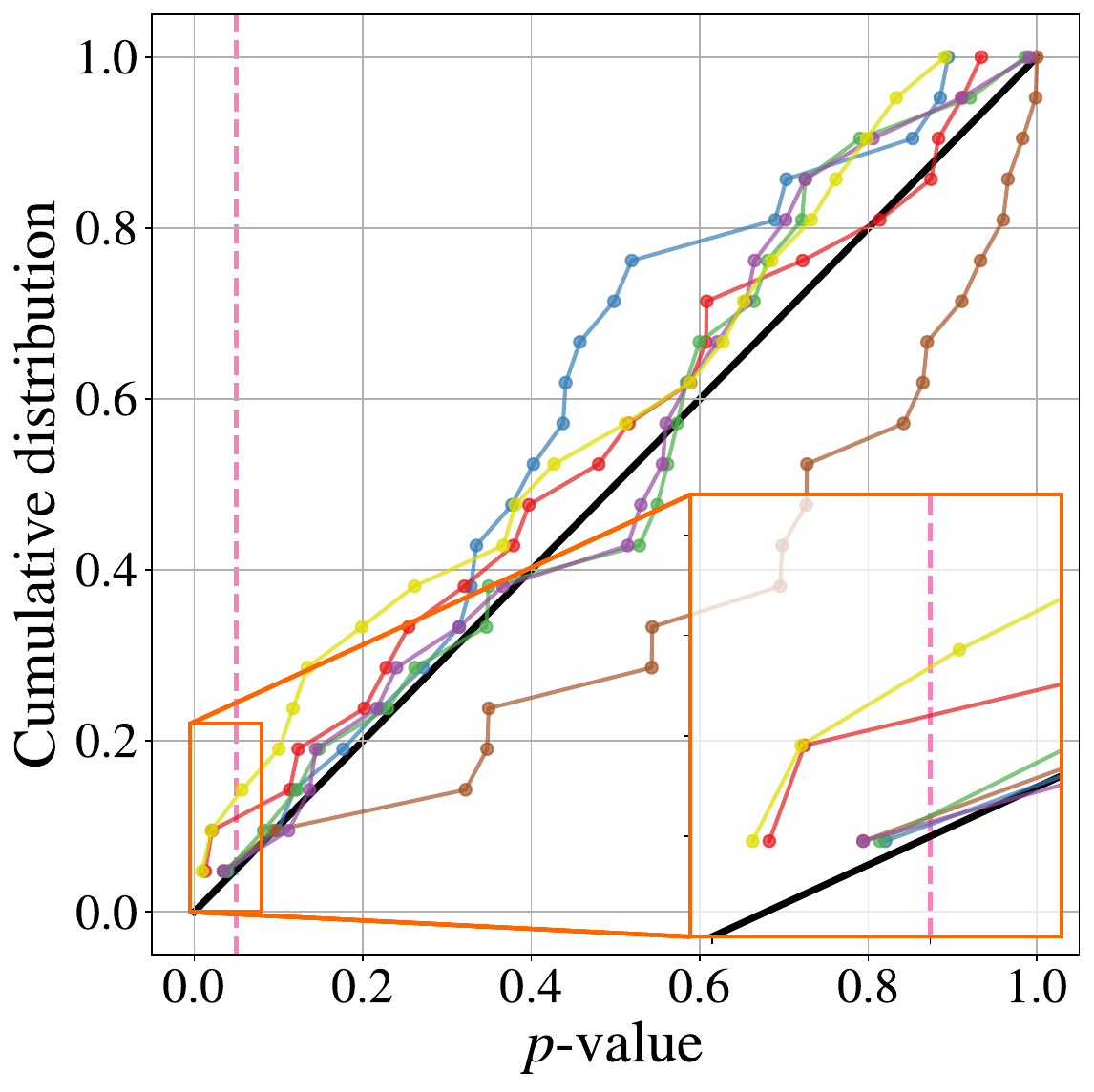}
\caption{The cumulative $p$-value distributions for the different metrics for lopsidedness applied to the isolated MATLAS (left) and ELVES (right) host environments. All $p$-values were calibrated from the comparison with 10$^{5}$ isotropic satellite distributions which have the same number of satellites and the same radial distribution as the dwarfs in the data. The pink dashed lines mark the significance threshold of 0.05, the orange boxes zoom in on the significant regions. The black diagonal lines show the expected distribution from random isotropic systems.}
\label{fig:cumulative_pvals}
\end{figure*}

\subsection{Lopsidedness in isolated MATLAS systems}

We apply the six metrics for lopsidedness described in Section \ref{sec:metrics} to our selected sample of 47 host environments in the MATLAS survey and present these different metrics in columns 6-10 of Table \ref{tab:matlas_prop}. We compare these metric values to the ones found in 10$^{5}$ MC realizations of each system and thus obtain a $p$-value of how common such degrees of lopsidedness are in isotropic satellite distributions. For the wedge metric, we have an additional dimension in the most populated wedge at every wedge opening angle $\theta$. For this metric, we use the minimum frequency of observation-like systems ($N_{\theta_{mc}} \geq N_{\theta_{data}}$) in the MC realizations across all wedge opening angles as an initial $p$-value. As described in Section \ref{sec:wedge_metric} we obtain a $p$-value for each isotropic realization in the same way and calculate a meta-$p$-value for the observed system quantified by how rare the observed $p$-value is compared to the isotropic ones. The $p$-values of the six applied metrics are shown in columns 2-7 of Table \ref{tab:matlas_pvals}. Statistically significant lopsidedness ($p$-value $\leq$ 0.05, i.e. 2$\sigma$) is highlighted in boldface. We show the cumulative $p$-value distribution for the MATLAS systems in the left panel of Figure \ref{fig:cumulative_pvals}. The area of statistical significance is highlighted by an inset zoom-in plot and the expected distribution from random isotropic systems is shown. For our chosen significance criterion of $p$ $\leq$ 0.05 we would expect 5\% of the systems to show $p$-values $\leq$ 0.05 if there is no overall lopsided distribution in the sample. We note no excess of small $p$-values for the max angle and hemisphere metrics, while the centroid, pairwise, MRL, and wedge metrics show such an excess. The excess is most prominent under the wedge metric where 19.1\% of the systems show a statistically significant degree of lopsidedness at the 2$\sigma$ level. The pairwise and MRL metrics follow with $\sim$ 17\% lopsided systems. 

\subsection{Lopsidedness in isolated ELVES systems}

We repeat the analysis by applying the six metrics and determining the statistical significance of varying degrees of lopsidedness to the 21 isolated ELVES host environments. The metric values are presented in columns 6-8 of Table \ref{tab:elves_prop} and the associated $p$-values from the comparison with MC realizations are shown in Table \ref{tab:elves_pvals}. The cumulative $p$-value distribution of the ELVES systems is shown in the right panel of Figure \ref{fig:cumulative_pvals}. In this sample, we find 9.5\% (2/21) of the systems to be significantly lopsided under the wedge metric when compared to isotropic MC systems. In contrast to the MATLAS sample, we note no clear excess of low $p$-values in any of the tested metrics. Two systems are found to be unusual under the non-wedge metrics, one of which is also picked up by the wedge metric. The other system is barely above the significance threshold under the wedge metric (0.057).

The discrepancy between the number of lopsided systems in MATLAS and ELVES is particularly interesting in light of recent results from the literature. Both observational and simulation-based evidence suggest that lopsidedness is particularly prevalent among blue hosts \citep{2020ApJ...898L..15B,2023ApJ...947...56S,2021ApJ...914...78W}. Our findings suggest the opposite trend, since MATLAS only surveyed red early-type hosts and ELVES mostly studied blue hosts. 

\begin{figure*}[!htb]
\centering
\includegraphics[width=\linewidth]{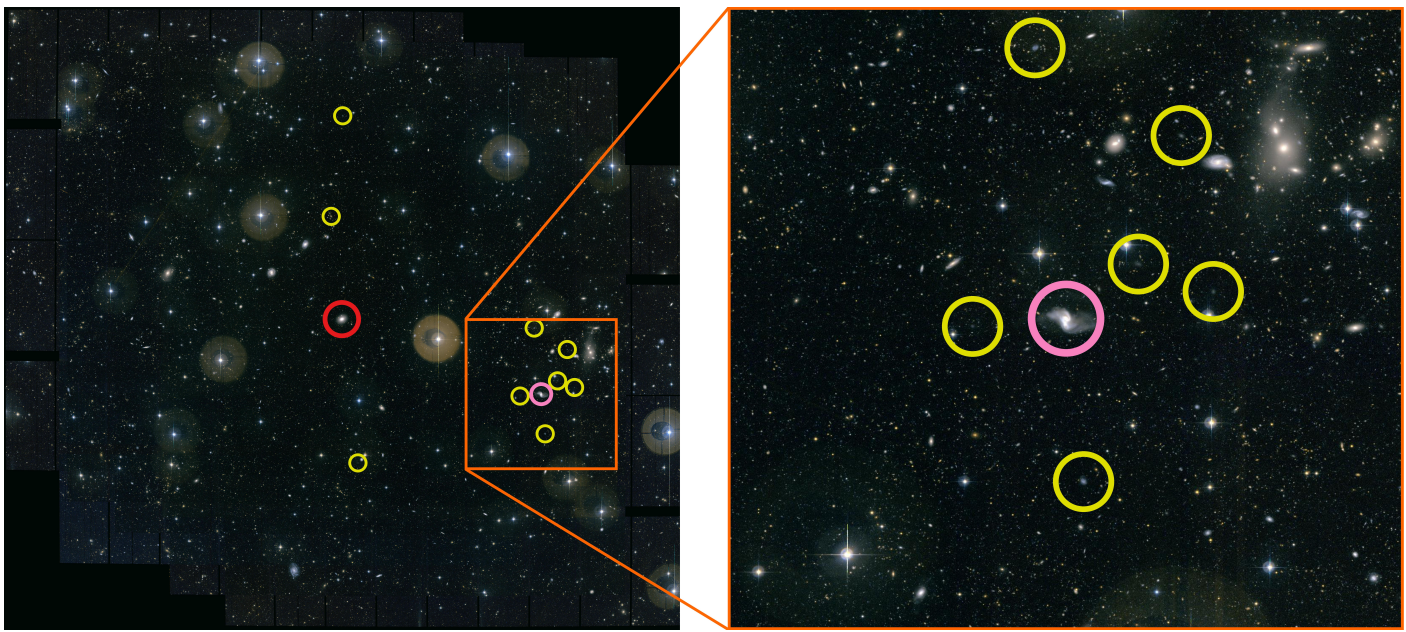}
\caption{Distribution of dwarf galaxies surrounding the MATLAS host NGC3694. Left: MATLAS color image centered on NGC3694 (red circle). The identified dwarf galaxies from \citet{2020MNRAS.491.1901H} are circled in yellow. The background galaxy Mrk 423 is highlighted by the pink circle. Right: zoom-in on the region (orange box) around Mrk 423, showing the spatial clustering of the identified dwarfs around it. Image from the MATLAS survey \citep{2013IAUS..295..358D}.}
\label{fig:example_interlopers}
\end{figure*}

The fact that we find more lopsided systems in the MATLAS sample may in part be caused by the fact that only the ELVES sample offers satellite systems where membership can be established through distance estimates. At this time we cannot be certain that all lopsided dwarf distributions in the MATLAS systems are genuine and do not arise due to interlopers. For instance, in the absence of distance measurements for the MATLAS sample, a scenario may arise where a group of dwarfs appears highly clustered on one side of the host. However, in reality, these dwarfs may be satellites of a background galaxy that happens to be in the line of sight. In Figure \ref{fig:example_interlopers} we show a potential example of this scenario from NGC3694, one of the MATLAS host environments in the studied sample. A group of six dwarf galaxies is highly clustered toward the south-west (bottom-right in the figure) of NGC3694 which has a recessional velocity of 2243\,km/s \citep{2011MNRAS.413..813C}. A zoom-in on this dwarf group reveals that they are spatially distributed around an ongoing galaxy merger Mrk 423, which has a measured line-of-sight velocity of 9549\,km/s \citet{2022ApJS..261....2K}. This puts Mrk 423 at a Hubble distance of $\sim$ 140\,Mpc compared to the 35.2\,Mpc of NGC3694. In this context, it is possible that some -- if not all -- of these six dwarfs are associated with the background galaxy and the lopsided signal arises because of this potential association error.

\begin{figure*}[!htb]
\centering
\includegraphics[width=\linewidth]{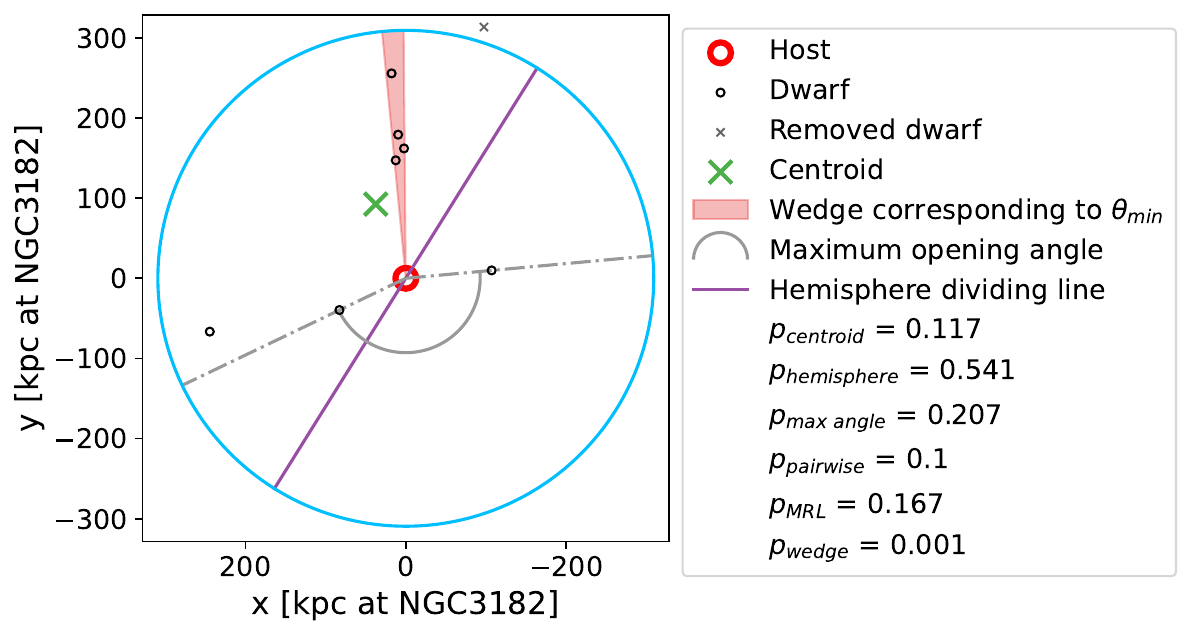}
\includegraphics[width=\linewidth]{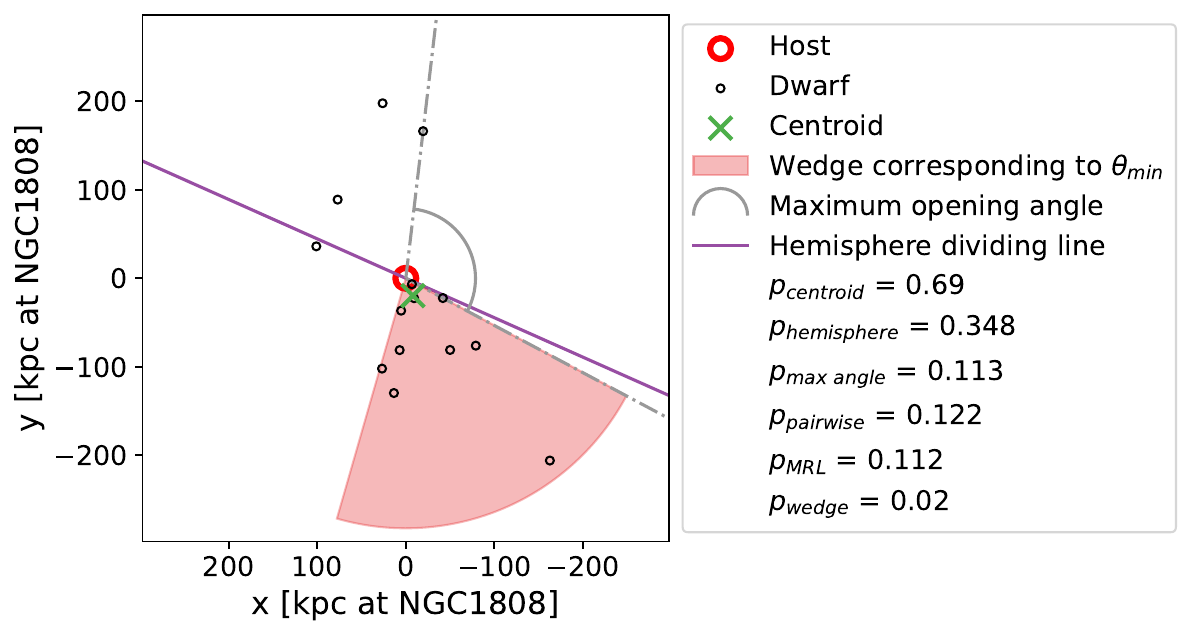}
\caption{Illustration of different behaviors of the six tested metrics on the same systems. We show one example from the MATLAS hosts (NGC3182, top) and one from the ELVES sample (NGC1808, bottom). The satellite systems are shown in a coordinate system with the respective host (red circle) at the origin and the satellites (black circle) at the projected relative distance in kpc from the host. The blue circle in the MATLAS field shows the largest possible circle centered on the host that fits into the MATLAS FoV. We exclude dwarfs beyond this circle (gray cross) due to possible radial incompleteness. The green cross marks the centroid of the satellite system, the red shaded area is the wedge containing the most unusually high number of dwarfs compared to the respective randomized systems, the gray arc shows the maximum opening angle in the system that is devoid of satellites and the purple line illustrates the most extreme possible hemisphere dividing line. We further show the $p$-values under the six different metrics.}
\label{fig:comparison_metric_behaviors}
\end{figure*}

\subsection{Intercorrelations between lopsidedness metrics}

Having tested six different metrics to classify the property of lopsidedness we now discuss our findings about the relationships between the metrics and the various features they are sensitive to. In this subsection, we consider all systems, irrespective of the survey they are found in. We do, however, exclude systems with less than three satellites since otherwise meaningful values cannot be assigned to all metrics.

It is interesting to note that the seven systems (five in MATLAS and two in ELVES) that show $p$ $\leq$ 0.05 in either the max angle or hemisphere metric are also found to be significant by the pairwise, MRL, or wedge metrics. This result can be explained by the fact that both of them can be seen as special cases of the pairwise, MRL, and wedge metrics. All of these quantify the angular distribution and ignore the distance between the satellites and their host. The max angle metric can be excluded as a suitable metric to describe a system's degree of lopsidedness since a single outlier can potentially have a large impact on the value of this metric. While intuitive and straightforward, the hemisphere metric is also redundant in this context since it is a special case of the wedge metric with a constant wedge opening angle of 180$^\circ$ and cannot accurately classify satellite distributions that show an unusual degree of angular clustering. 

The centroid metric is the only metric we tested that takes the projected distance between the satellites and the host into account. Seven systems show a significant satellite centroid shift with respect to their host position. Even though the centroid metric shows a strong correlation with the other metrics, it may not be well suited to describe a system's degree of lopsidedness as a standalone method. A few outliers at large radial distances from the host and on the opposite side of any unusually lopsided satellite distribution would significantly shift the centroid toward the center, diminishing the lopsided signal. 

The pairwise and MRL metrics find the same nine systems to show a significant degree of lopsidedness. Seven of these are also picked up by the wedge metric. One of the other two is just above the significance threshold under the wedge metric (0.057) and the other one has a low $p$-value (0.13) but is well above the threshold.

The wedge metric finds another three to be unusually distributed about their hosts, that are not significant under any of the other metrics. We inspect these cases on an individual basis and point out the following: In two systems, namely NGC3182 and NGC5611, the low $p$-values are based on a low number of satellites (four and three) that are highly clustered in angular space around the host. Some of these dwarfs may stem from a compact group and are falling in together on the host halo. This would be in line with the literature \citep[see e.g.,][]{2020ApJ...898L..15B,2023ApJ...947...56S,2021ApJ...914...78W}, since the lopsided signal in satellite systems both in observations and simulations is, in part, driven by recently infalling dwarf galaxies. The third system that is only picked up by the wedge metric is NGC1808. For this system the pairwise and MRL metrics show low but not statistically significant $p$-values. This could be because the satellites form a flattened structure around the host with ten on one side and four on the other. Because these four are located directly opposite to the bulk of the satellite population, they introduce maximal angular separations to the other ten dwarfs that drive the lopsided distribution. This increases the mean pairwise angle and decreases the MRL, thus diminishing the lopsided signal. Hence, these two metrics, similar to the centroid metric, also suffer from such outliers. Since the centroid metric also factors in satellite distances to the host and most of the satellites on one side are close to the host, the opposed dwarfs pull the centroid further toward the host, erasing the lopsided signal. We, therefore, note that the wedge metric for the most part encompasses the capabilities of the pairwise and MRL metrics while being less sensitive to outliers. 

In Figure \ref{fig:comparison_metric_behaviors} we show two of these systems that are only found to be statistically significant under the wedge metric to illustrate how the tested metrics applied to the same systems can yield different results given their individual sensitivities to various forms of lopsidedness. In the top panel of Figure \ref{fig:comparison_metric_behaviors} we show NGC3182, an example from the MATLAS systems. As discussed before, the system features four satellites that are almost aligned. Unlike the other metrics, the wedge is not influenced by the remaining three dwarfs and can extract this unusual feature. In the bottom panel, we show NGC1808, an example from the ELVES hosts. This example illustrates the sensitivity of the centroid metric to outliers on the opposite side of the lopsided distribution particularly well, as it yields a high $p$-value of 0.69. The distribution would therefore be considered consistent with being isotropic under this metric. The wedge metric classifies the distribution as very unusually lopsided with a $p$-value $\sim$0.02. As mentioned in Section \ref{sec:pairwise}, we have tested the pairwise metric by using the median instead of the mean and noticed overall no significant changes to the $p$-values. In the few systems where the bulk of the distribution is clustered together and a few outliers are located on the opposite side of the host, the metric delivers lower $p$-values. The only system that becomes significant under the pairwise metric with this change is NGC1808, which is discussed above. Many other systems that are considered significant under the centroid, MRL, or wedge metrics, obtain on average slightly higher $p$-values. This effect is most noticeable at lower satellite numbers. We therefore use the mean for this metric and rely on a combination with the wedge metric to extract these types of lopsided configurations.

It should be noted that for the hemisphere metric the metric values we use to test the significance of lopsidedness can only take discrete values from zero to one. The number of these values varies according to the number of satellites in a given system. Due to the relatively low number of dwarfs in our analyzed systems, there are effects of this discrete nature on the $p$-value distribution. We determine the $p$-values by counting equal or more lopsided systems in the isotropic samples. The equally lopsided configurations are particularly emphasized by discrete metrics which naturally leads to higher $p$-values compared to metrics that take continuous values. We have tested the implications on random samples of satellite systems and see a discrete $p$-value distribution that depends on the number of satellites considered. This differs from the expectation we have in the case of the other metrics in which real numbers are used to compare with MC realizations. For these other metrics, we expect the $p$-values to be distributed uniformly under the null hypothesis that the analyzed samples are consistent with following an overall isotropic distribution. The peak of realizations close to a $p$-value of one for the hemisphere metric can be explained by the low number of satellites per system in these bins. With very few satellites it is fairly unlikely -- or impossible, depending on the number of dwarfs and the distribution in the data -- to find lines in MC realizations that split the population in a less extreme way compared to the observations. 

\begin{figure*}[!htb]
\centering
\includegraphics[width=1\linewidth]{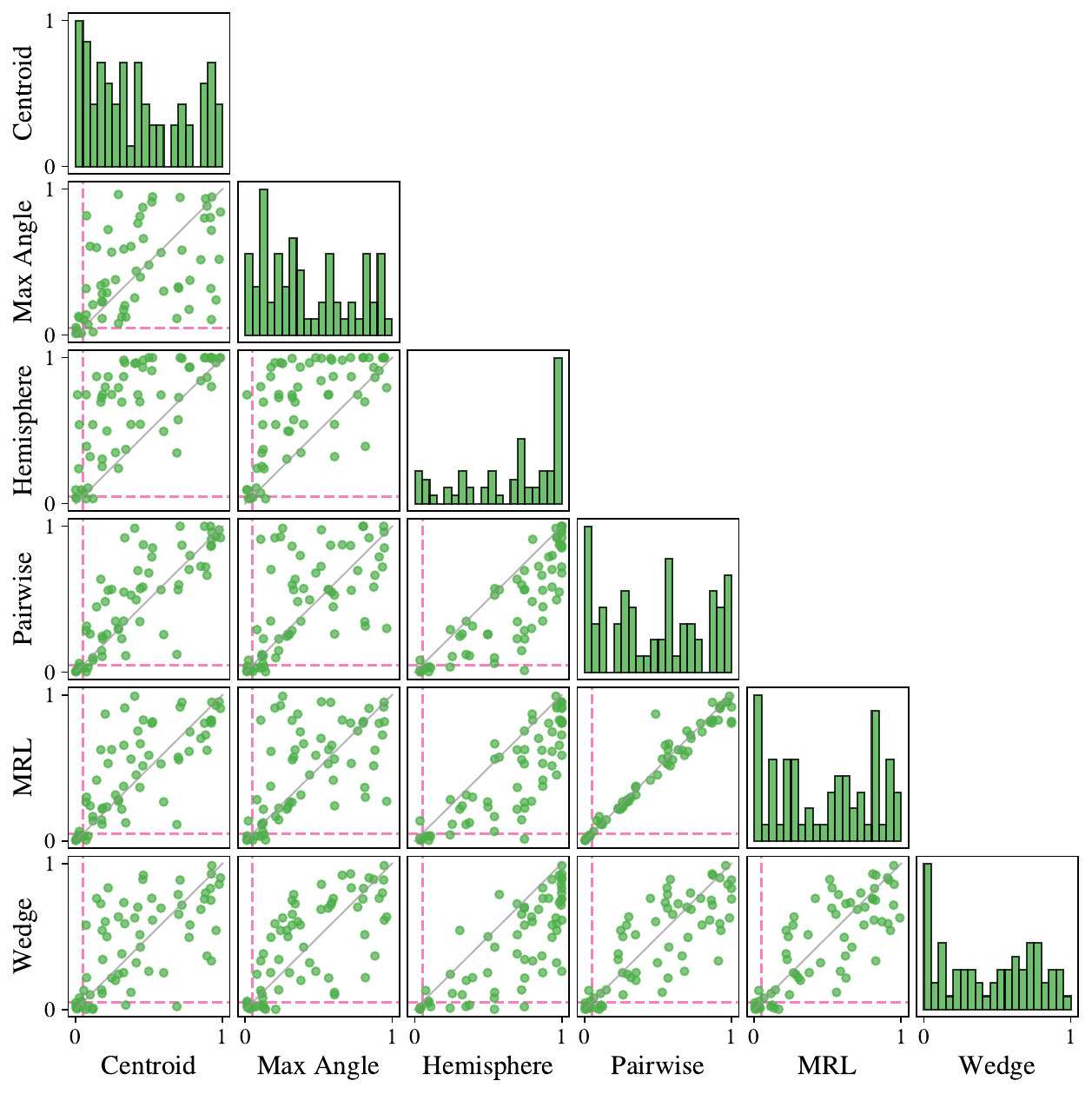}
\caption{Relationships between the $p$-values of the six applied metrics on all systems. Each off-diagonal subplot illustrates the $p$-values for a pair of metrics. The pink dashed lines show the $\alpha$ = 0.05 significance threshold to indicate statistically significant lopsidedness under either or both metrics. The gray diagonal line shows the one-to-one relation. The on-diagonal subplots show the normalized $p$-value distributions under the different metrics. The bin size was chosen to be 0.05 in line with the significance threshold. For this plot we only consider systems with at least three satellites.}
\label{fig:scatter_matrix}
\end{figure*}

To visualize the correlations between the different metrics for lopsidedness and to better understand the features of satellite distributions they are sensitive to, we plot the $p$-values of the metrics against one another in a scatter matrix in Figure \ref{fig:scatter_matrix}. We note a rather large scatter between the $p$-values of most metric pairs. There are pairs with no apparent correlation between them such as the max angle and the centroid metrics. In most cases some degree of correlation is present. The hemisphere metric shows correlations with the other metrics but consistently shows higher $p$-values compared to the other metrics. This may be a result of the effect discussed before. The most significant correlation is present between the MRL and the pairwise metrics, with a Spearman correlation coefficient $\rho$ = 0.97, and the weakest one is observed between the max angle and centroid metrics with $\rho$ = 0.54.

\subsection{Considerations of low number statistics}

Some of the satellite systems we analyze in this work only feature very few dwarf galaxies ($\sim$2-3). We choose to consider all systems equally without making any cuts at some specific number of dwarf galaxies since our MC simulations intrinsically handle low-number statistics. To showcase the effect the satellite population has on the potential statistical significance of a given system, we plot the $p$-values from the wedge metric for a given system against its satellite population. The results are shown in the top panel of Figure \ref{fig:pval_correlations}. In general, only starting from a population of seven satellites per system statistically significant lopsidedness is observed. Only a single system (NGC3032) below this population threshold, where all three satellites in the FoV are clustered to a small angular space, yields a statistically significant $p$-value.

\subsection{Varying degrees of radial coverage}

\begin{figure}[!htb]
\centering
\includegraphics[width=\linewidth]{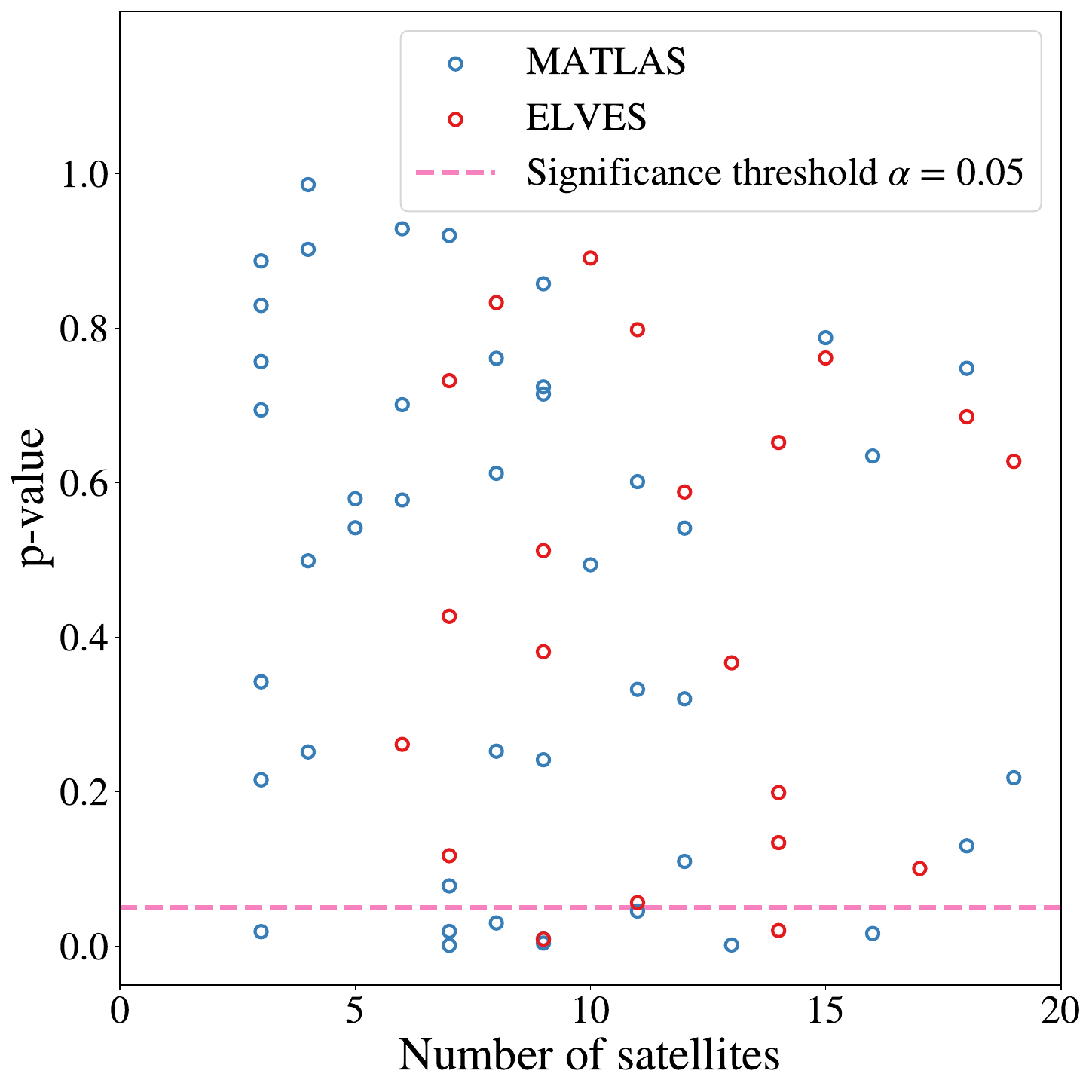}
\includegraphics[width=\linewidth]{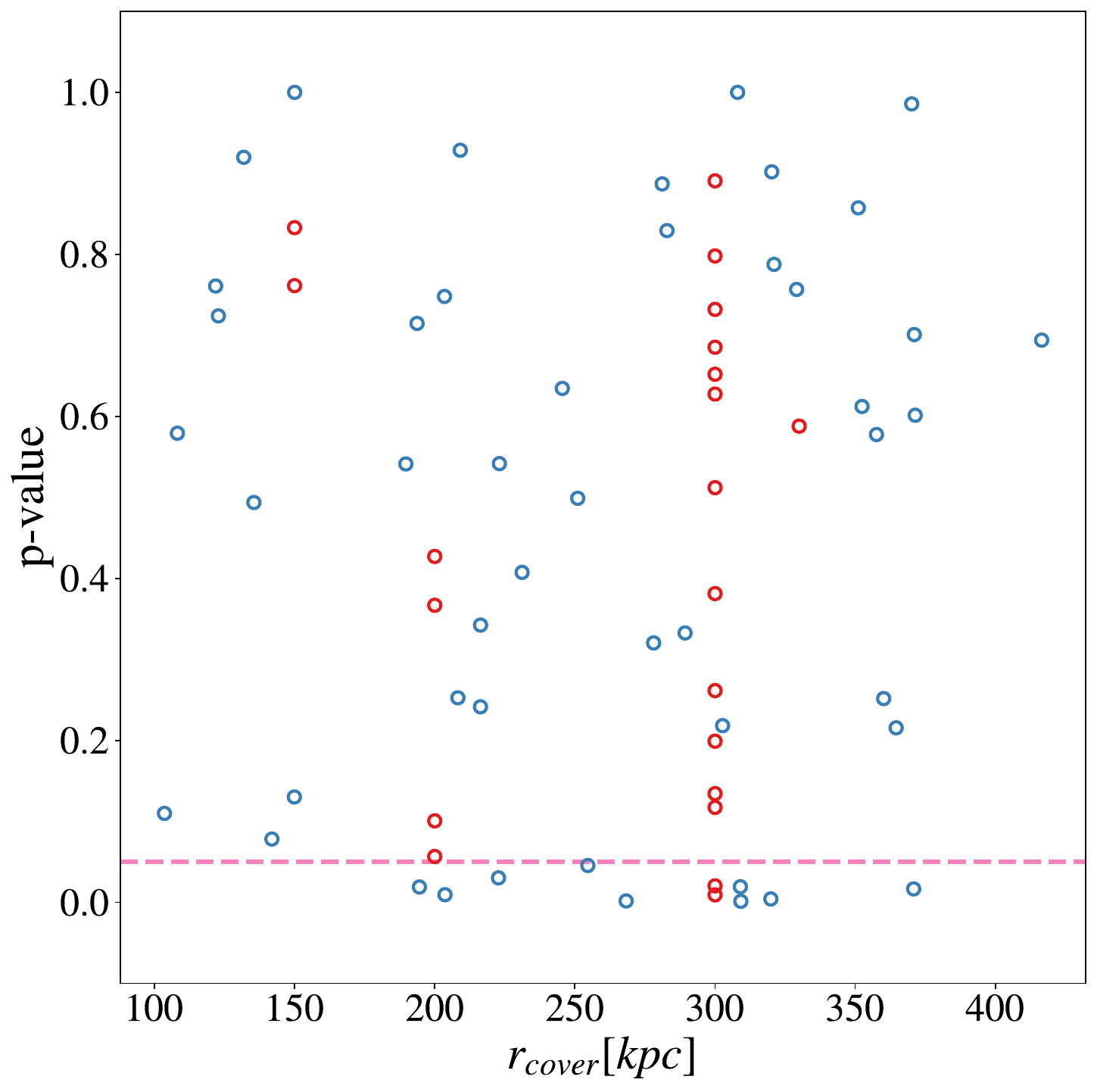}
\caption{Correlations between wedge metric $p$-values and satellite system properties. Top: wedge metric $p$-values as a function of the number of satellites in a given satellite system. Bottom: wedge metric $p$-values as a function of the surveyed radius around the host galaxy in kpc at the distance of the host.}
\label{fig:pval_correlations}
\end{figure}

As mentioned in Section \ref{sec:sample_selection}, the satellite systems were surveyed with varying radial coverage around the hosts. In the MATLAS survey, this coverage is determined by the physical extent of the largest circle around the observed ETG that fits into the rectangular MATLAS FoV. This extent depends on the distance to the respective host. We estimate the virial radii $r_{vir}$ of the ETGs in our sample by calculating the stellar mass from the $K_{s}$ magnitude reported in \citet{2011MNRAS.413..813C}, converting it to halo mass following \citet{2021NatAs...5.1069C} for red hosts and using the virial theorem. With these calculations, we obtain a mean virial radius of $\sim$340\,kpc. Most of the MATLAS hosts ($\sim$80\%) are covered to 170\,kpc ($\sim r_{vir}$/2), $\sim$20\% are covered to the estimated mean virial radius. In ELVES all hosts were surveyed to at least half of the estimated virial radius ($\sim$150\,kpc) and $\sim$70\% to the virial radius. Since previous studies \citep[][]{2020ApJ...898L..15B,2021ApJ...914...78W} report a correlation between lopsidedness and satellites at larger distances to their host ($\sim$300\,kpc or roughly the virial radius), we investigate the presence of such an effect in this study. In the bottom panel of Figure \ref{fig:pval_correlations} we show the $p$-values from the wedge metric as a function of the covered radius around the host in kpc. Consistent with the hypothesis that lopsidedness is driven by satellites located at the outer edges of the systems, our results show no statistically significant lopsidedness in systems where observational coverage did not exceed 195\,kpc ($\sim$ 2/3$r_{vir}$). To further quantify this effect, we restrict the radius around the host galaxies and only include dwarfs that fall within a projected 200\,kpc at the distance of the host. We repeat our analysis on this radius-limited dataset and obtain corresponding $p$-values for the six tested metrics. The outcome of this test is presented in Table \ref{tab:perc_full_restricted}, where we compare the percentage of significantly lopsided systems in the original and in the restricted datasets. The frequency of significantly lopsided systems is considerably diminished across all metrics. The percentage of systems that are lopsided under any of the metrics drops from 20.6\% using the full available coverage to 14.7\% on the restricted dataset. \\

\begin{table*}
\captionsetup{width=.8\linewidth}
\caption{Percentages of statistically significant lopsided systems across the entire dataset under six tested metrics, comparing full available coverage and coverage restricted to a radius of 200\,kpc at the distance of the host.}\label{tab:perc_full_restricted}
\centering
\begin{threeparttable}
\begin{tabular}{llllllll}
\toprule
& Centroid & Max angle & Hemisphere & Pairwise & MRL & Wedge & Combined \\
 & (\%) & (\%) & (\%) & (\%) & (\%) & (\%) & (\%) \\
 & (1) & (2) & (3) & (4) & (5) & (6) & (7) \\
\midrule
Full cover & 8.8 & 7.4 & 5.9 & 13.2 & 13.2 & 16.2 & 20.6 \\
Restricted cover & 7.4 & 4.4 & 0.0 & 7.4 & 8.8 & 7.4 & 14.7 \\
\bottomrule
\end{tabular}
Columns: (1-6) percentage of statistically significant lopsided distributions for the different metrics: centroid, maximum opening angle, hemisphere, mean pairwise angle, mean resultant length (MRL), and wedge, (7) percentage of systems that are found significant under any of the metrics. \\
\end{threeparttable}
\end{table*}

\section{Summary and conclusions}
\label{sec:summary}

We have tested the 2D distribution of the satellites around 68 isolated host environments in the MATLAS and ELVES surveys (complete down to an absolute magnitude of $M\sim-9$) for signs of lopsidedness using six different metrics. These metrics are described in Section \ref{sec:metrics}. Most of these methods are sensitive to particular characteristics of satellite distributions. We find that the wedge metric, the maximum number of satellites contained in a wedge centered on the host, in general, is the most versatile, capturing the features seen in other metrics. This manifests such that statistically significant lopsided dwarf distributions in any other metric are also picked up by the wedge metric in 8/11 ($\sim$73\%) cases. In two out of the three remaining systems, the wedge $p$-value is just above the significance threshold. We obtain a measure of the statistical significance of the varying degrees of lopsidedness under the different metrics by calibrating the results with MC simulations using 10$^{5}$ isotropic realizations of our satellite systems. These isotropic systems are generated by keeping the radial distribution of the satellites around a given host and assigning a random new position angle for every dwarf. The $p$-values are obtained by counting MC realizations that match or exceed the degree of lopsidedness found in the data. Our main results are as follows:

\begin{enumerate}
    \item For the 47 MATLAS host environments 6/47 ($\sim$12.7\%), 3/47 ($\sim$6.4\%), 3/47, 8/47 ($\sim$17.0\%), 8/47 and 9/47 ($\sim$19.1\%) satellite systems are statistically significant ($p$-value $\leq$ 0.05, i.e. 2$\sigma$ level) under the centroid, max angle, hemisphere, pairwise, MRL and wedge metrics, respectively.
    \item For the 21 isolated ELVES systems we find 1/21 ($\sim$4.7\%), 2/21 ($\sim$9.5\%), 1/21, 1/21, 1/21 and 2/21 systems to be statistically significant. Contrary to results from the literature, we thus find more lopsided systems around red, rather than blue hosts.
    \item There are correlations of widely varying strength between the six metrics. We find the strongest relationship between the pairwise and MRL metrics, while the weakest link exists between centroid and max angle.
    \item We find that the max angle and hemisphere metrics are redundant in this context, as they are special cases of other metrics and do not provide additional information on particular features of lopsidedness. Either one of the pairwise and MRL metrics is sufficient in providing a good measure of lopsidedness, as the two metrics are so closely correlated. The centroid metric additionally provides valuable insight as it takes into account the distance of the satellites to the host. Overall, the wedge metric is most robust to outliers and thus performs the best in finding unusual angular distributions of satellites. In addition to quantifying the overall lopsidedness in a system, it is also capable of detecting highly clustered satellites in angular space that may fall in on the host halo as a group.
    \item By inspecting the relation between the number of satellites per system and the $p$-values under the wedge metric we find that -- with the exception of a single system (NGC3032) -- at least seven satellites are required to make a robust statement about the degree of lopsidedness in a given system.
    \item We find no statistically significant lopsidedness under the wedge metric in systems where the observations covered less than approximately two-thirds of the estimated virial radius ($\sim$300\,kpc). An additional test where we restrict the radius around each host to 200\,kpc ($\sim$2/3$r_{vir}$) reveals that the percentage of significantly lopsided systems notably drops across all tested metrics. In agreement with previous studies, this result suggests that satellites at larger projected distances from the host, likely recent additions to the systems, are the primary drivers of lopsidedness in these isolated satellite configurations.
\end{enumerate}

Overall we find that $\sim$20.6\% of our analyzed satellite systems show a significant degree of lopsidedness under any of the tested metrics. Under the wedge metric alone, 16.2\% are significantly lopsided. As illustrated in Figure \ref{fig:cumulative_pvals} the proportion of $p$-values falling below the significance threshold of 0.05 exceeds the 5\% typically expected in a sample with no inherent lopsidedness. This suggests a notable excess of lopsided satellite systems compared to what would be anticipated. The $p$-value distributions show an excess of low values for the centroid, pairwise, MRL, and wedge metrics. We would argue, however, based on our results, that the wedge metric is most suitable as a standalone metric to capture the features by which lopsidedness is characterized. Combining the centroid, wedge and either pairwise or MRL metrics likely delivers the most well-rounded picture of lopsidedness. On the other hand, the hemisphere metric is unable to detect an unusual concentration of satellites in a small range of position angles and the max angle metric can easily miss the true degree of lopsidedness since any outlier carries significant weight with this metric.

In a future study, we will compare these results with expectations from cosmological simulations to gain a clearer picture of how rare such lopsided distributions are in a $\Lambda$CDM context. 

\begin{acknowledgements} 
We thank the referee for the constructive report, which helped to clarify and improve the manuscript. O.M. and N.H. are grateful to the Swiss National Science Foundation for financial support under the grant number PZ00P2\_202104. N.H. thanks A. Kaindl for helpful inputs regarding the figures. M.S.P. and K.J.K. acknowledge funding via a Leibniz-Junior Research Group (project number J94/2020). This research made use of hips2fits,\footnote{https://alasky.cds.unistra.fr/hips-image-services/hips2fits} a service provided by CDS. This research has made use of "Aladin sky atlas" developed at CDS, Strasbourg Observatory, France. 
\end{acknowledgements}

\bibliographystyle{aa}
\bibliography{bibliographie}

\begin{appendix}

\section{Supplementary materials}
\label{appendix}

\subsection{System plots}

In this section, we show the system plots for all host environments studied in this work. In Figure \ref{fig:matlas_satellite_dists1} we show the systems from the MATLAS survey and in Figure \ref{fig:elves_satellite_dists1} the ELVES systems. Where applicable, we show the metric attributes for visualization purposes: the centroid of the distribution, the maximum opening angle devoid of satellites, the hemisphere dividing line that splits the system maximizing the population contrast, and the wedge that hosts the most unusually high population of satellites compared to the isotropic realizations.

\subsection{Wedge metric examples}

\begin{figure*}[ht]
    \centering
    \includegraphics[width=0.49\linewidth]{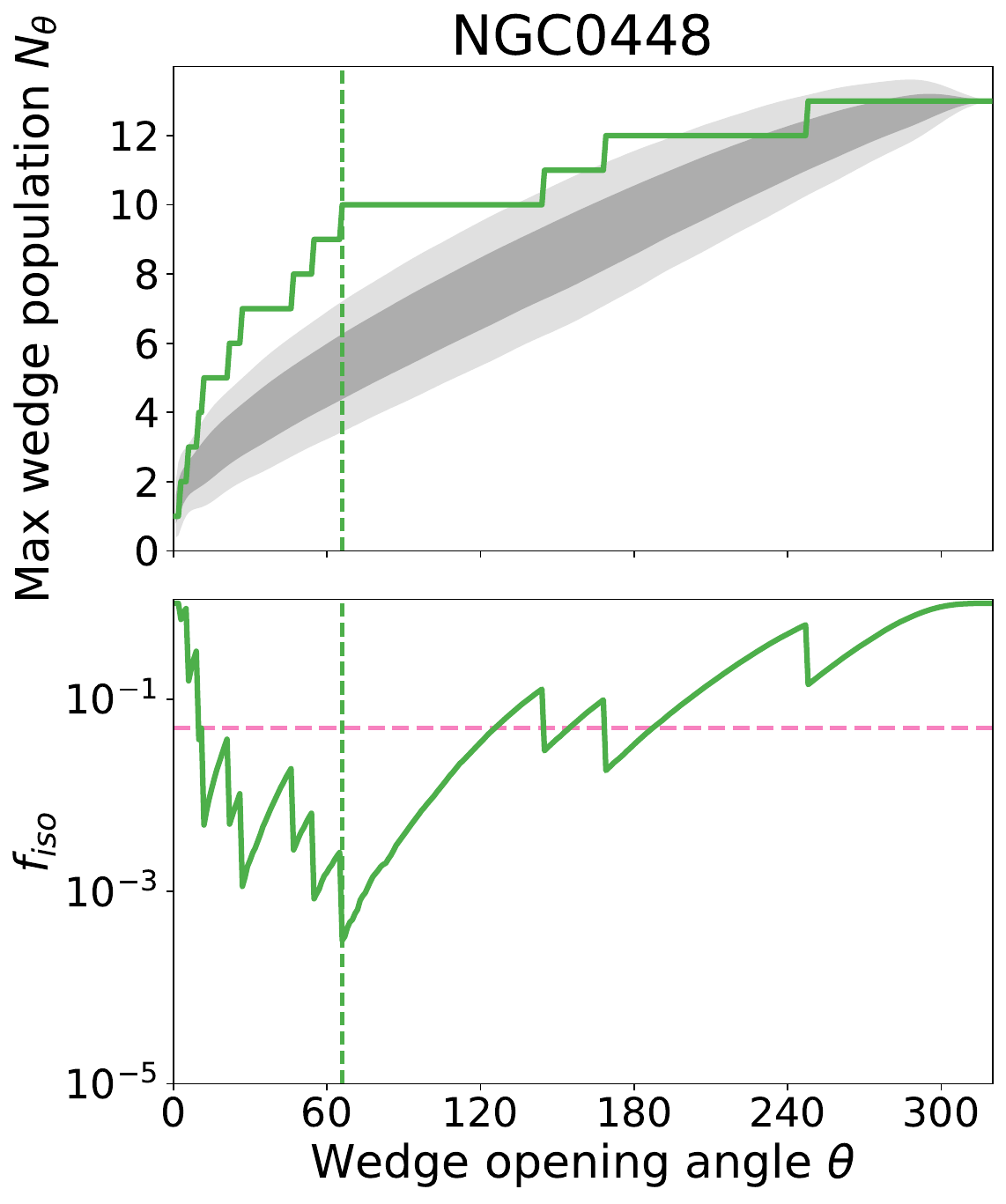}
    \includegraphics[width=0.49\linewidth]{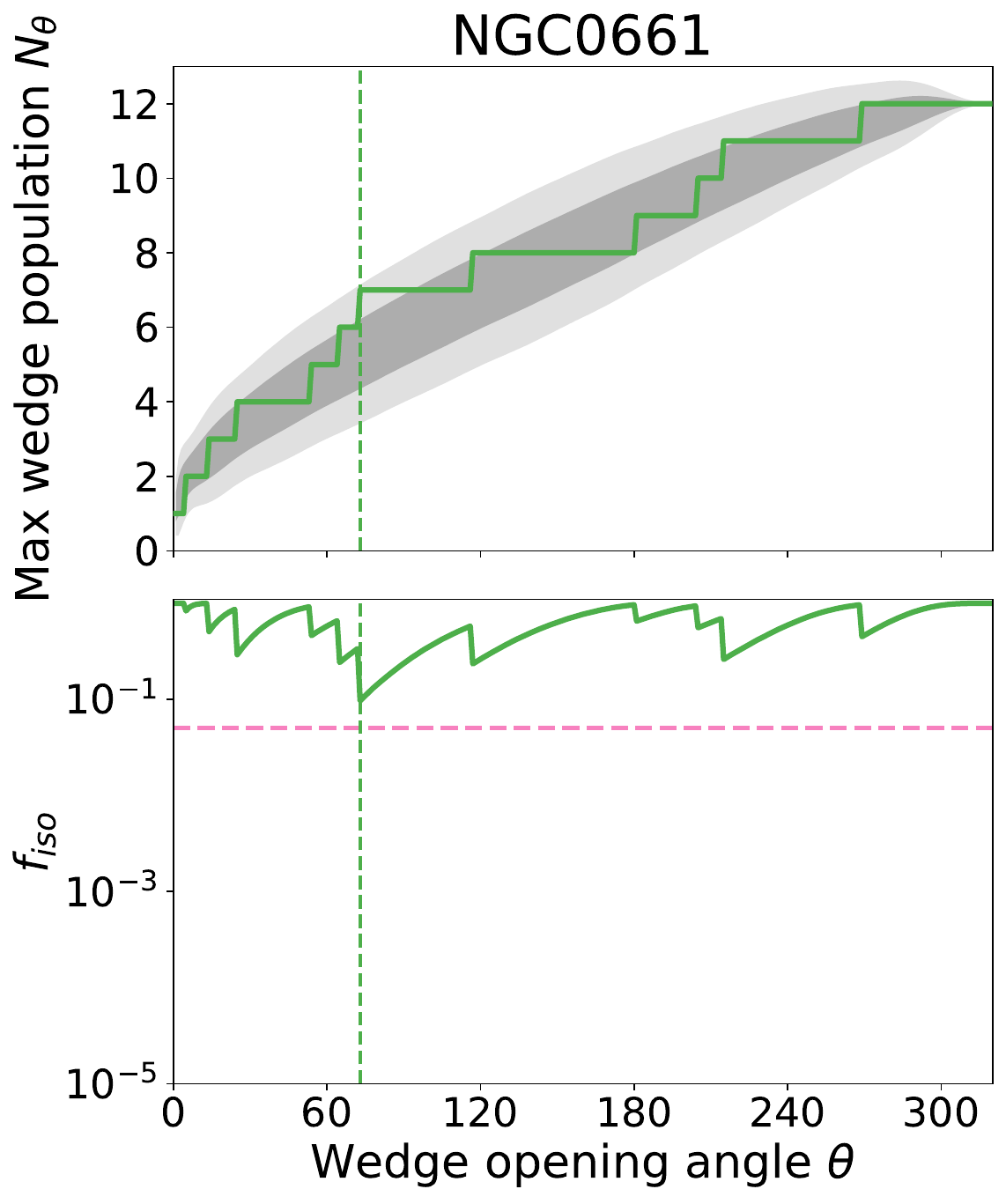}
\caption{Maximum wedge population $N_{\theta}$ (top) and isotropic frequency \textsf{f}$_{iso}$ (bottom) versus the wedge opening angle $\theta$ for two examples from the MATLAS survey. Top: the green solid line shows the wedge populations in the observed systems. The gray-shaded areas illustrate the results of the 10$^{5}$ isotropically distributed satellite systems with the same number of observed satellites. The dark and light gray areas mark the 1$\sigma$ and 2$\sigma$ intervals of the Monte Carlo realizations, respectively. The green dashed vertical lines mark the wedge opening angles for which the observed systems show an unusually high satellite population compared to the isotropic random realizations. Bottom: the green solid lines show the frequency \textsf{f}$_{iso}$ (logarithmic scale) at which a given wedge opening angle is equally or more populated in isotropic systems as it is in the observed one. The pink dashed horizontal lines show the significance threshold $\alpha$ = 0.05. Thus the dwarf configuration around NGC0448 (left) is found to be statistically significant. The one around NGC0661 (right) is not significantly unusual as the isotropic frequency does not dip below the 0.05 threshold. We extract the individual $p$-values for the specific observed configuration with these graphs. The overall meta-$p$-value for a given system is determined after repeating this procedure for all isotropic random systems and comparing the resulting $p$-values.
}
    \label{fig:wedge_metric}
\end{figure*}

In this section, we show the concept of the wedge metric using two examples from the MATLAS survey. For this purpose, we choose the same systems as in Figure \ref{fig:wedge_examples}: one statistically significant (NGC0448) and one non-significant (NGC0661) satellite distribution. In Figure \ref{fig:wedge_metric} we show in the upper panels the maximum wedge population $N_{\theta}$ as a function of the wedge opening angle $\theta$ and in the bottom panels the frequency of isotropic systems that match or exceed the observed $N_{\theta}$ at a given opening angle.

\begin{figure*}[ht]
    \centering
    \includegraphics[width=6cm]{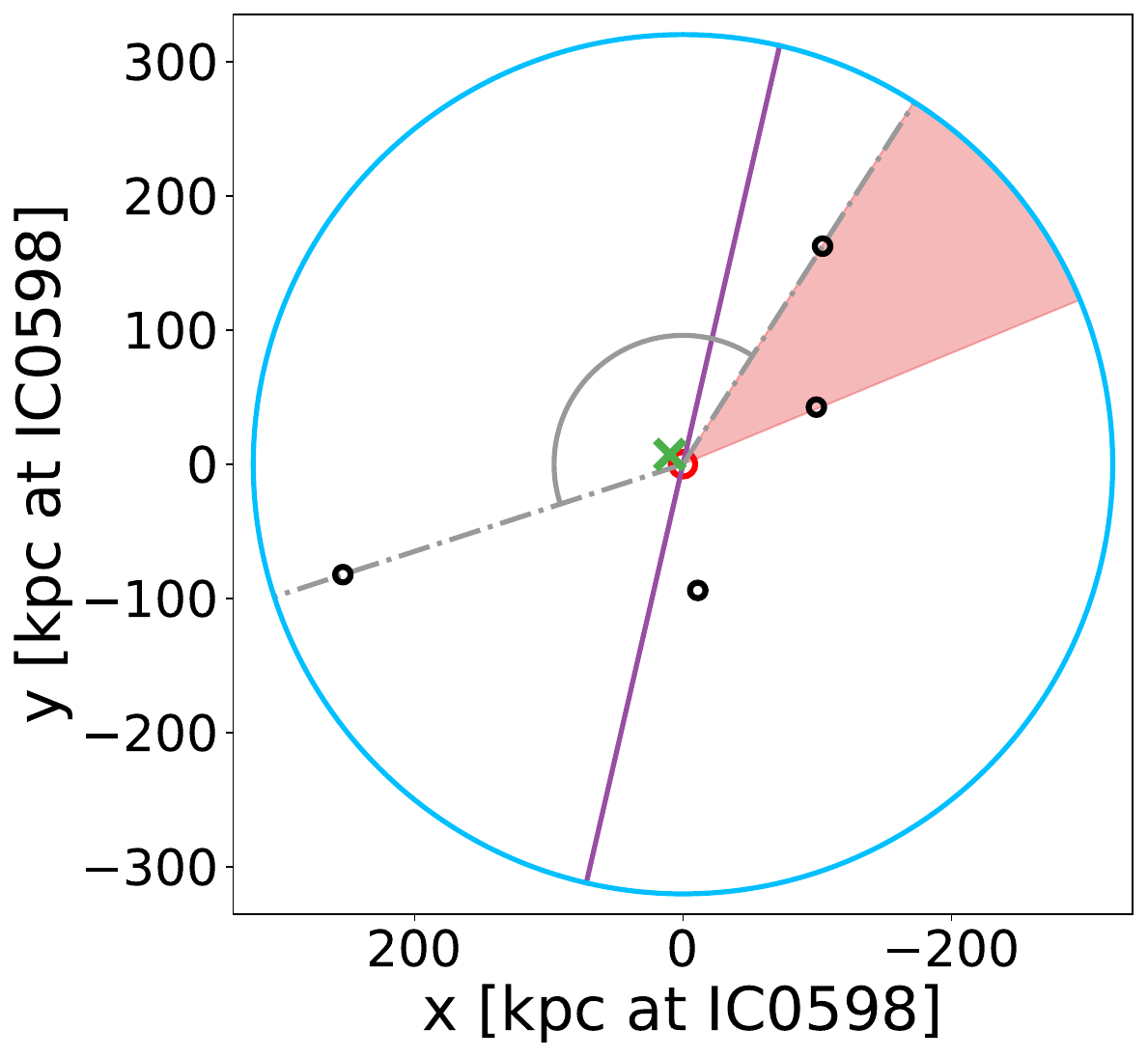}
    \includegraphics[width=6cm]{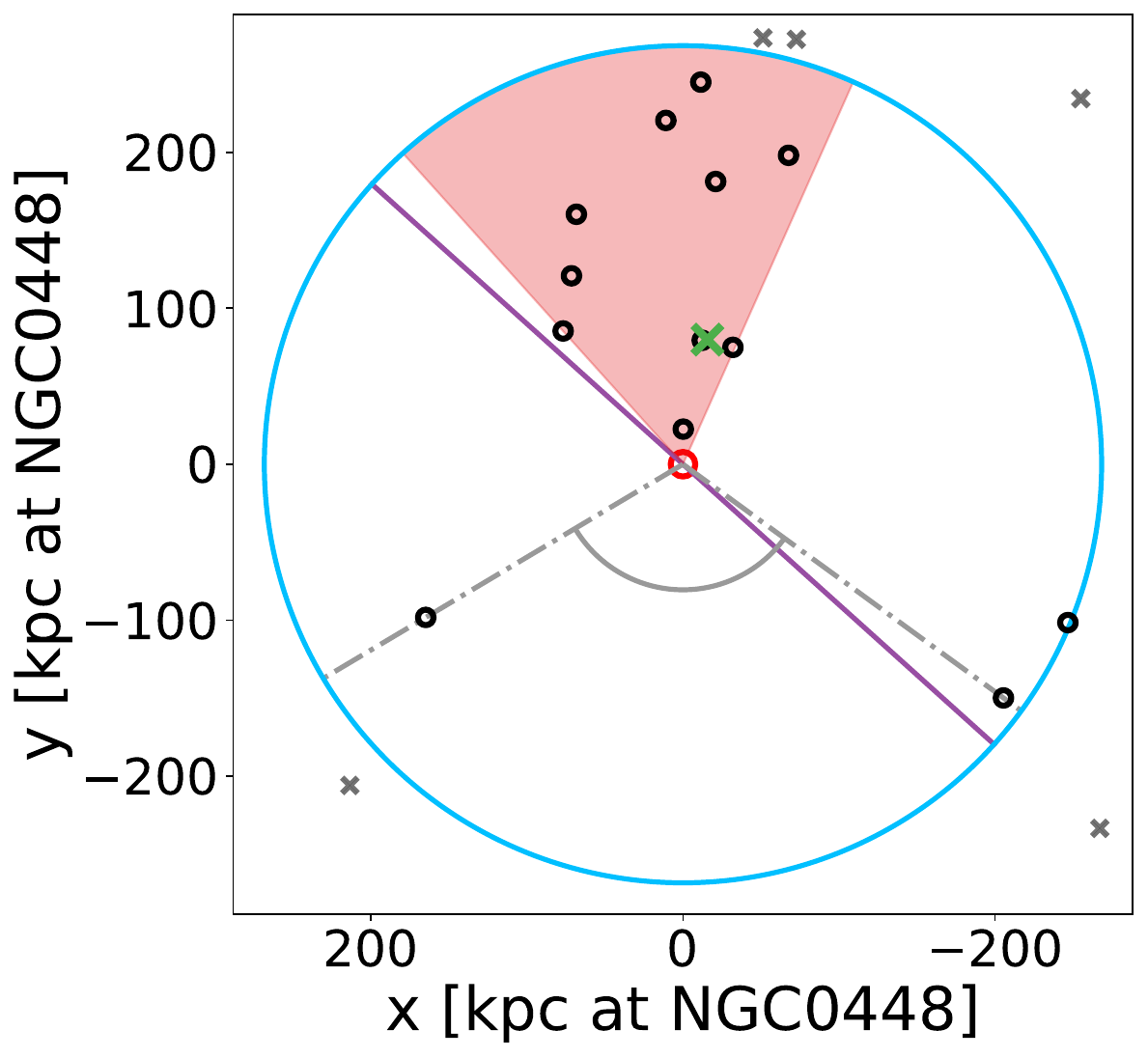}
    \includegraphics[width=6cm]{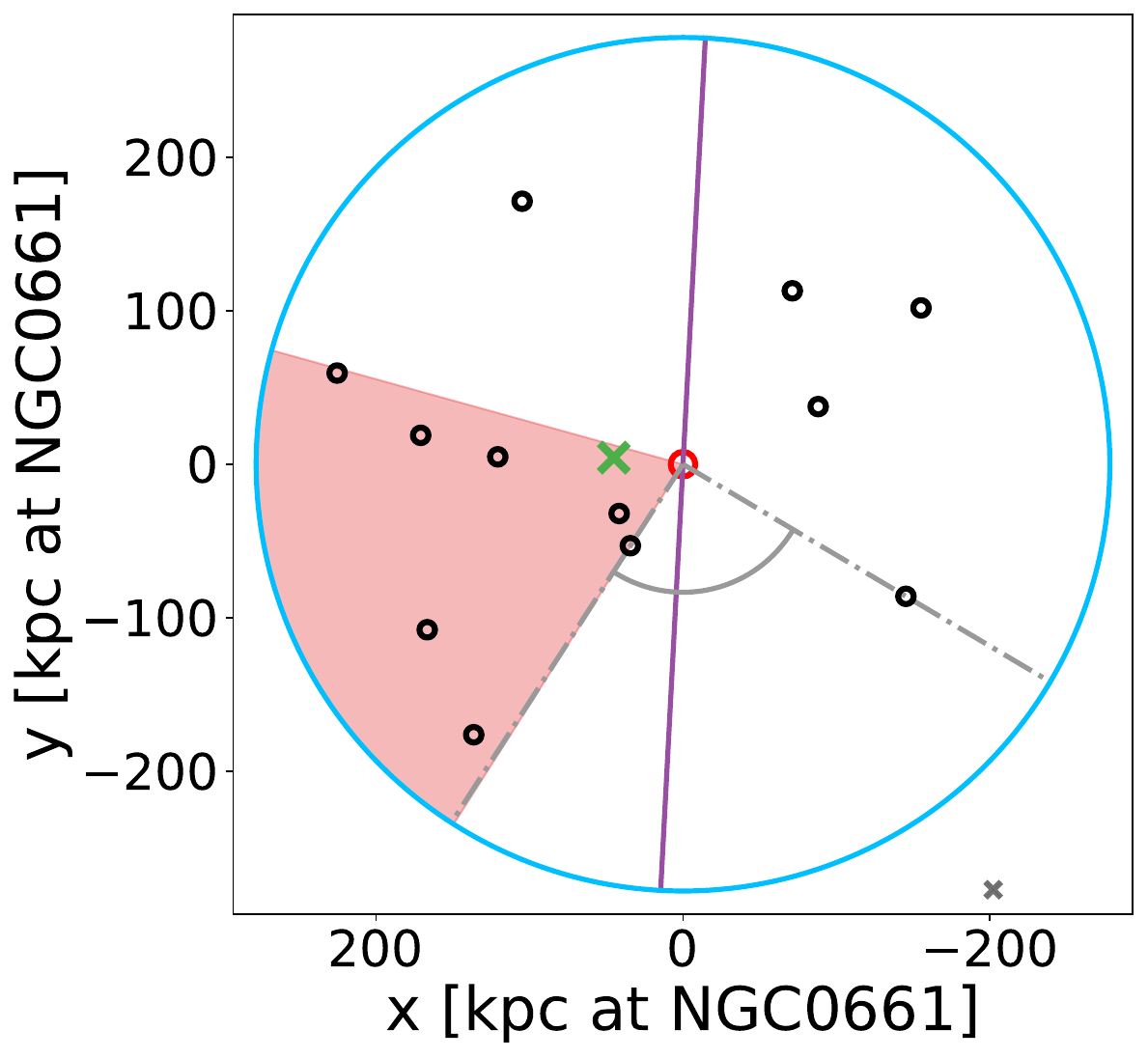}
    \includegraphics[width=6cm]{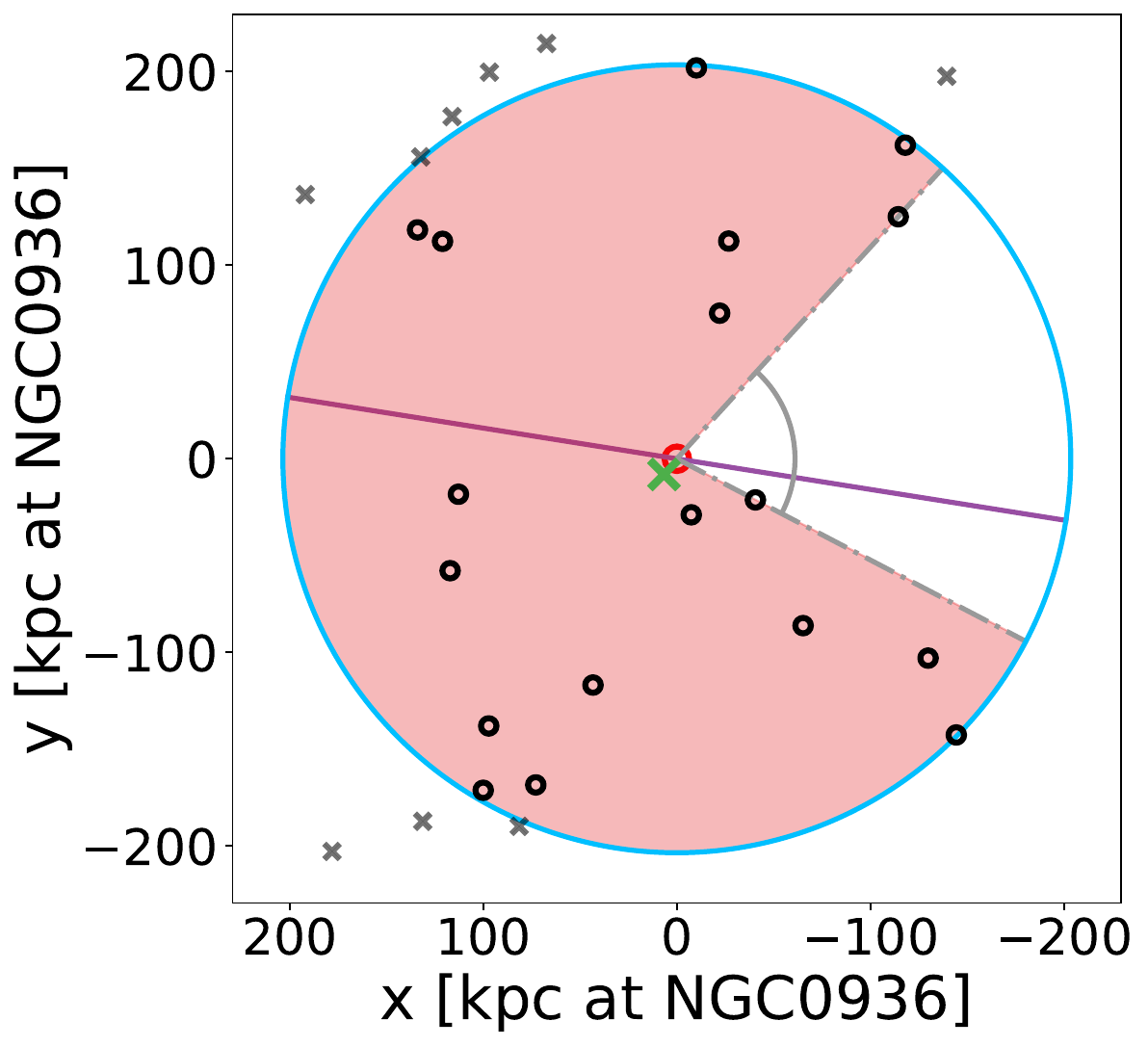}
    \includegraphics[width=6cm]{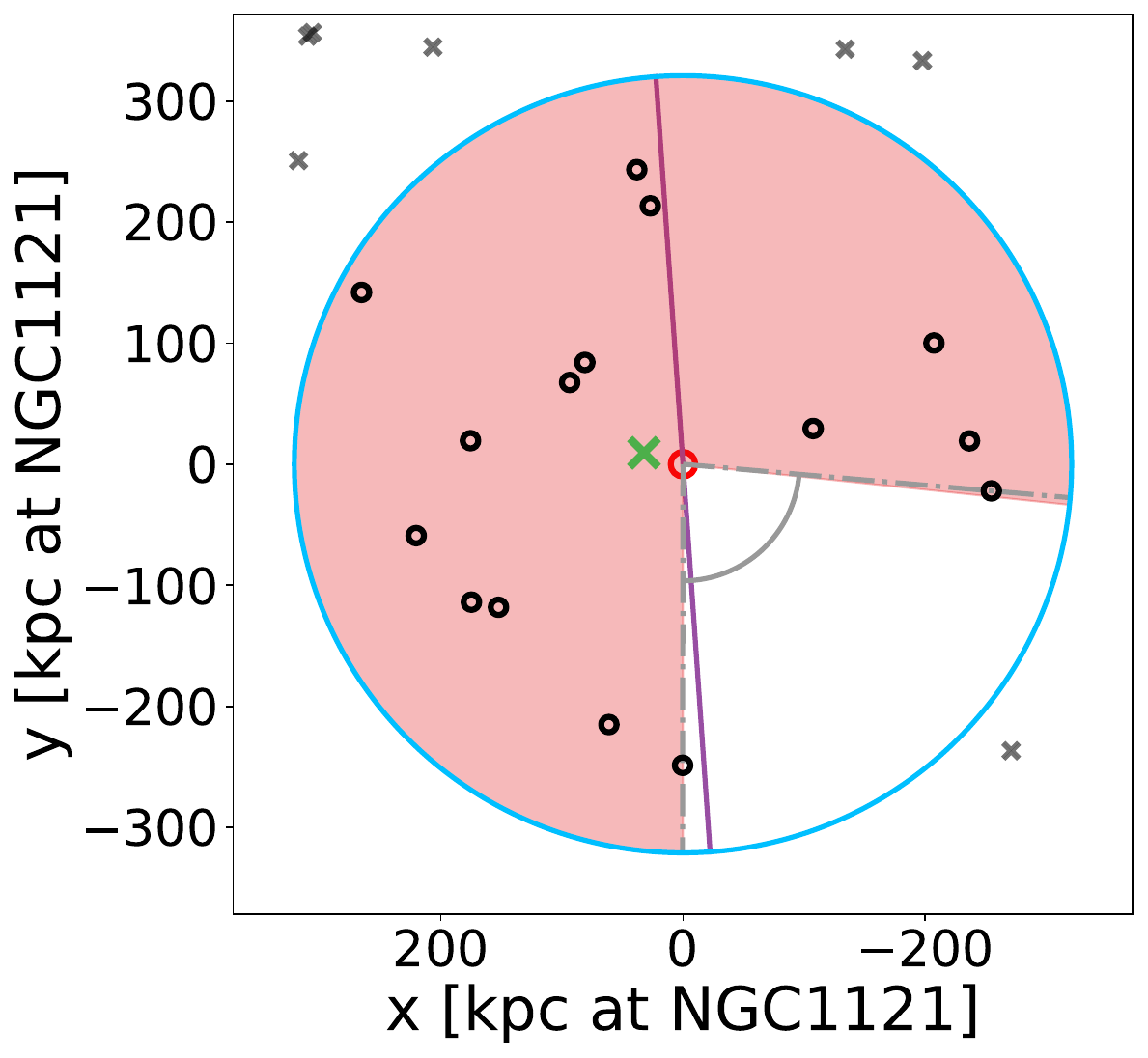}
    \includegraphics[width=6cm]{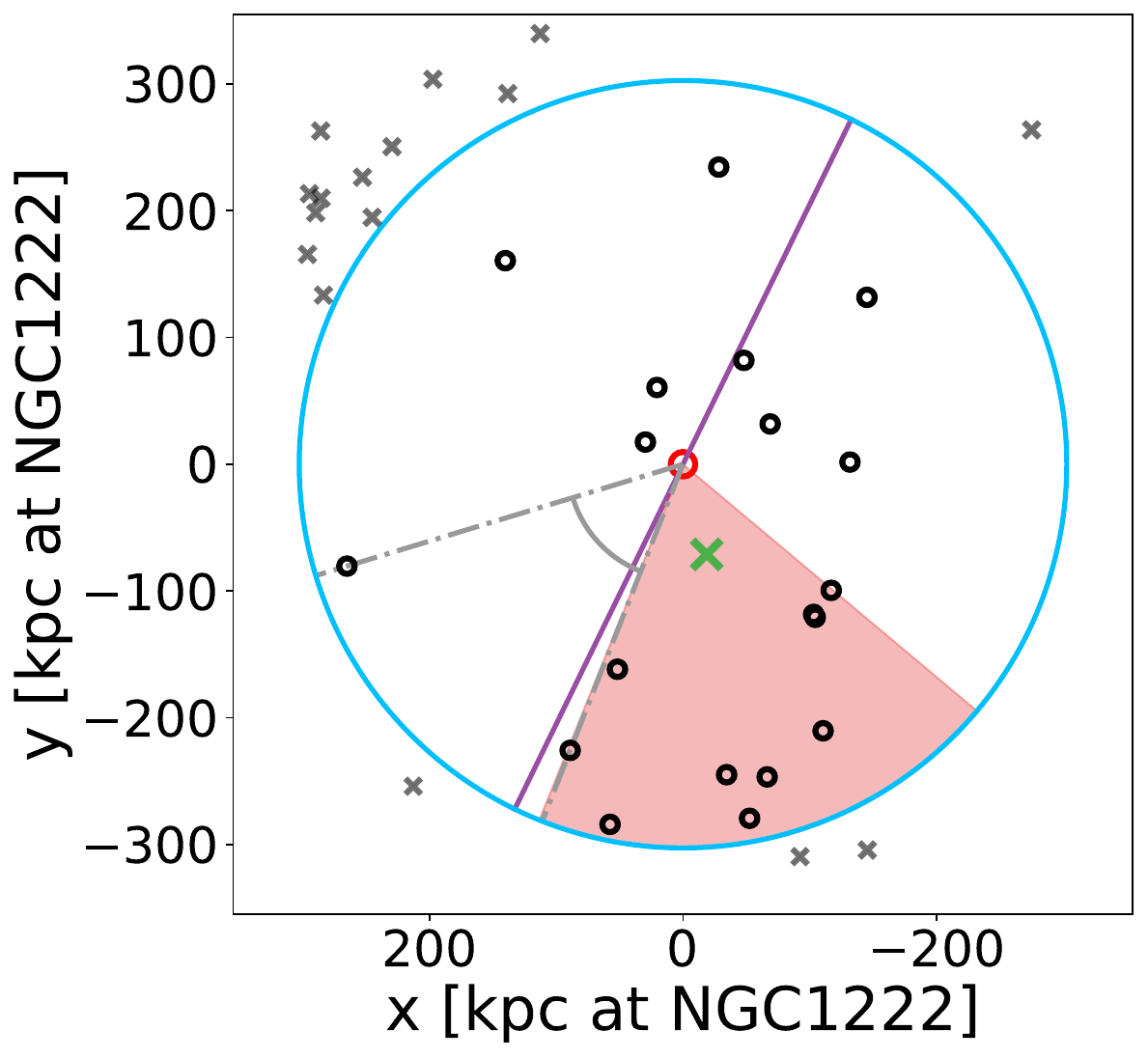}
    \includegraphics[width=6cm]{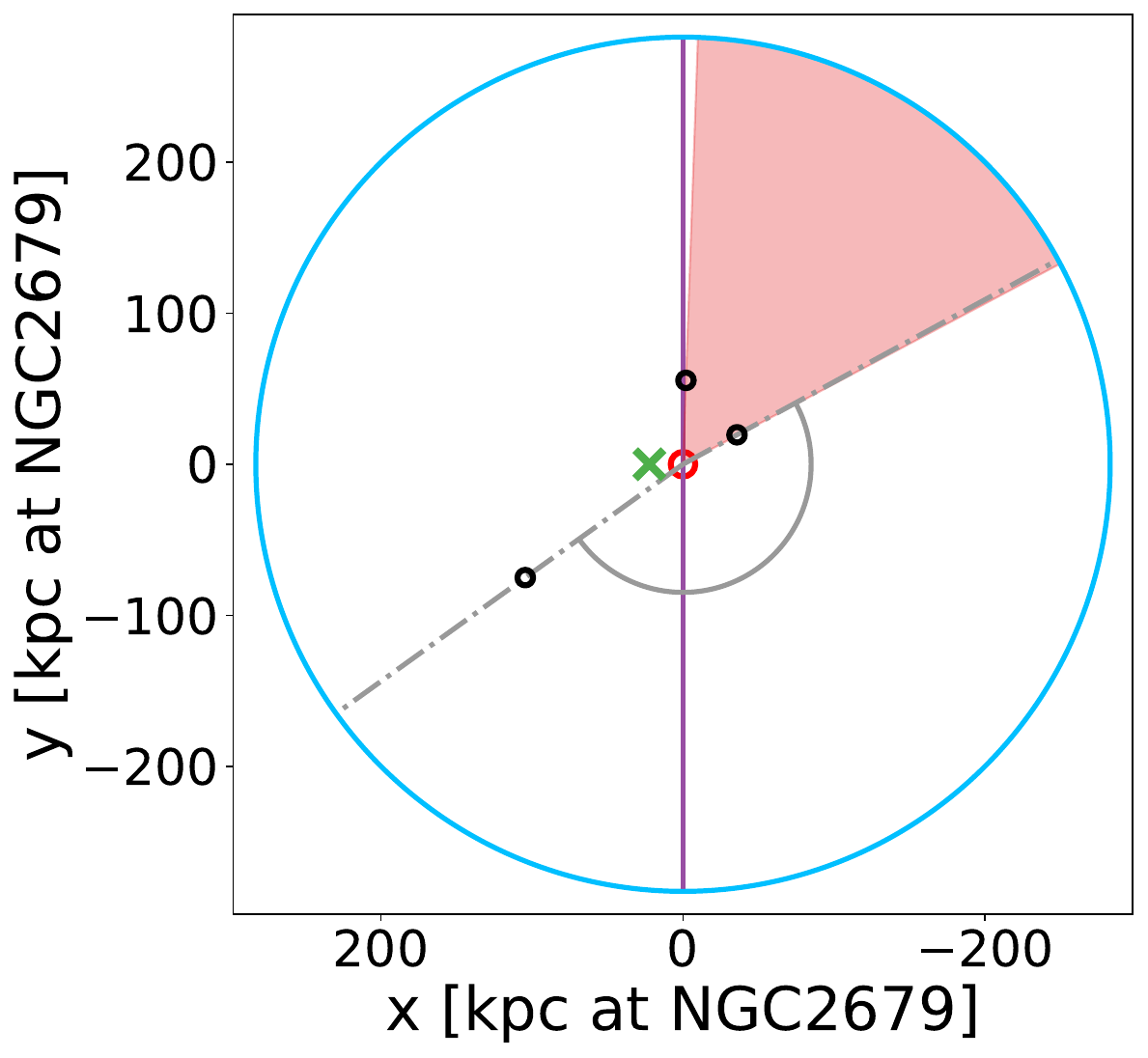}
    \includegraphics[width=6cm]{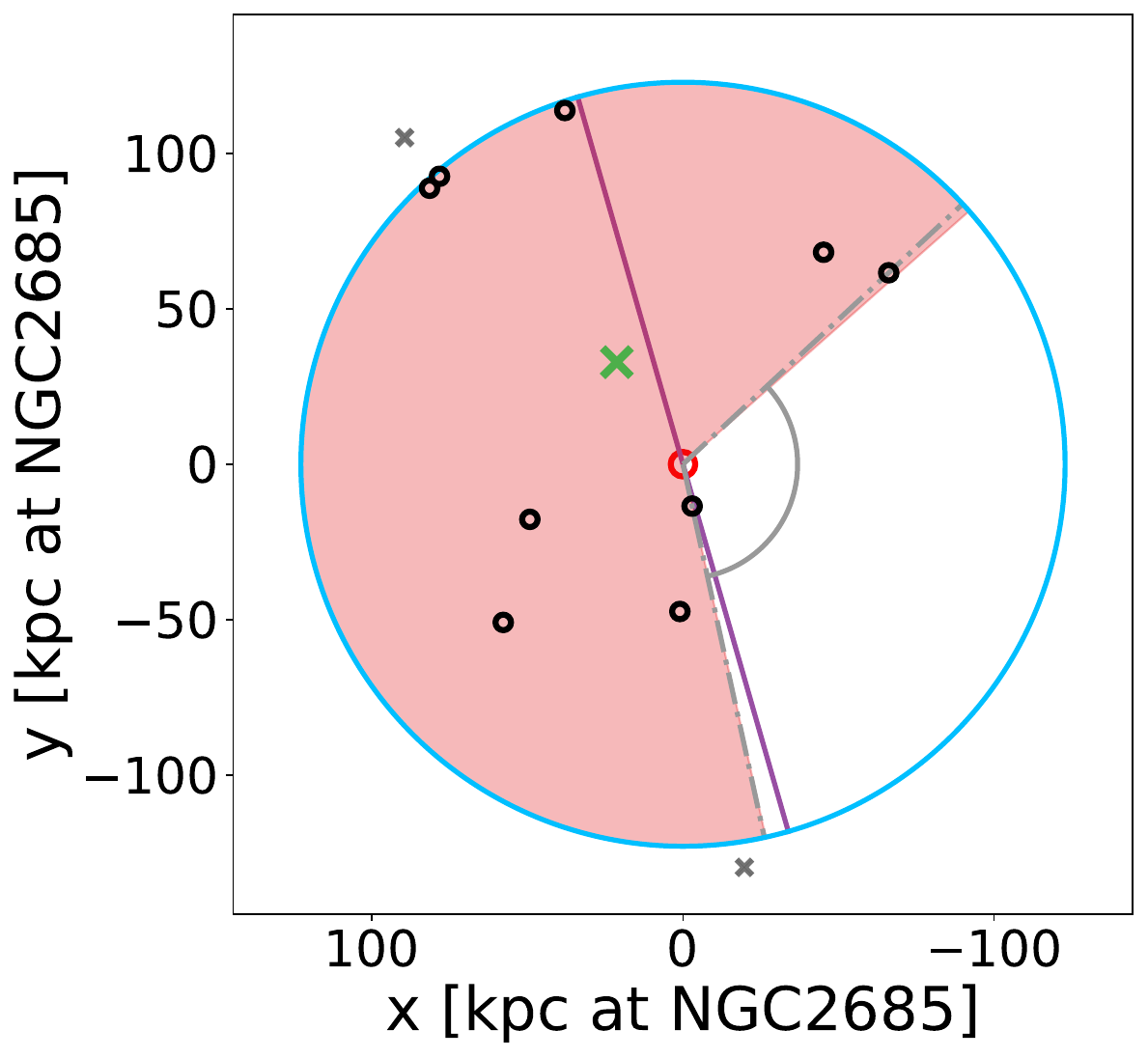}
    \includegraphics[width=6cm]{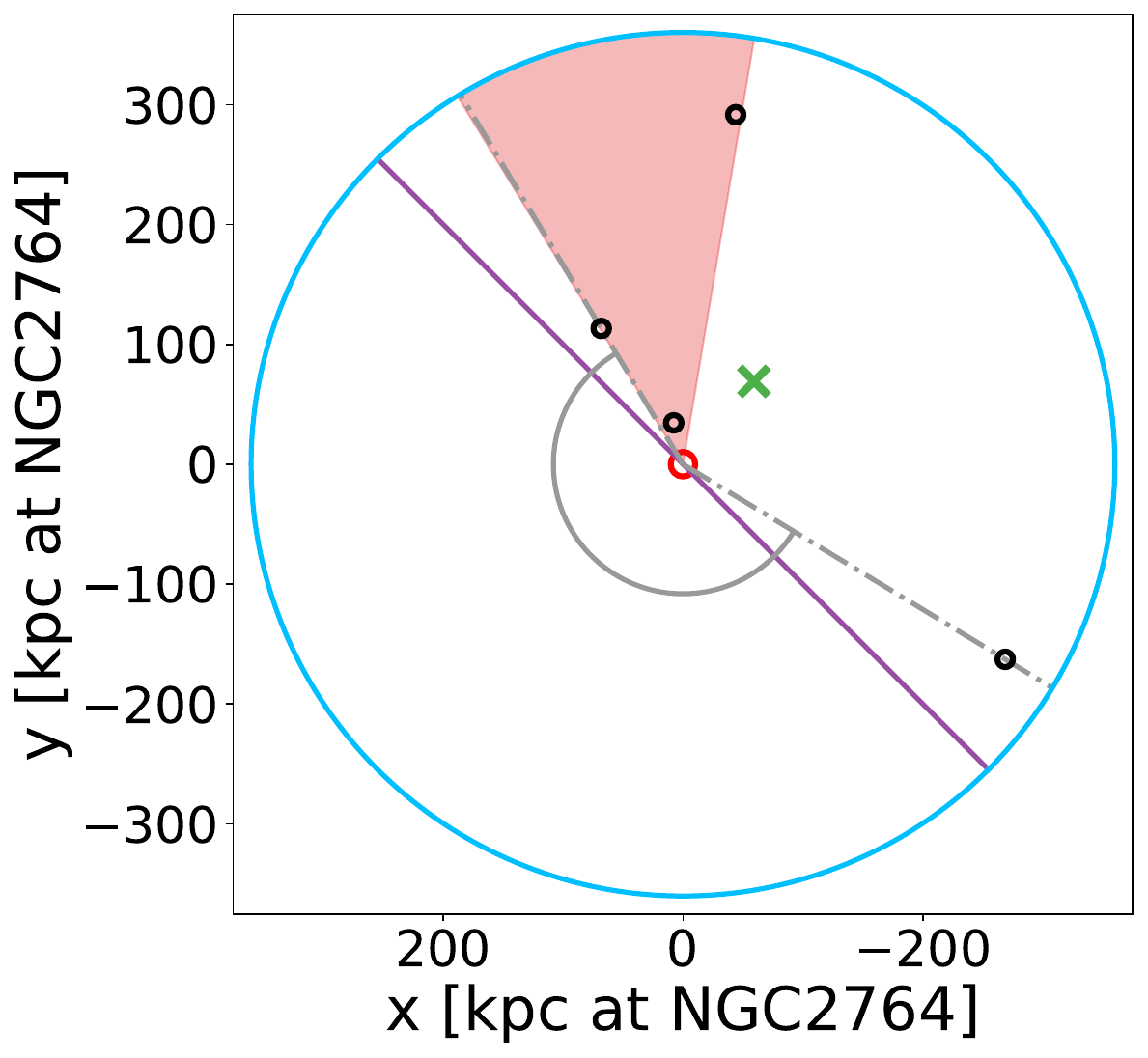}
    \includegraphics[width=6cm]{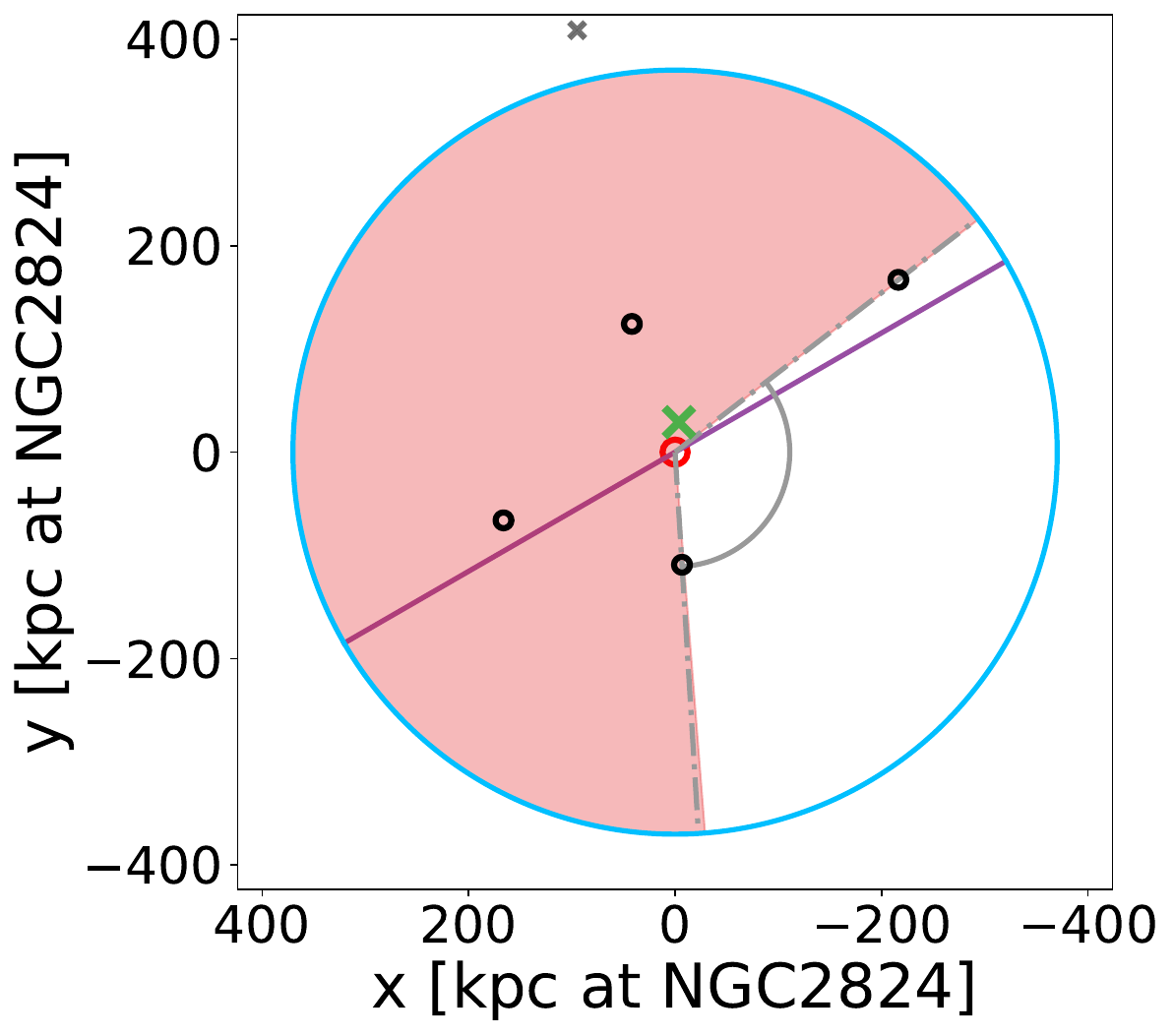}
    \includegraphics[width=6cm]{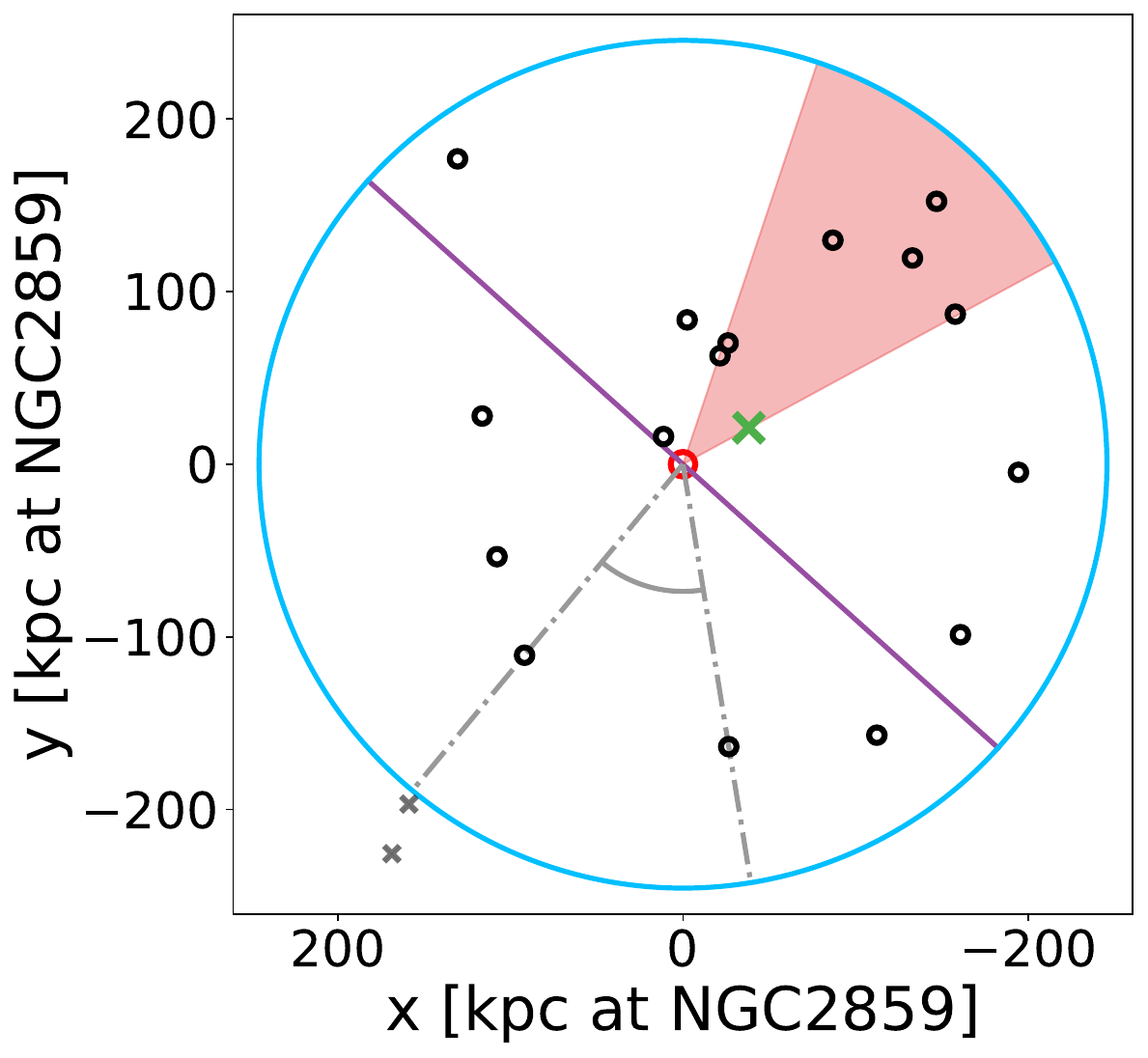}
    \includegraphics[width=6cm]{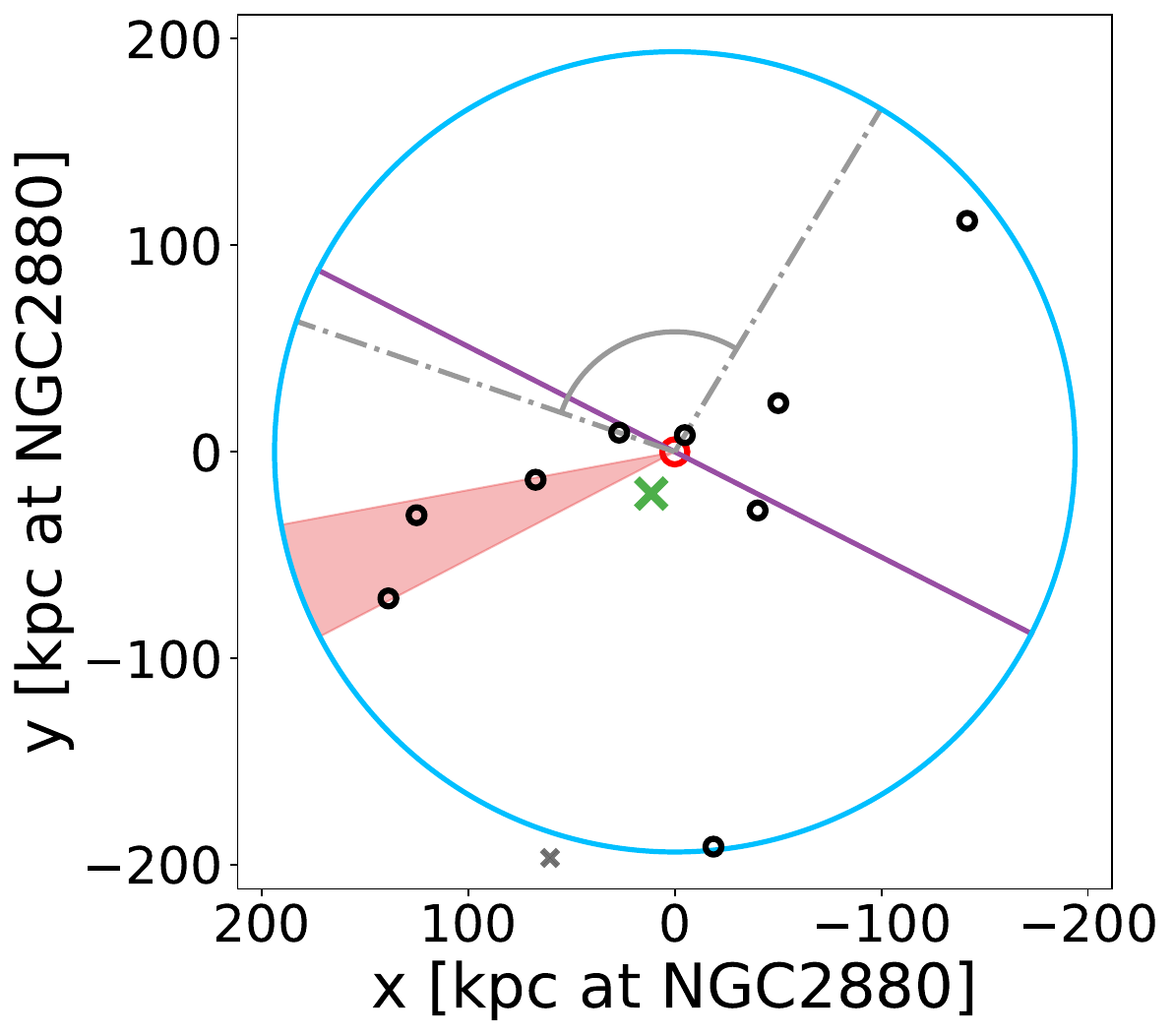}
\caption{Satellite distributions around the isolated ETGs from the MATLAS survey. The blue circle is the largest circle centered on the ETG that fits in the 1\,deg$^{2}$ FoV. Black circles are satellites within the blue circle. Gray crosses are satellites outside the circle but still within the FoV. The red circle is ETG. The purple line is the maximum hemisphere separation. The green cross is the centroid of satellite distribution. The two gray dotted lines, along with the gray arc show the largest wedge without any satellites. The red-shaded wedge hosts the most unusually high population of satellites compared the the isotropic realizations in the Monte Carlo simulation.
}
\label{fig:matlas_satellite_dists1}
\end{figure*}

\begin{figure*}[ht]
    \centering
    \ContinuedFloat
    \includegraphics[width=6cm]{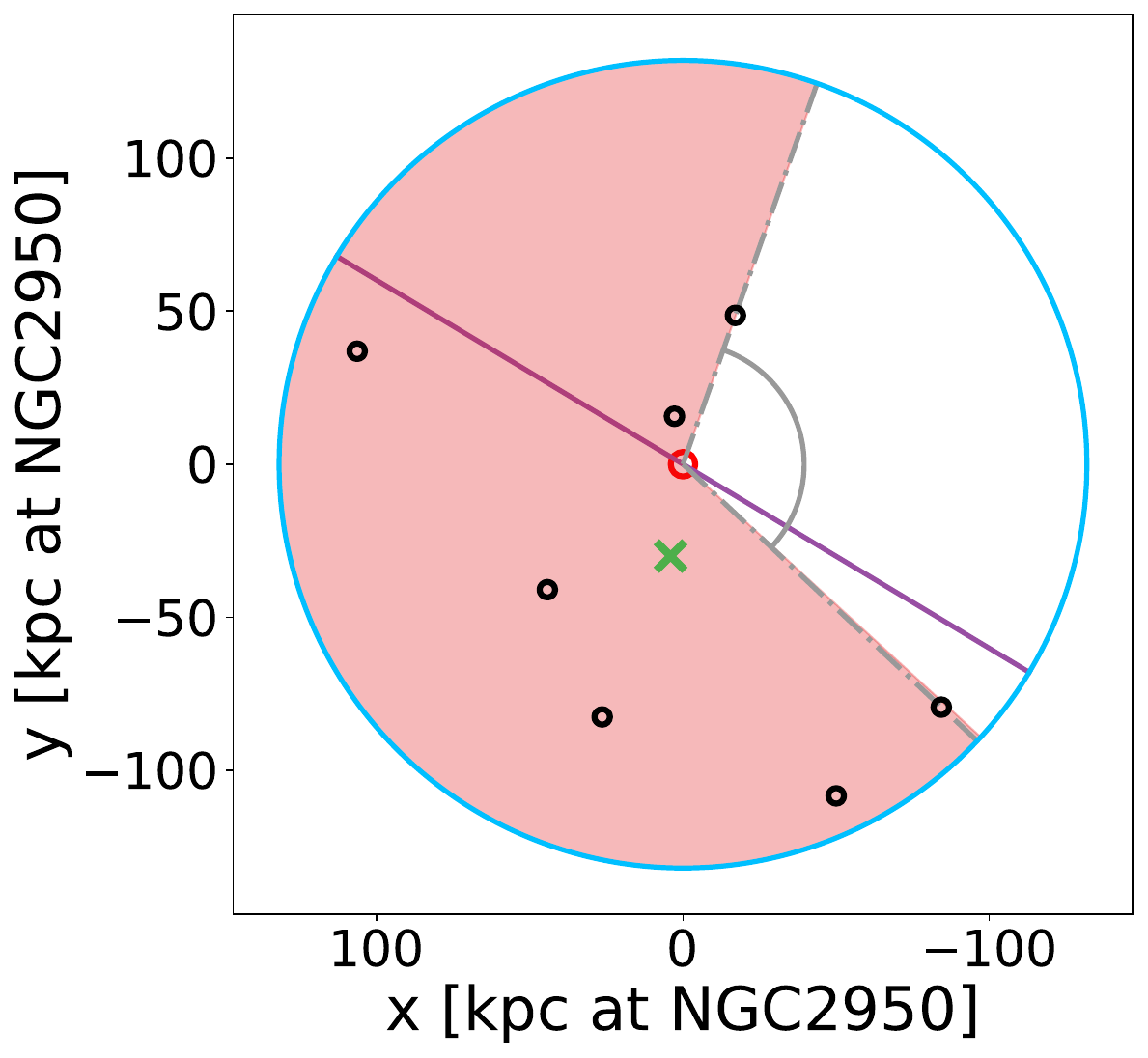}
    \includegraphics[width=6cm]{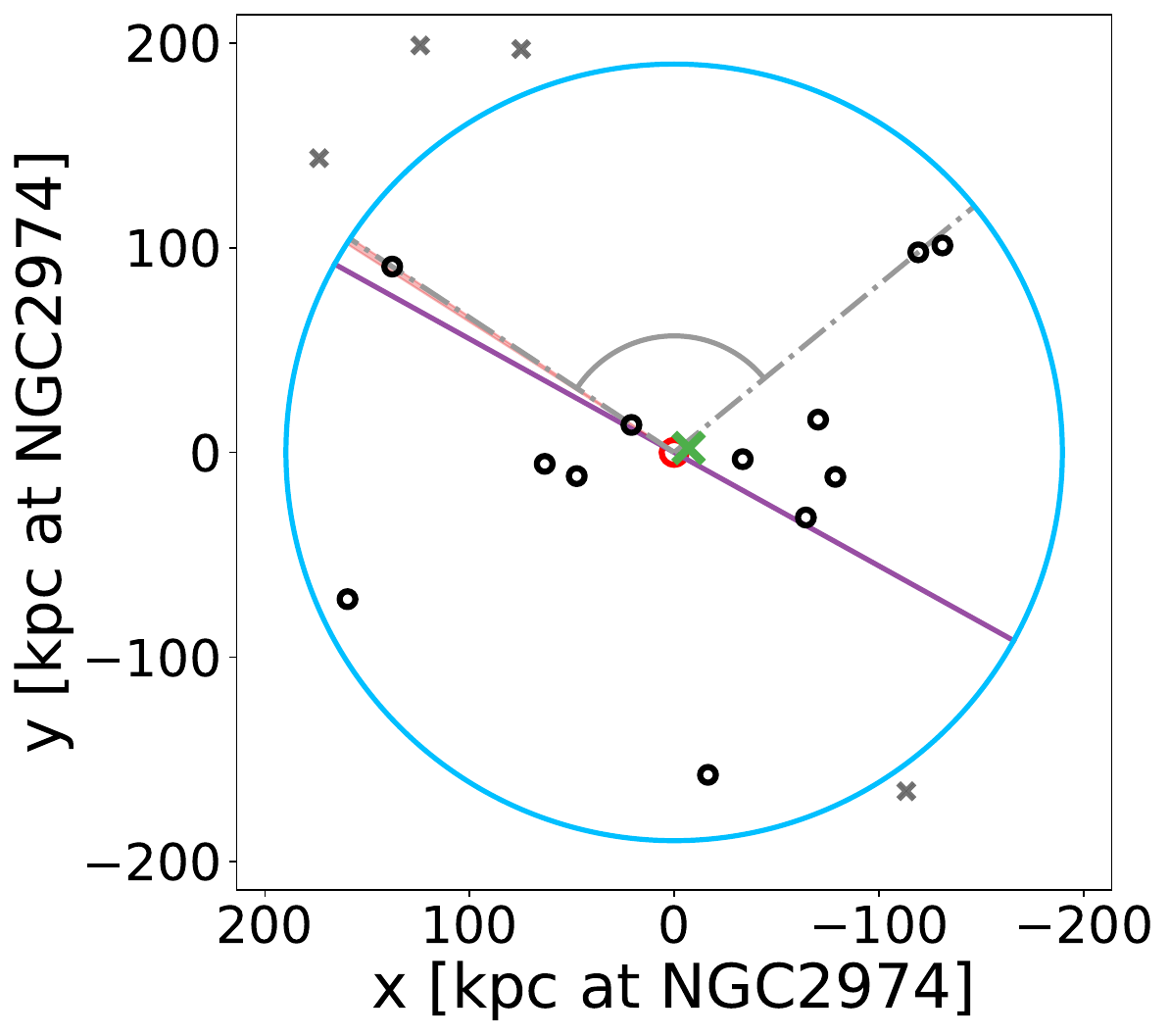}
    \includegraphics[width=6cm]{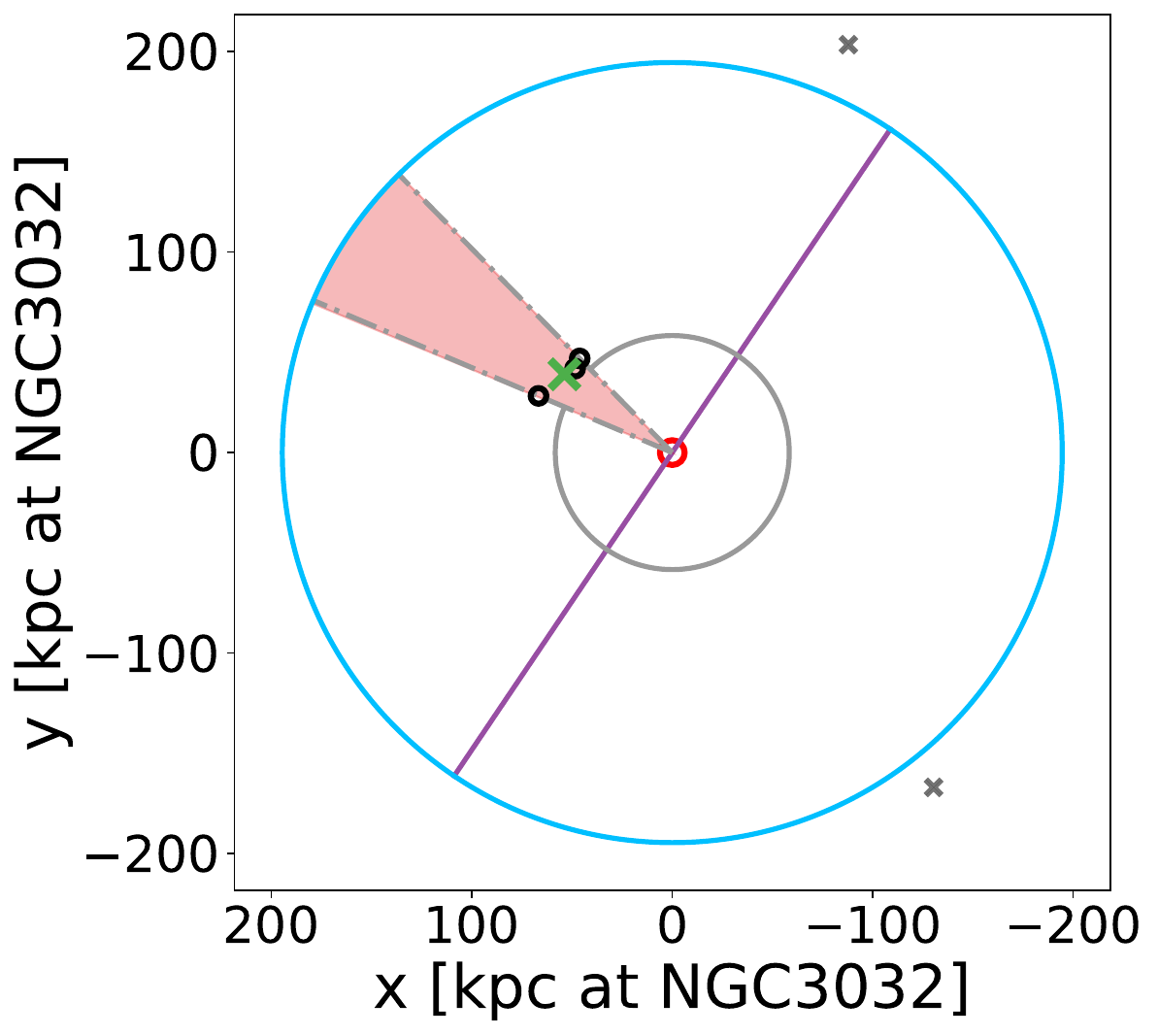}
    \includegraphics[width=6cm]{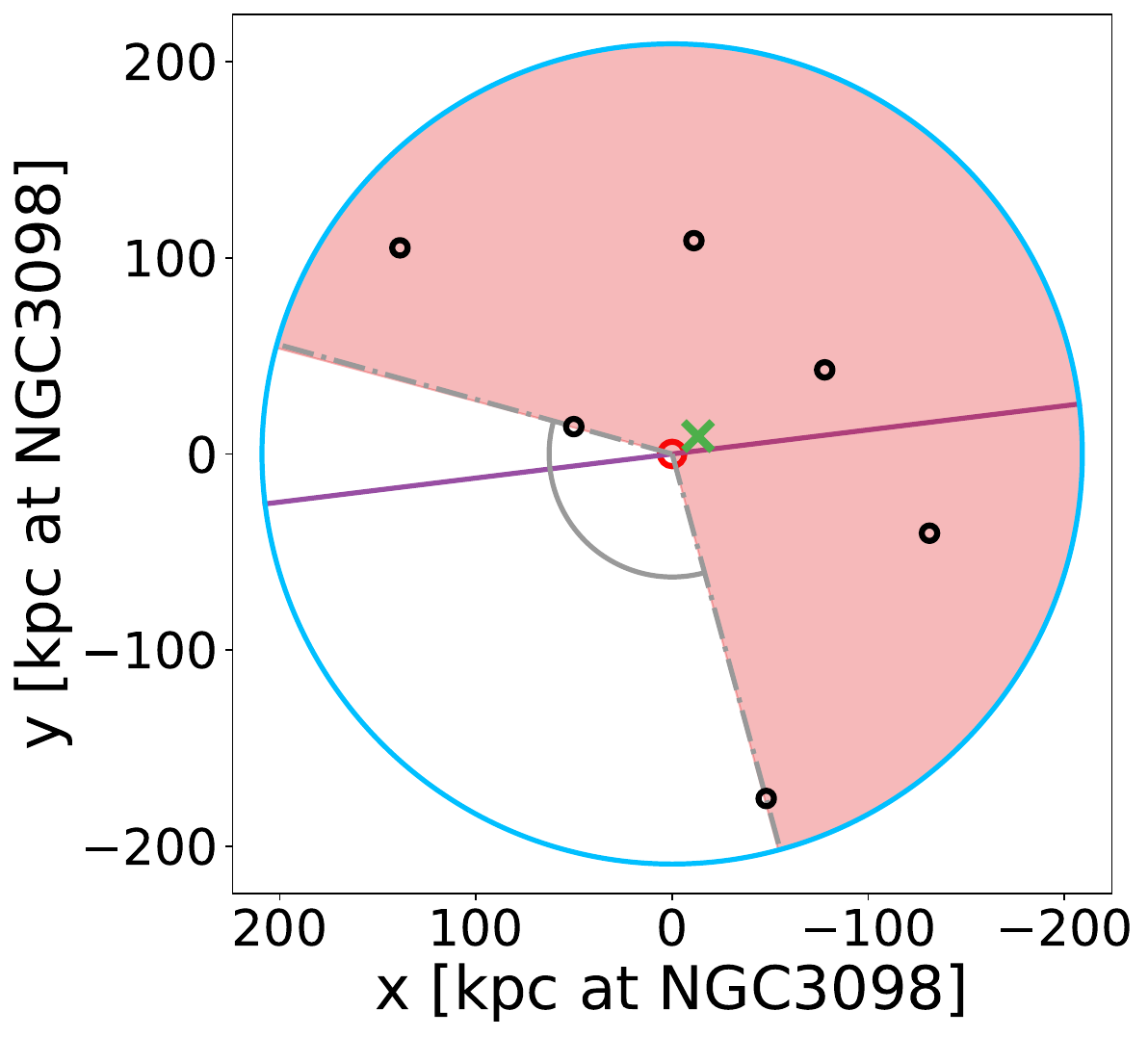}
    \includegraphics[width=6cm]{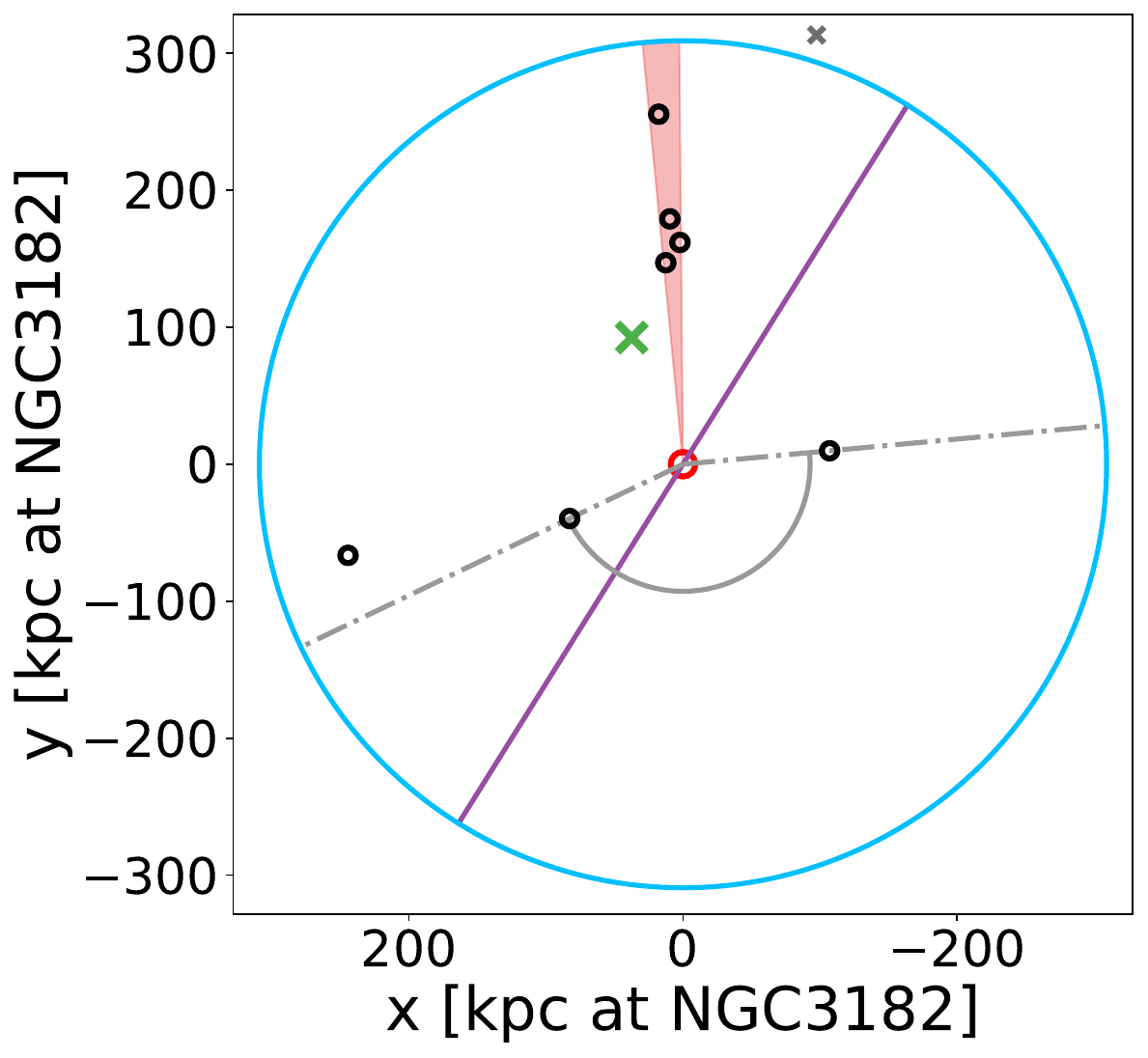}
    \includegraphics[width=6cm]{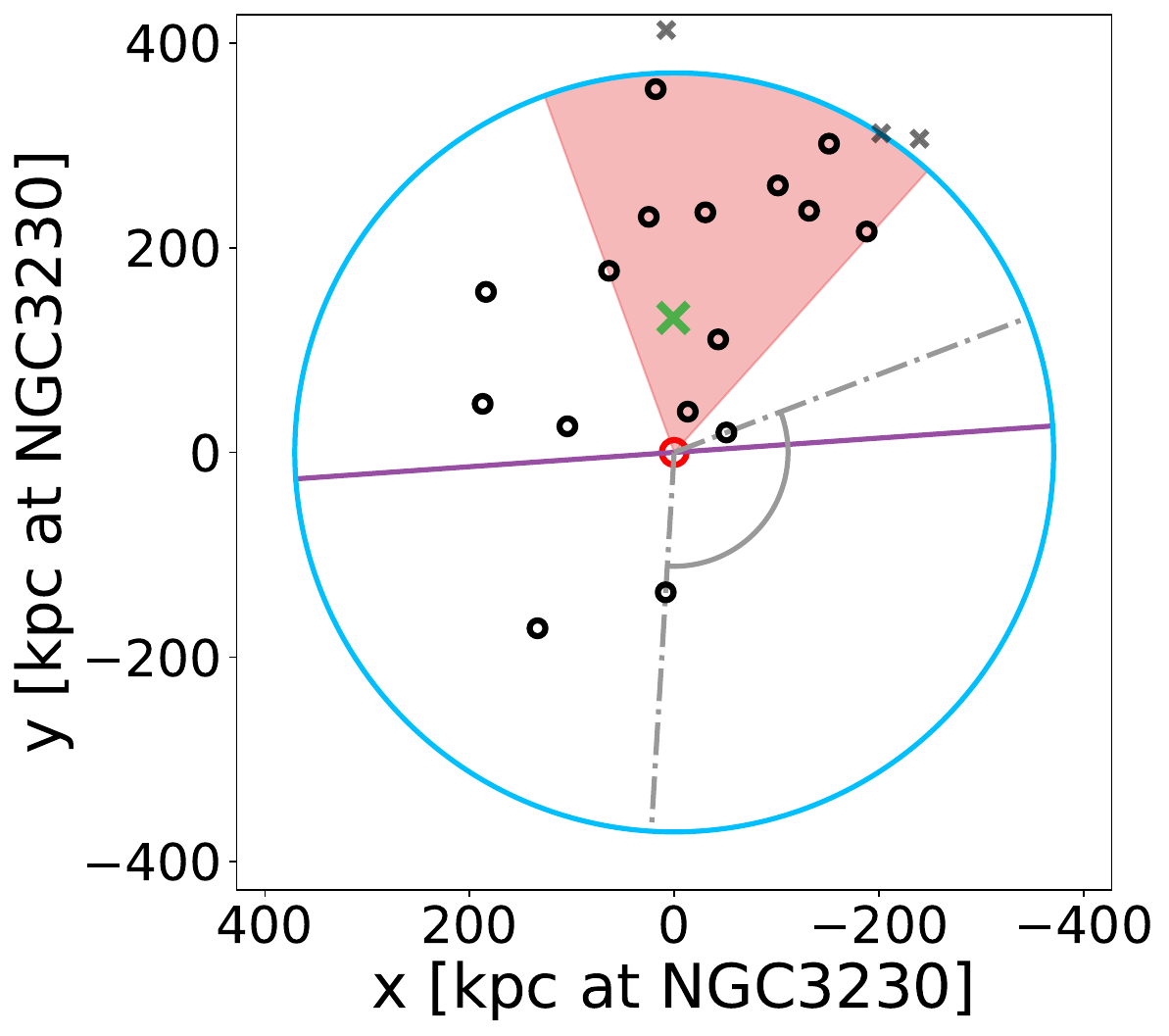}
    \includegraphics[width=6cm]{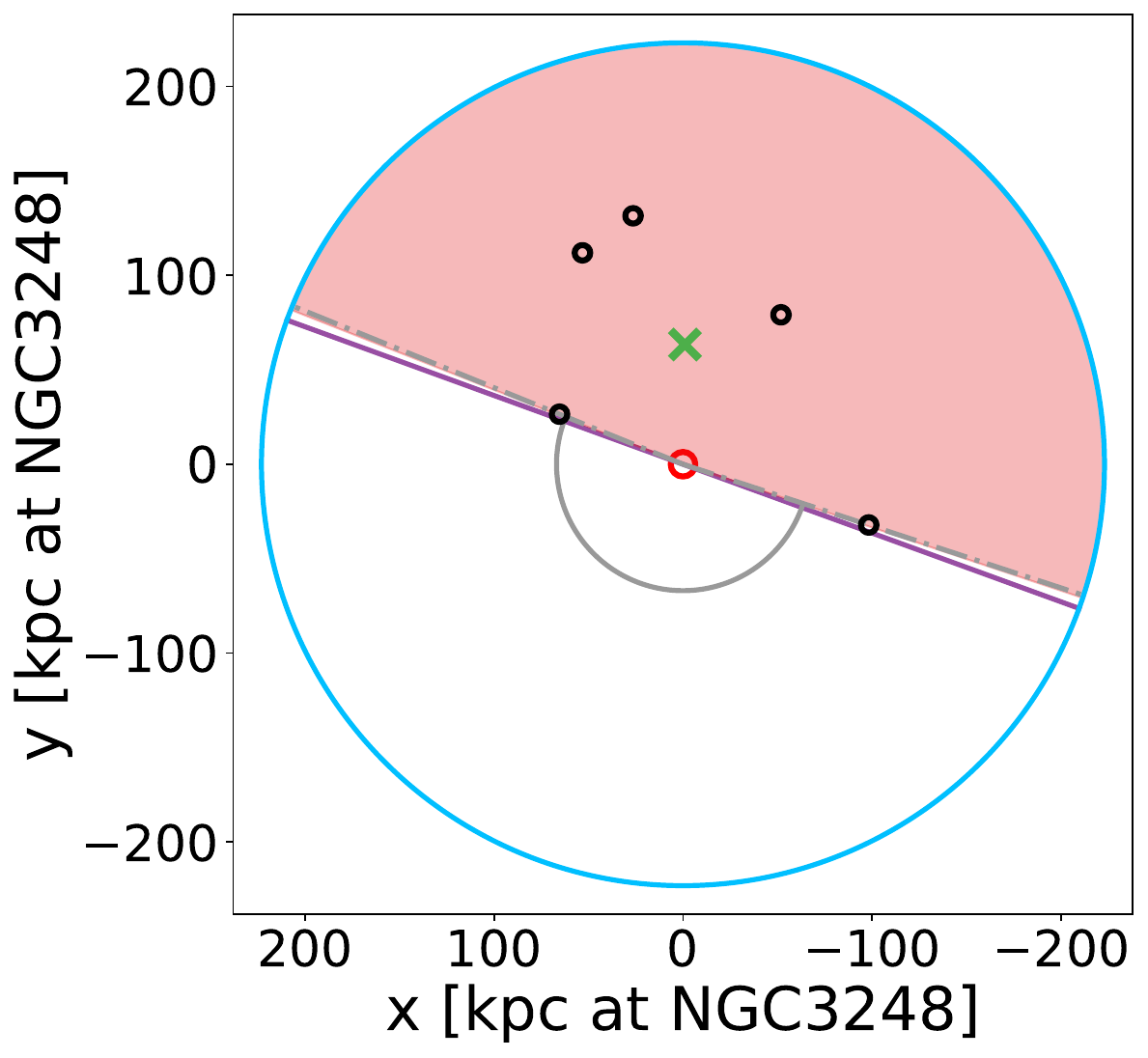}
    \includegraphics[width=6cm]{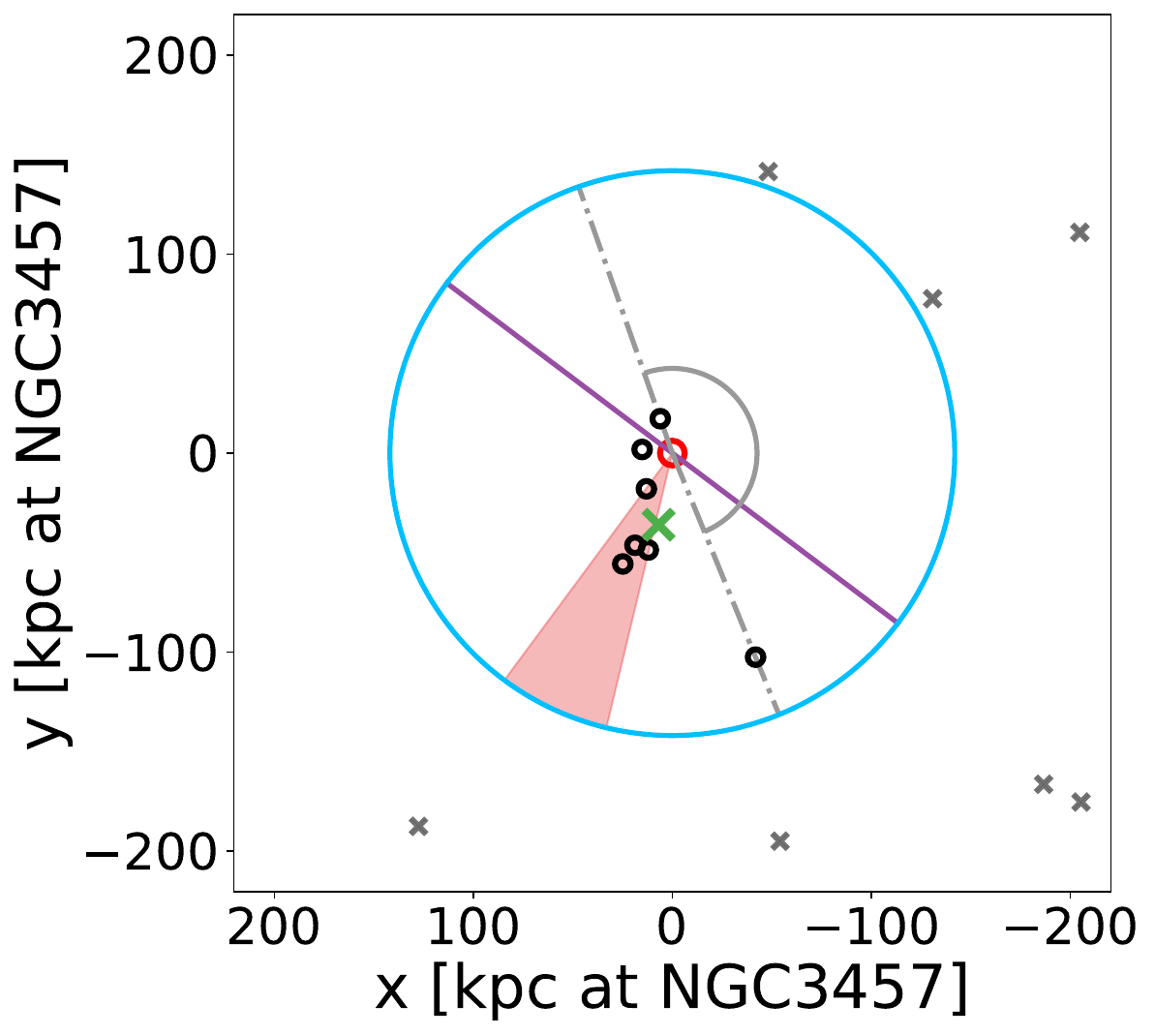}
    \includegraphics[width=6cm]{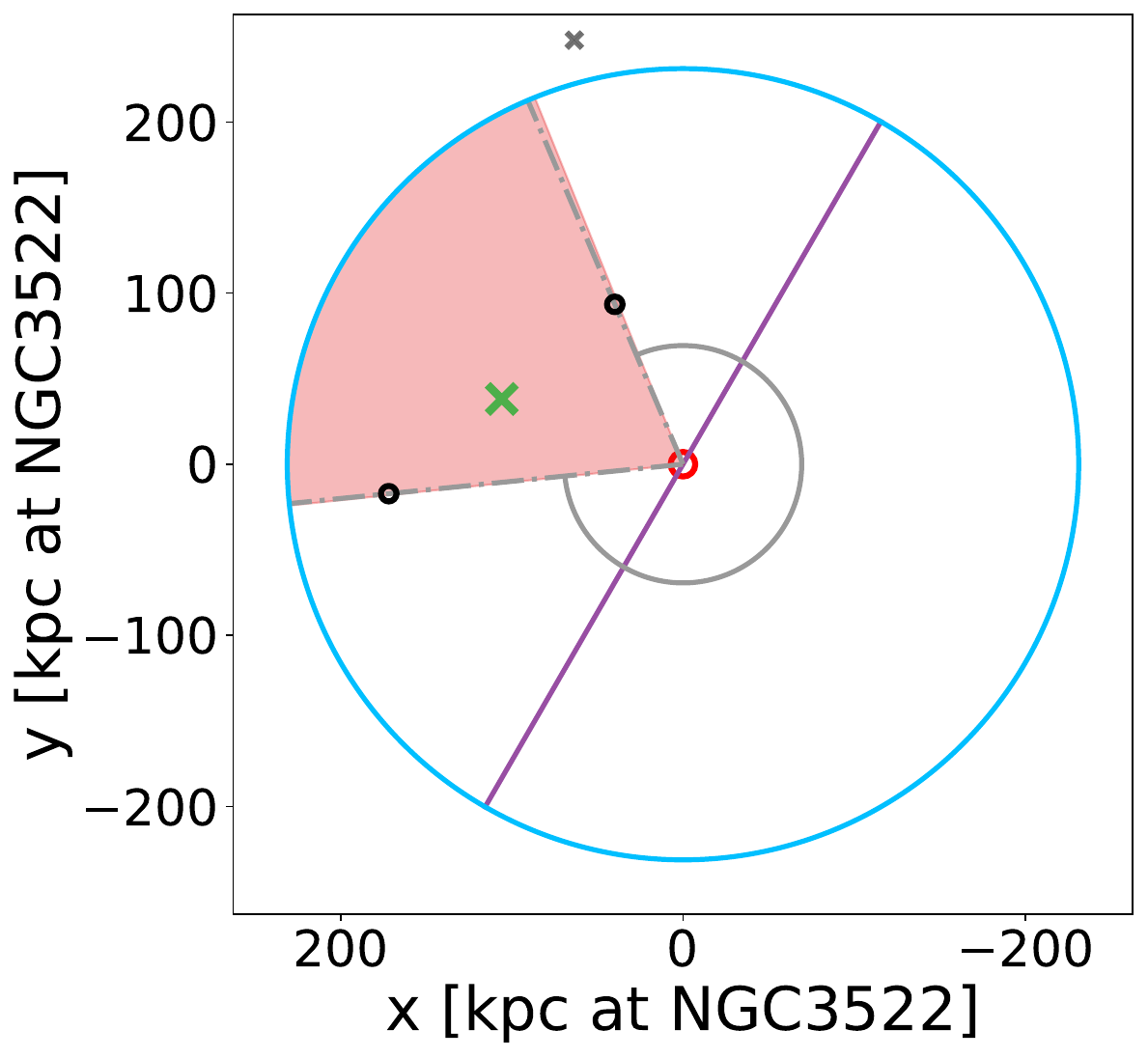}
    \includegraphics[width=6cm]{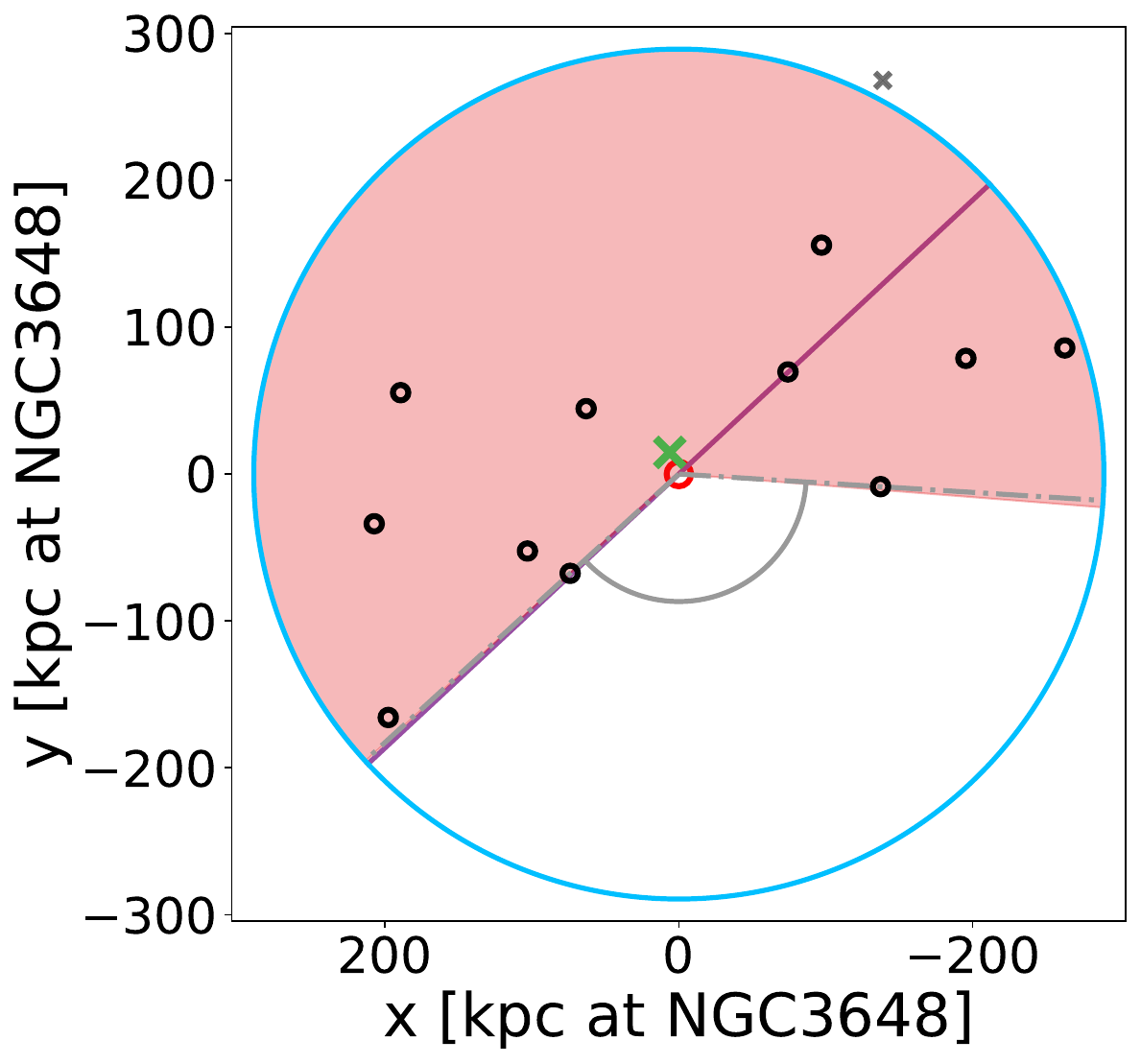}
    \includegraphics[width=6cm]{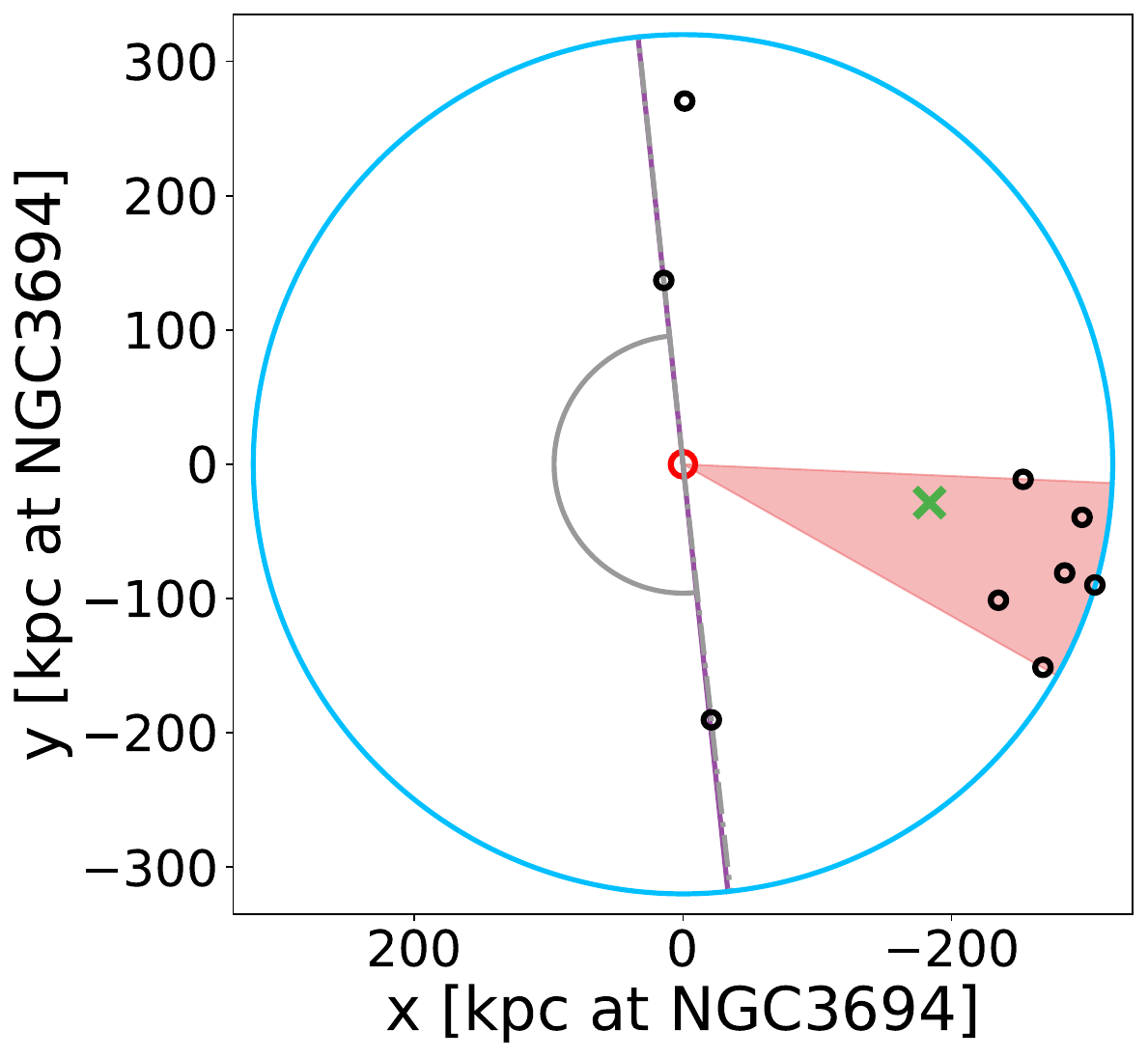}
    \includegraphics[width=6cm]{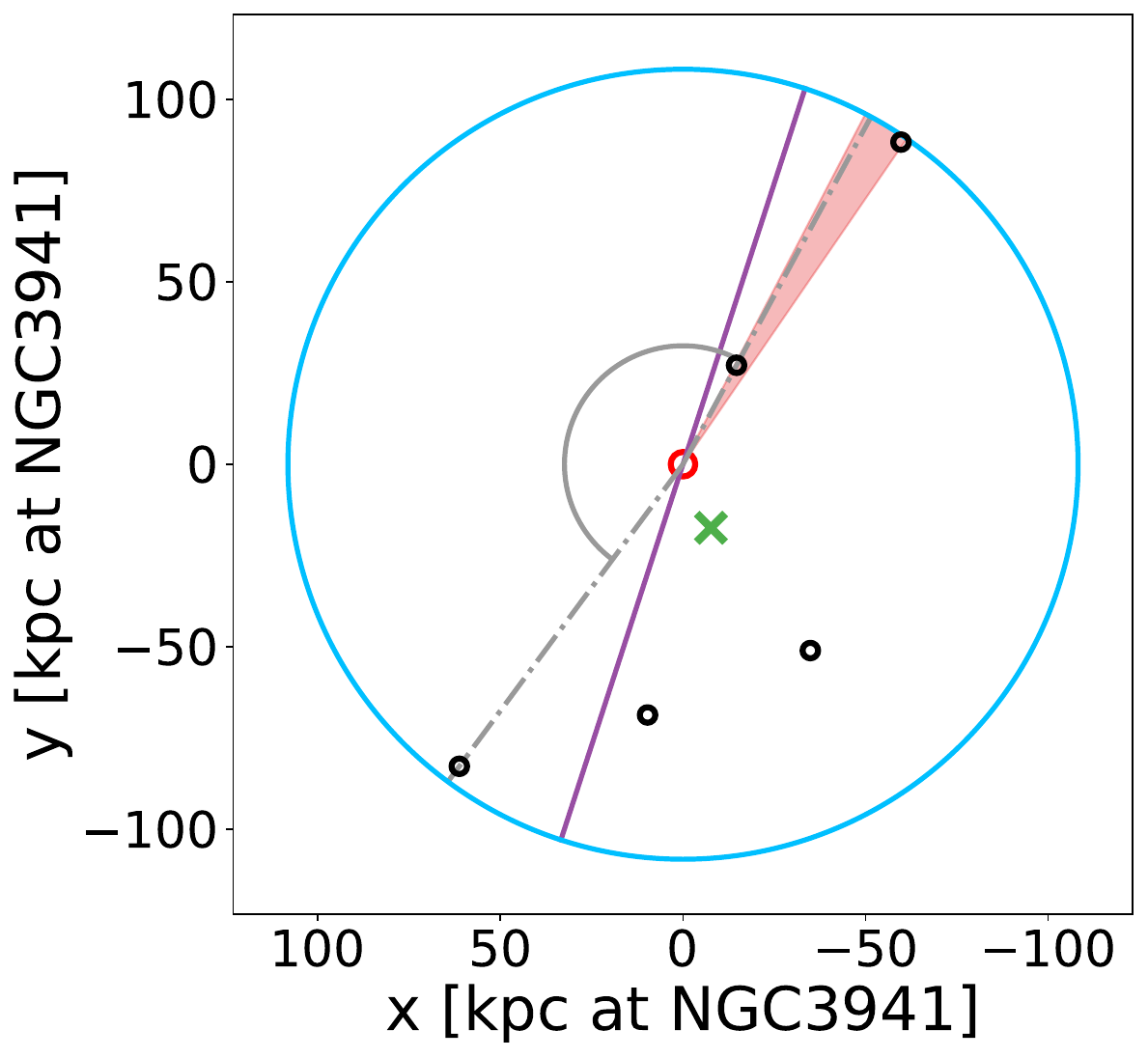}
\caption{Continued.
}
    \label{fig:matlas_satellite_dists2}
\end{figure*}

\begin{figure*}[ht]
    \centering
    \ContinuedFloat
    \includegraphics[width=6cm]{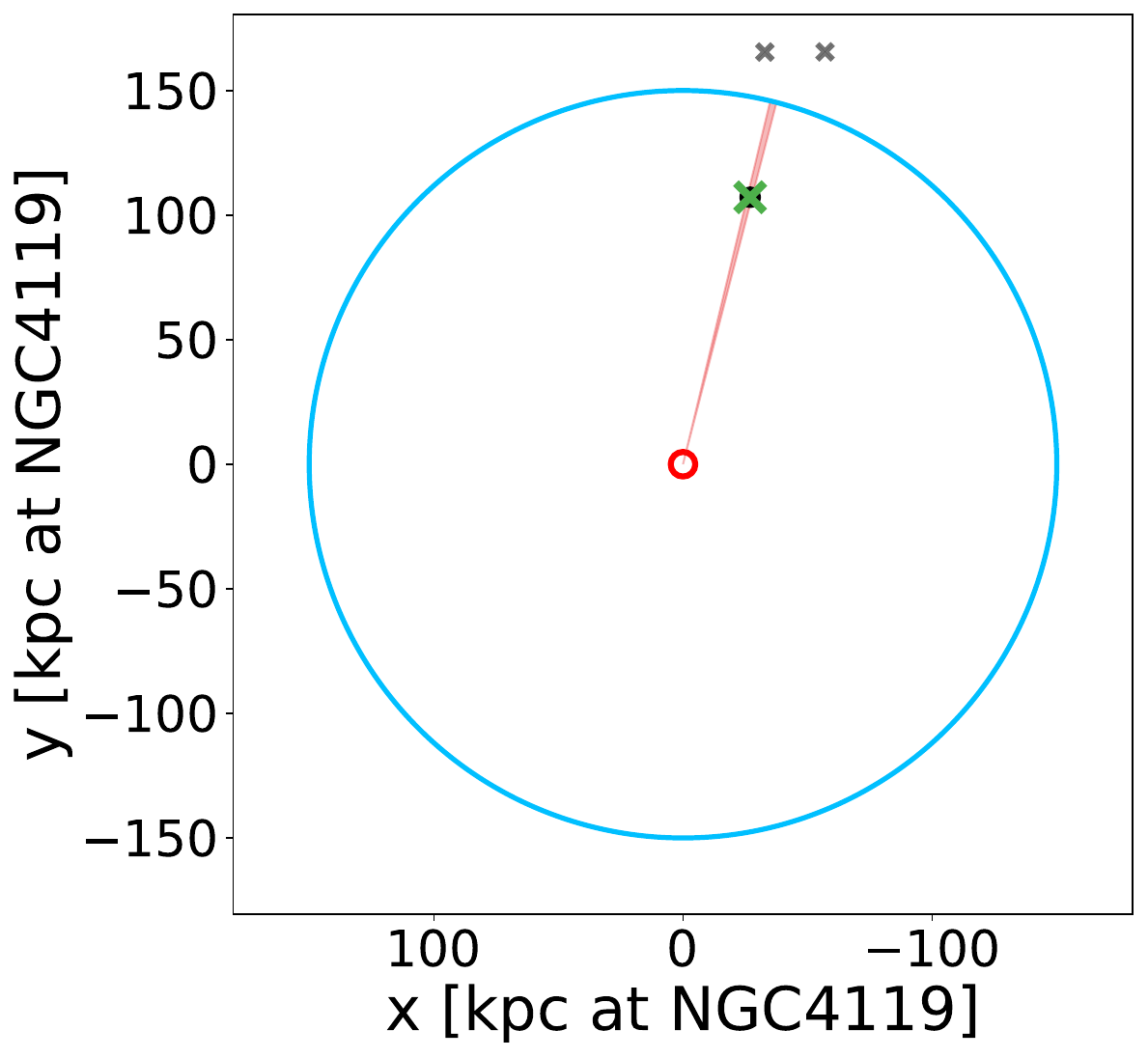}
    \includegraphics[width=6cm]{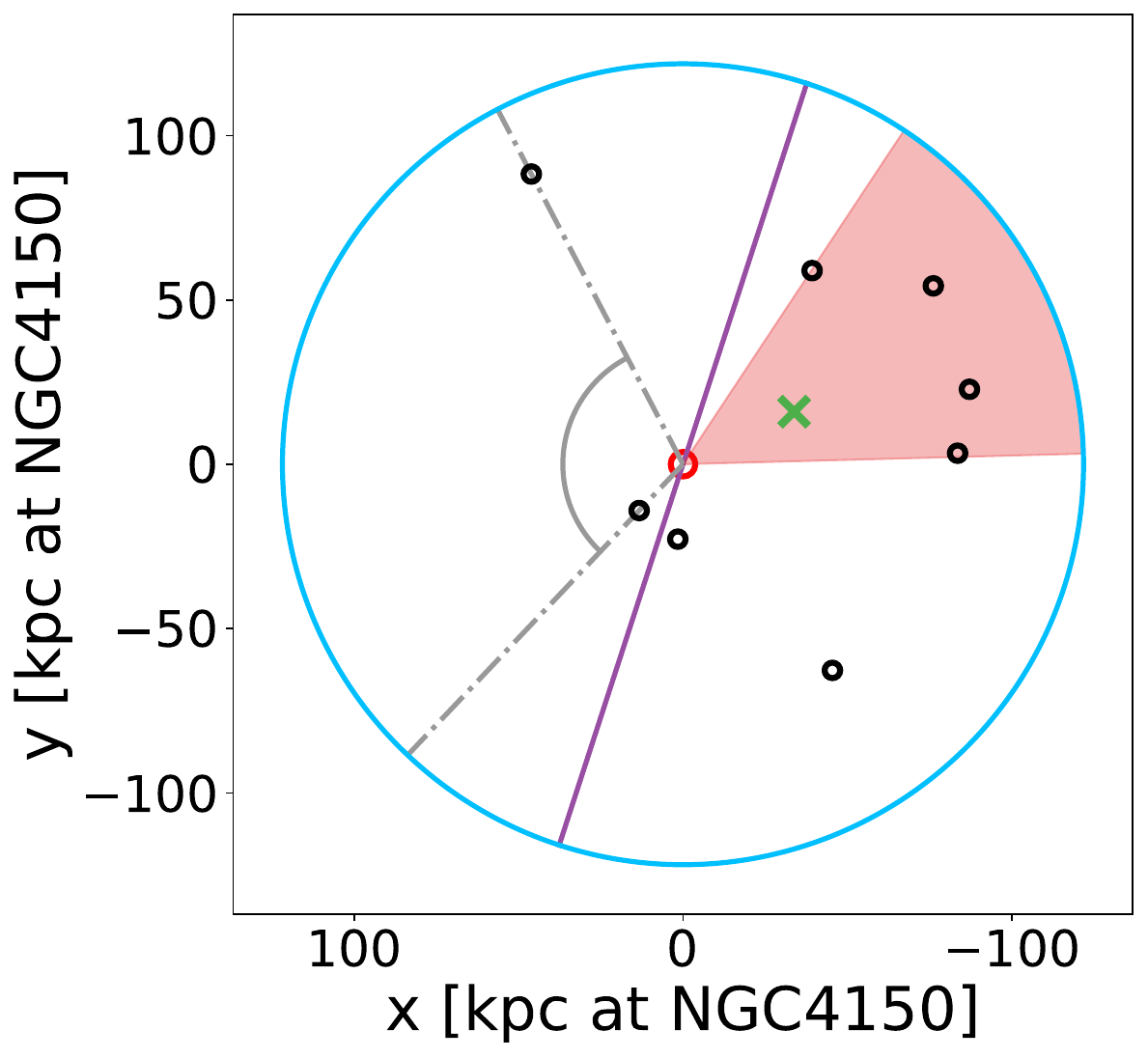}
    \includegraphics[width=6cm]{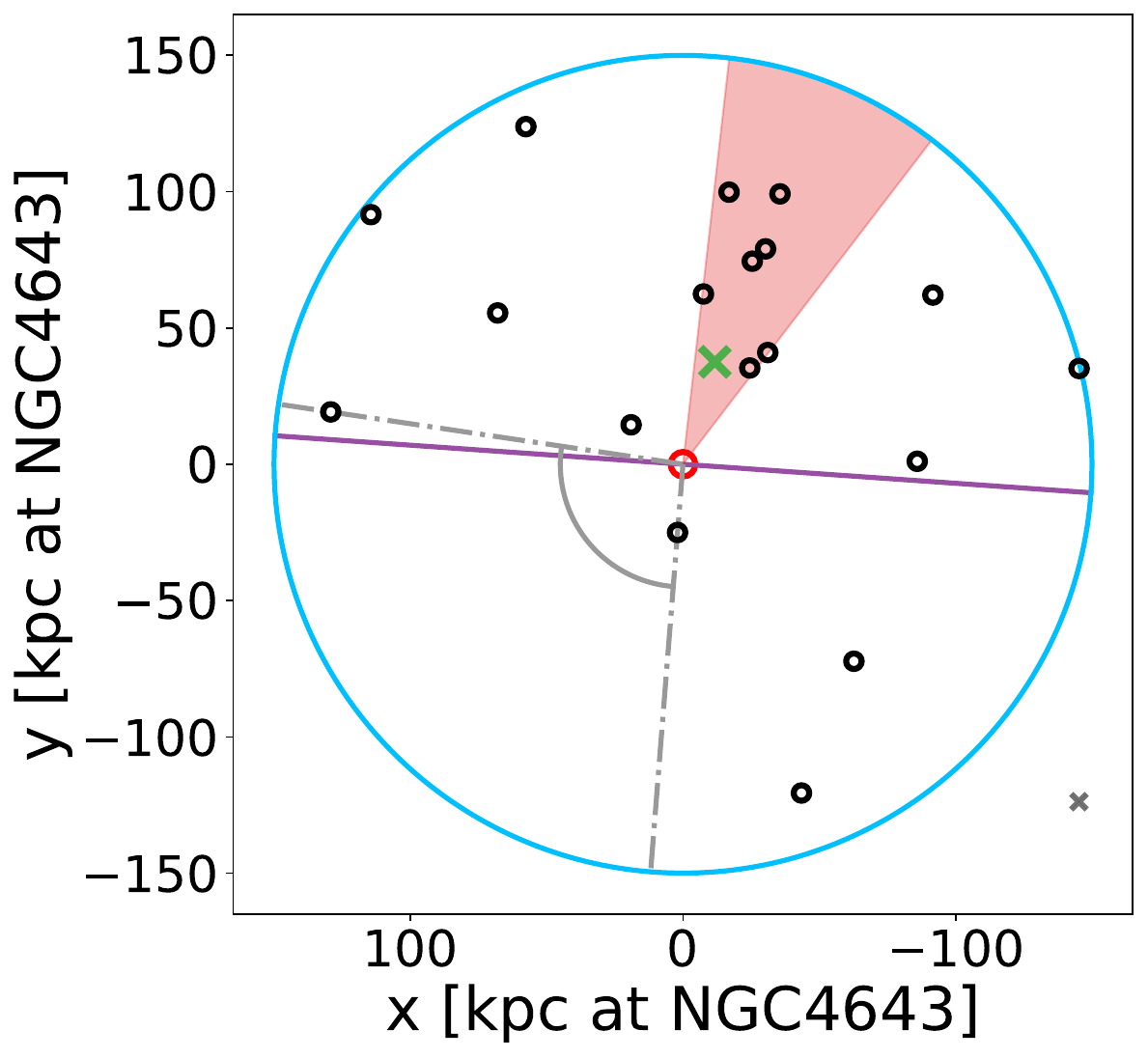}
    \includegraphics[width=6cm]{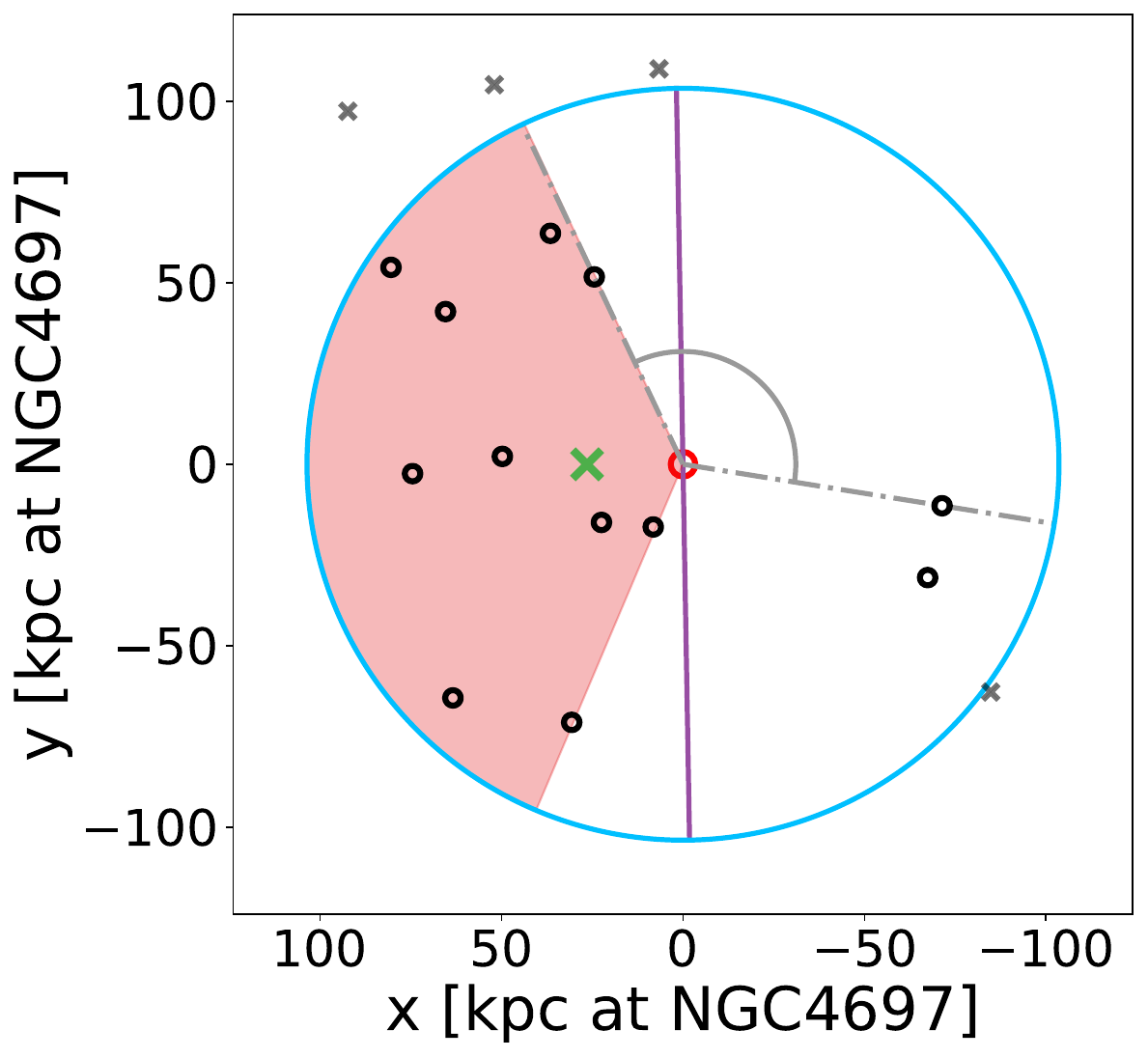}
    \includegraphics[width=6cm]{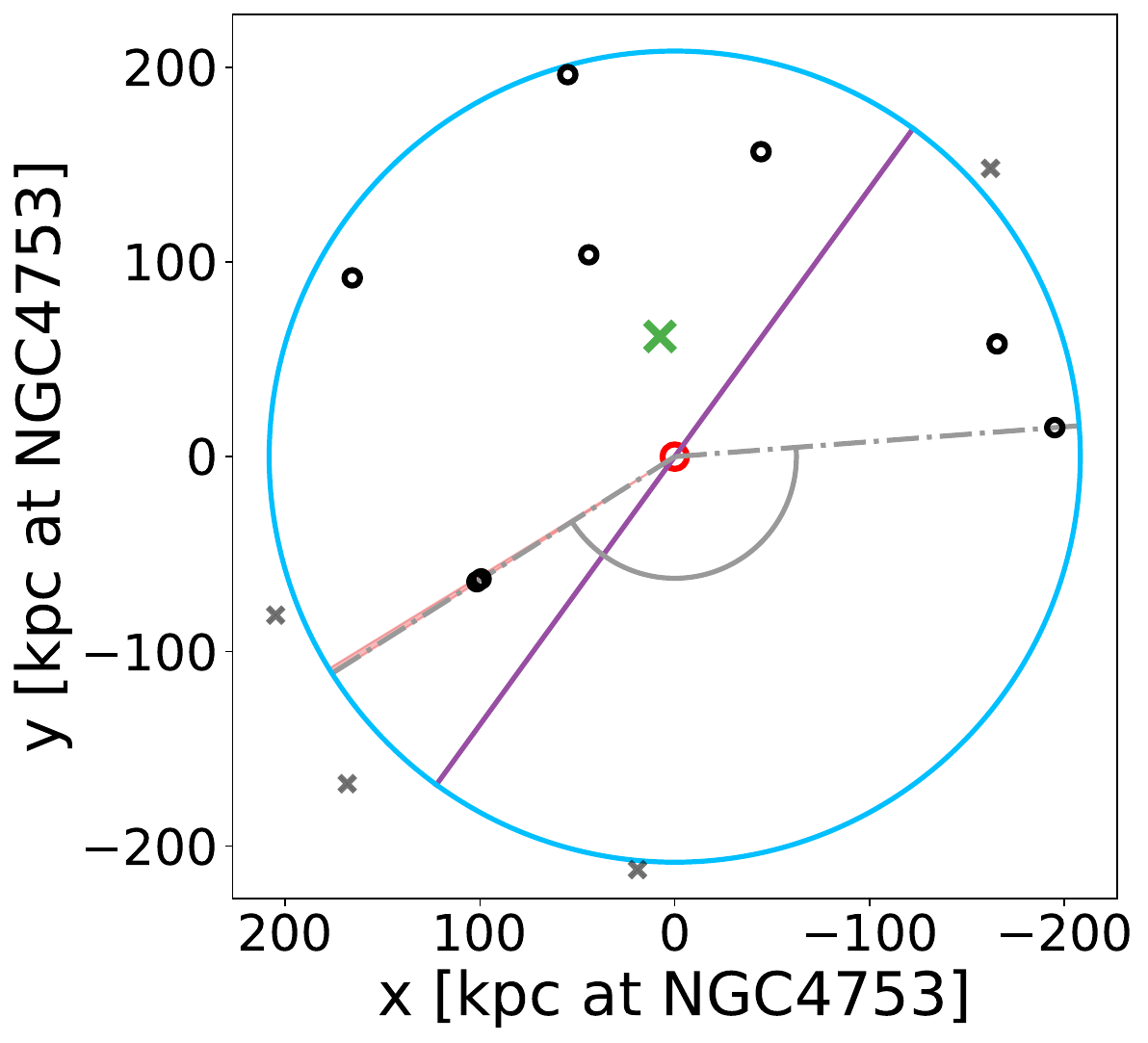}
    \includegraphics[width=6cm]{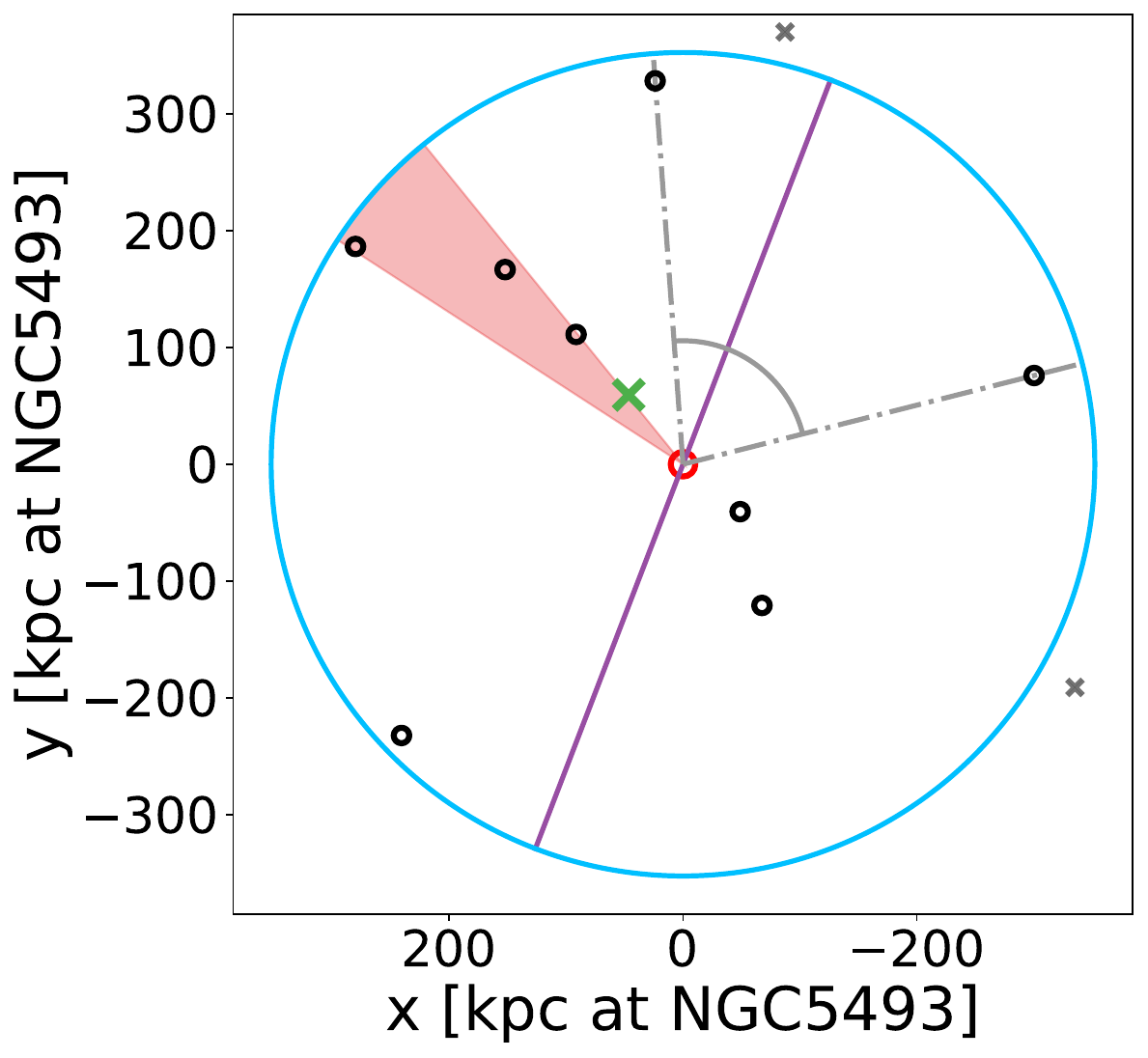}
    \includegraphics[width=6cm]{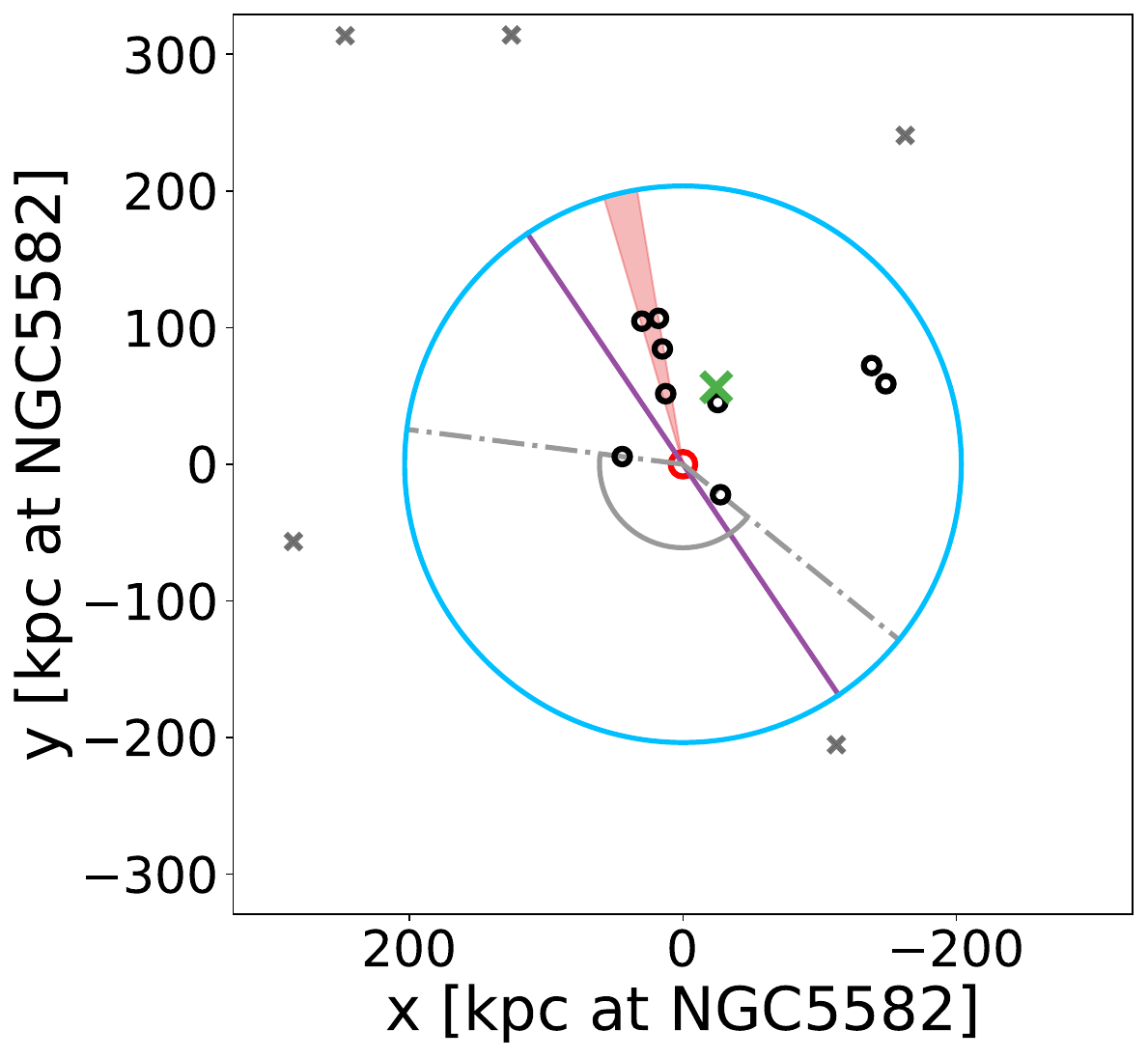}
    \includegraphics[width=6cm]{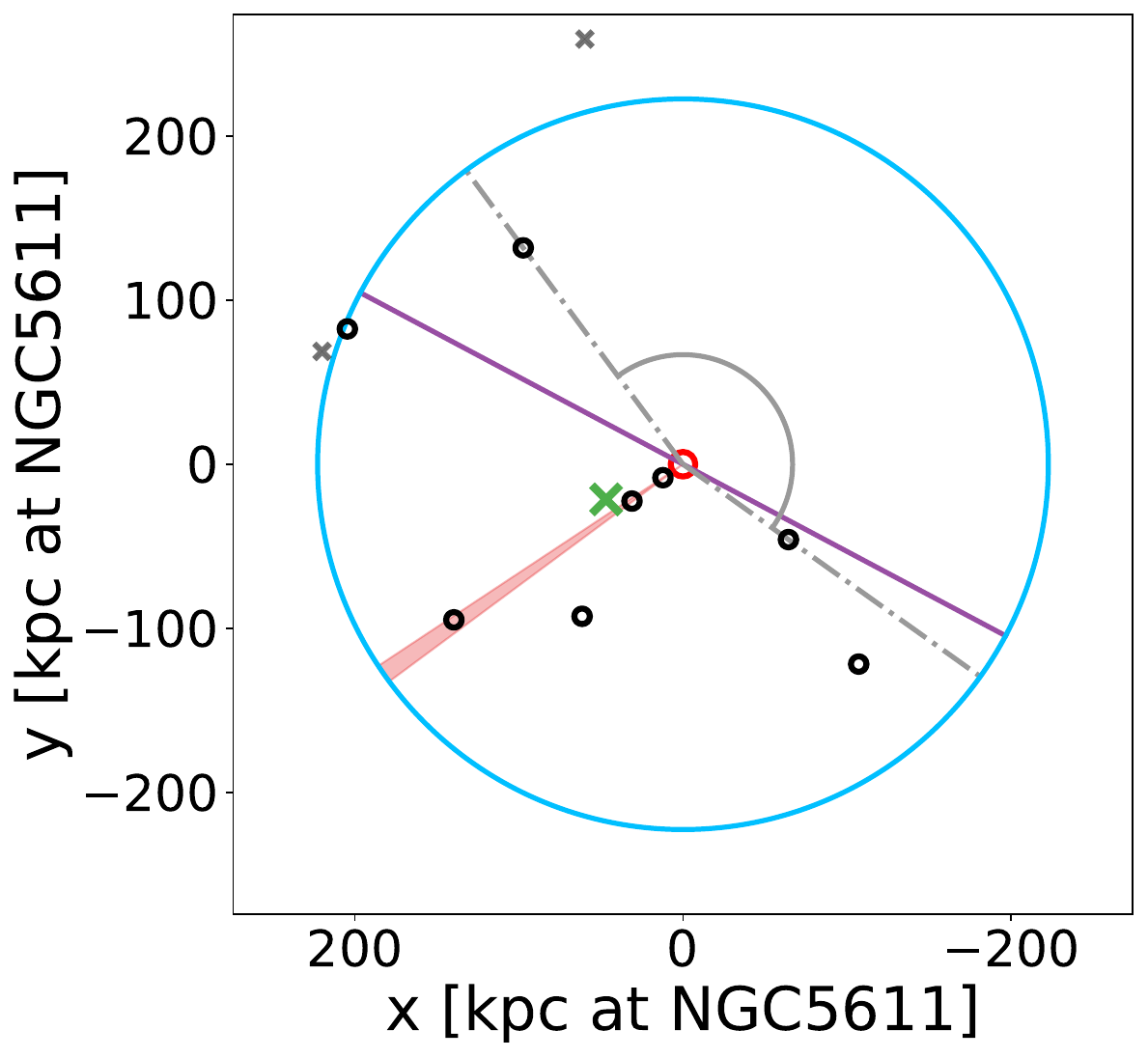}
    \includegraphics[width=6cm]{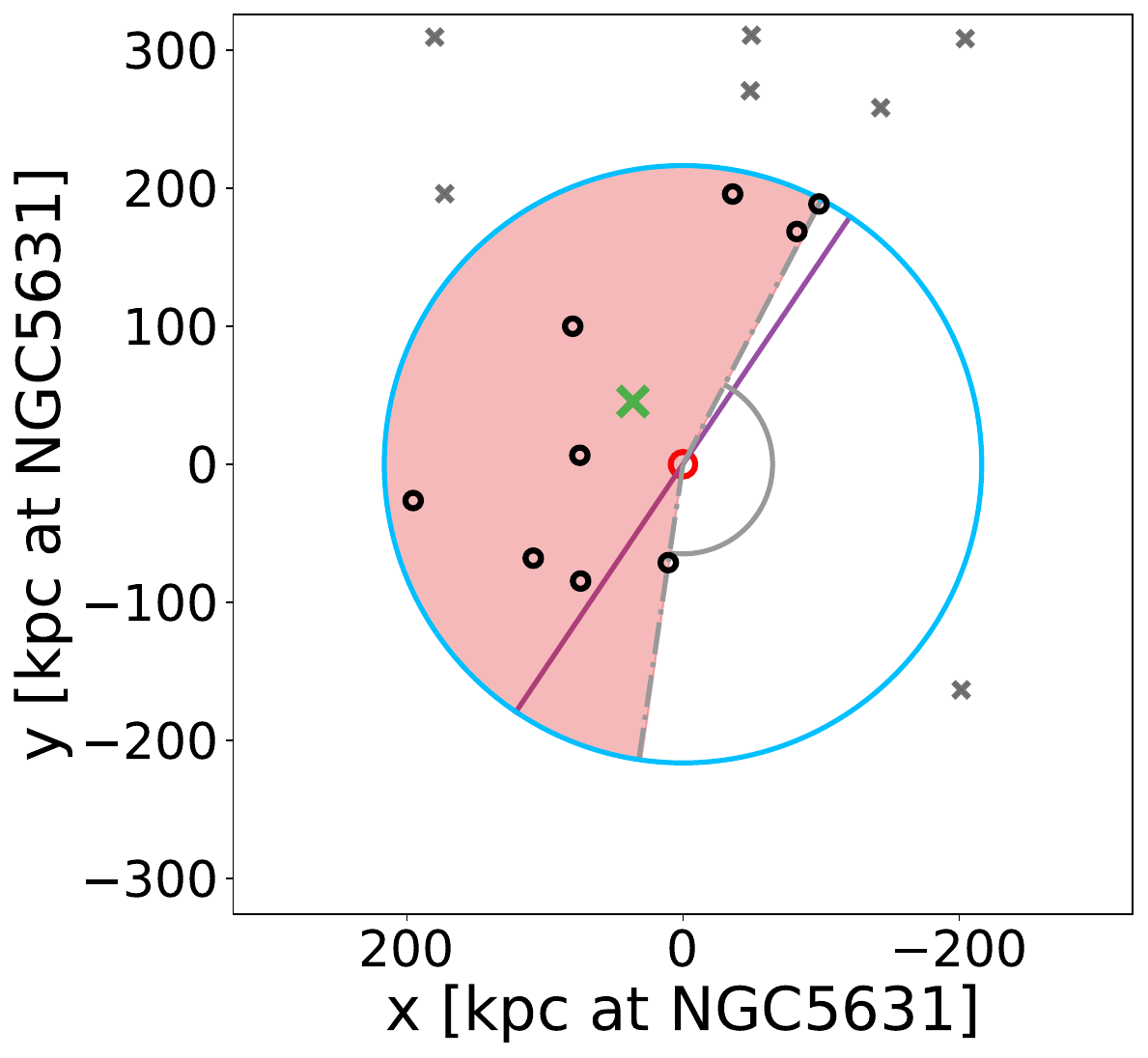}
    \includegraphics[width=6cm]{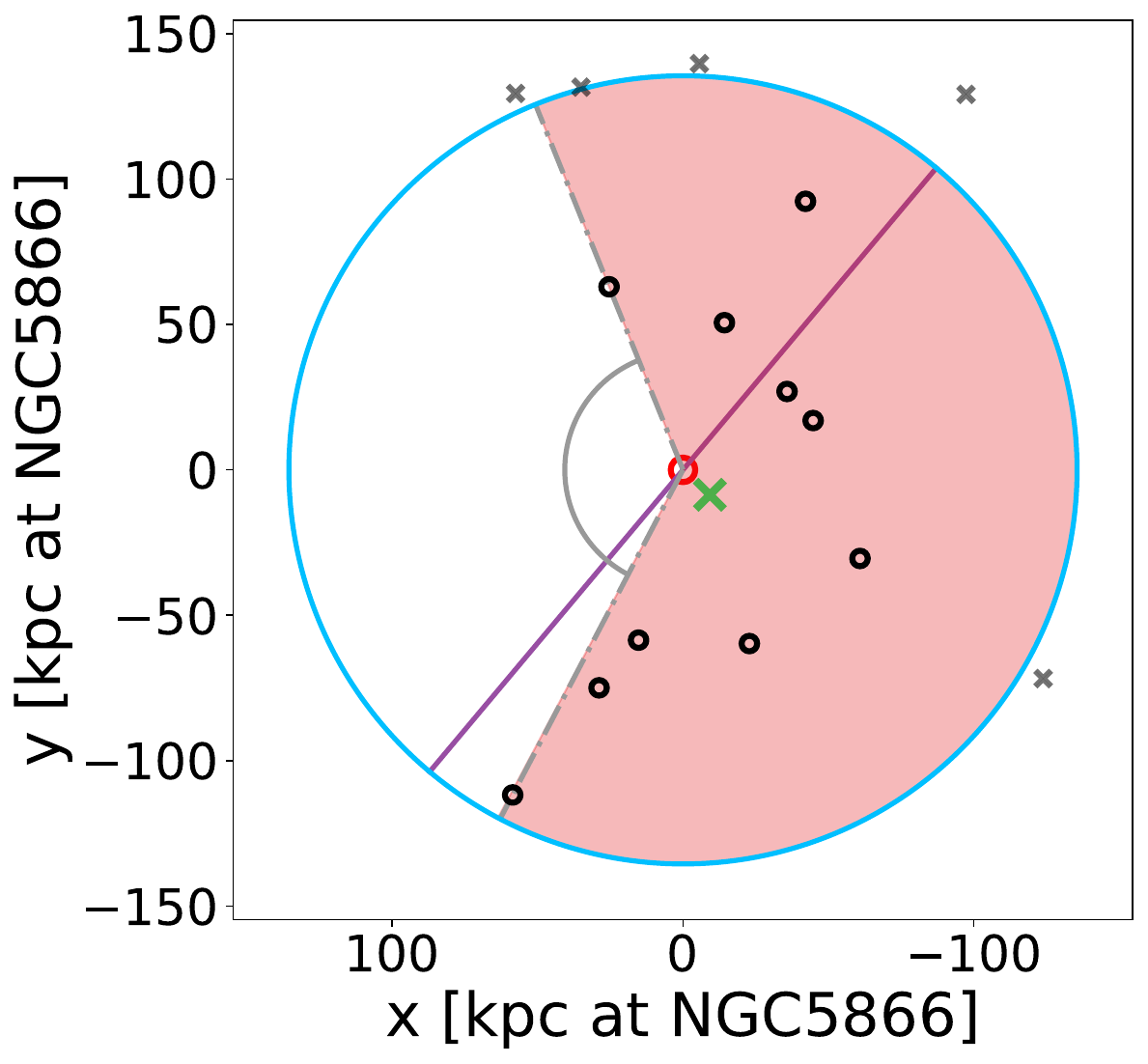}
    \includegraphics[width=6cm]{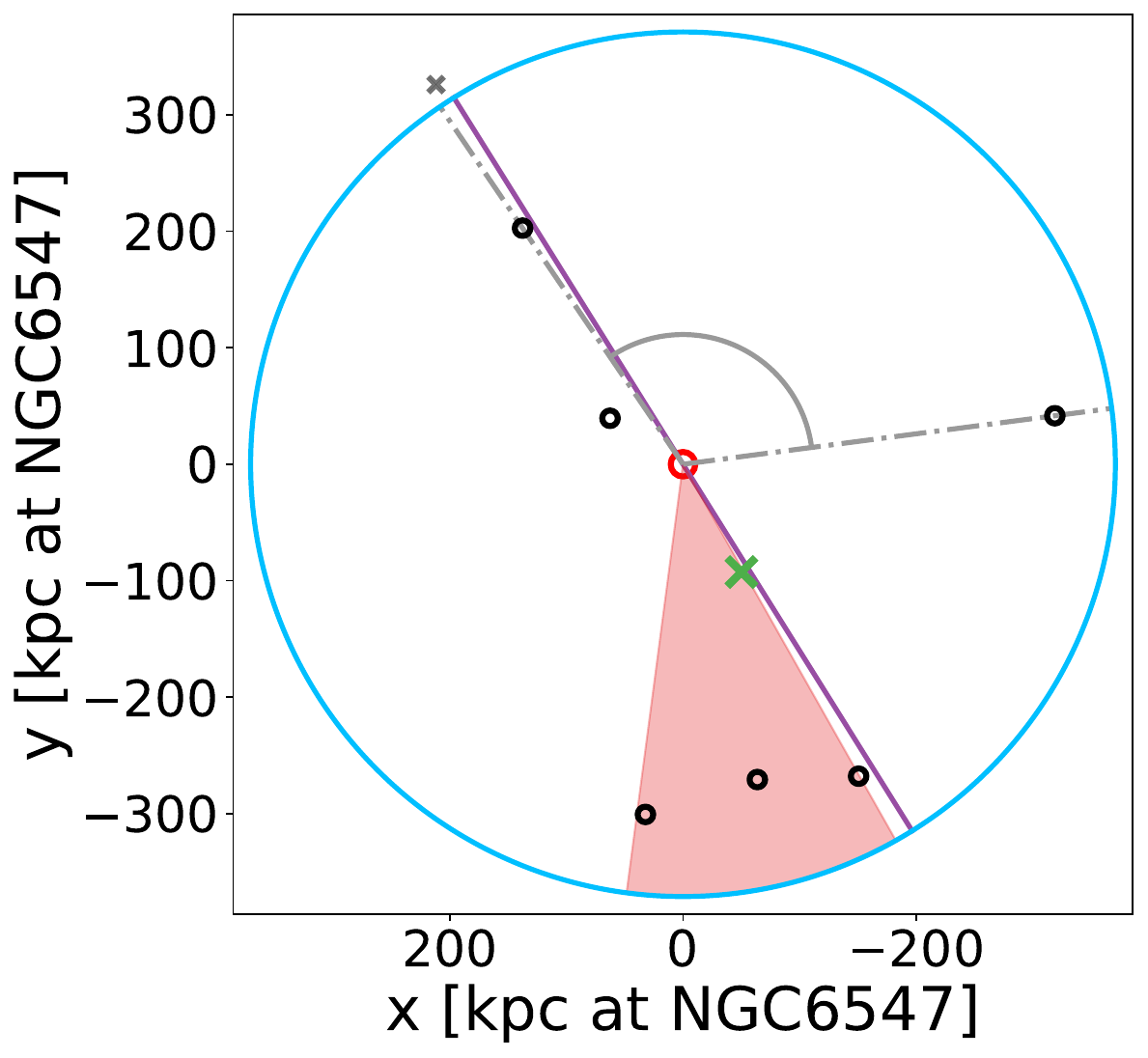}
    \includegraphics[width=6cm]{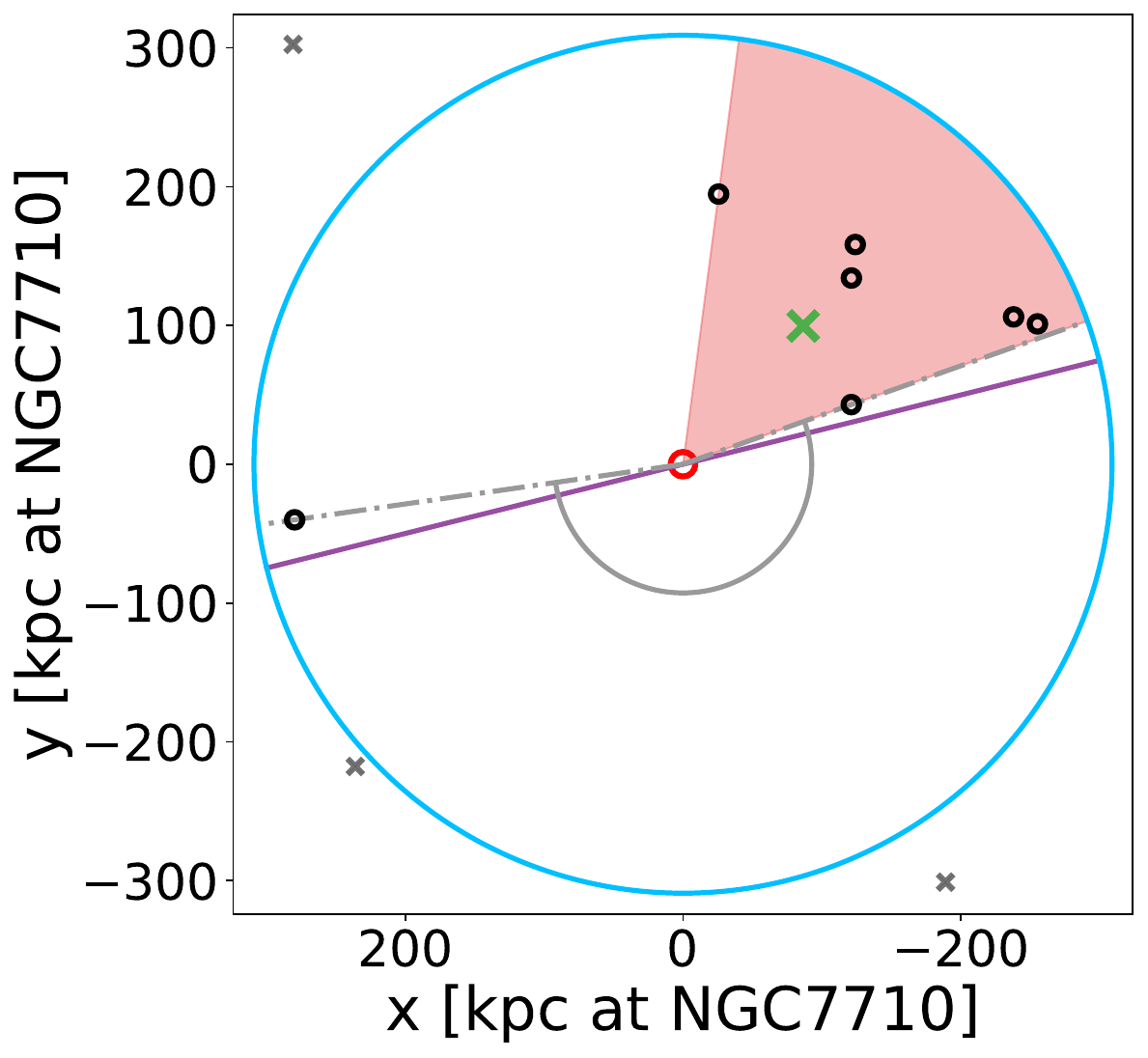}
\caption{Continued.}
    \label{fig:matlas_satellite_dists3}
\end{figure*}

\begin{figure*}[ht]
    \centering
    \ContinuedFloat
    \includegraphics[width=6cm]{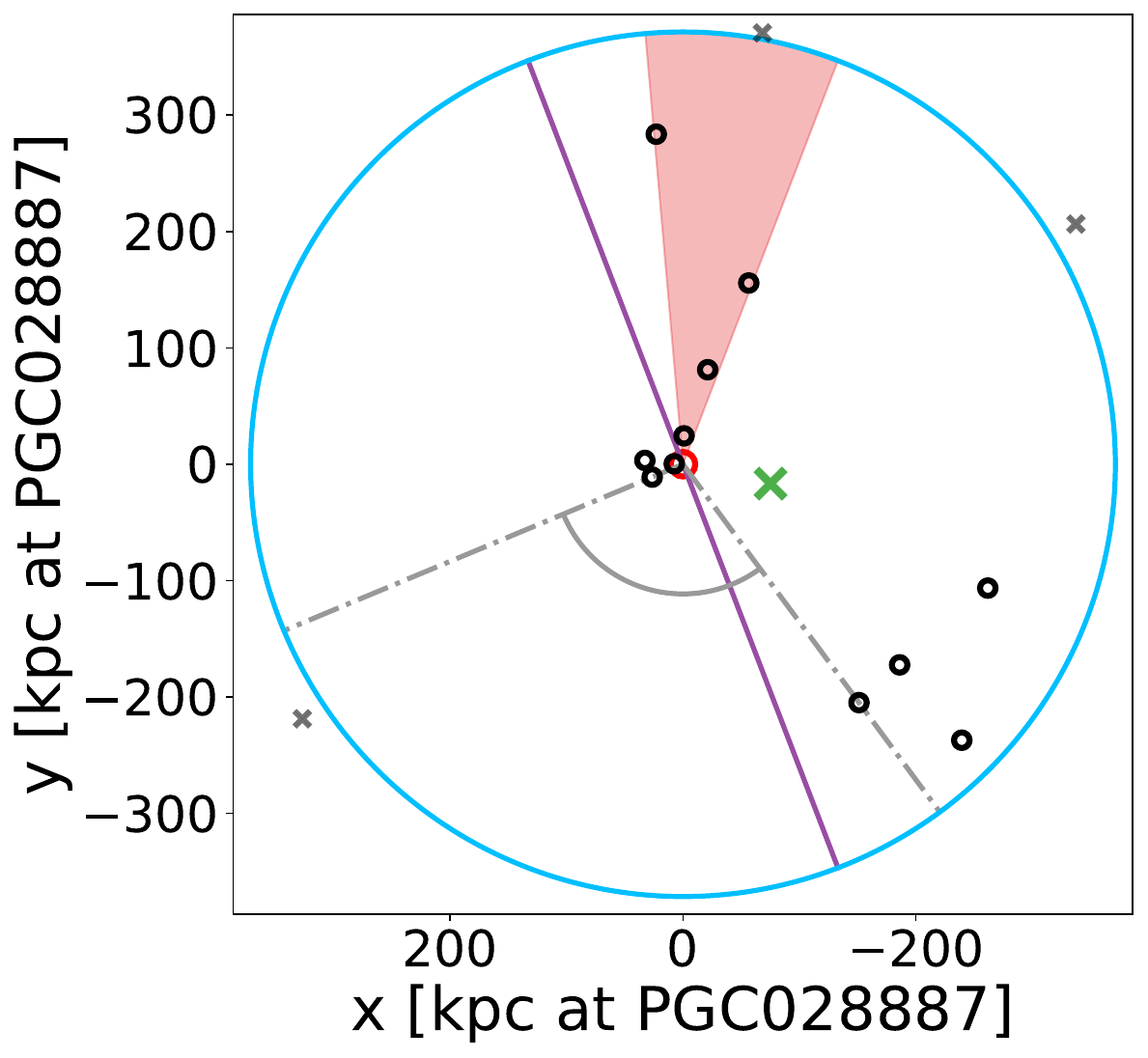}
    \includegraphics[width=6cm]{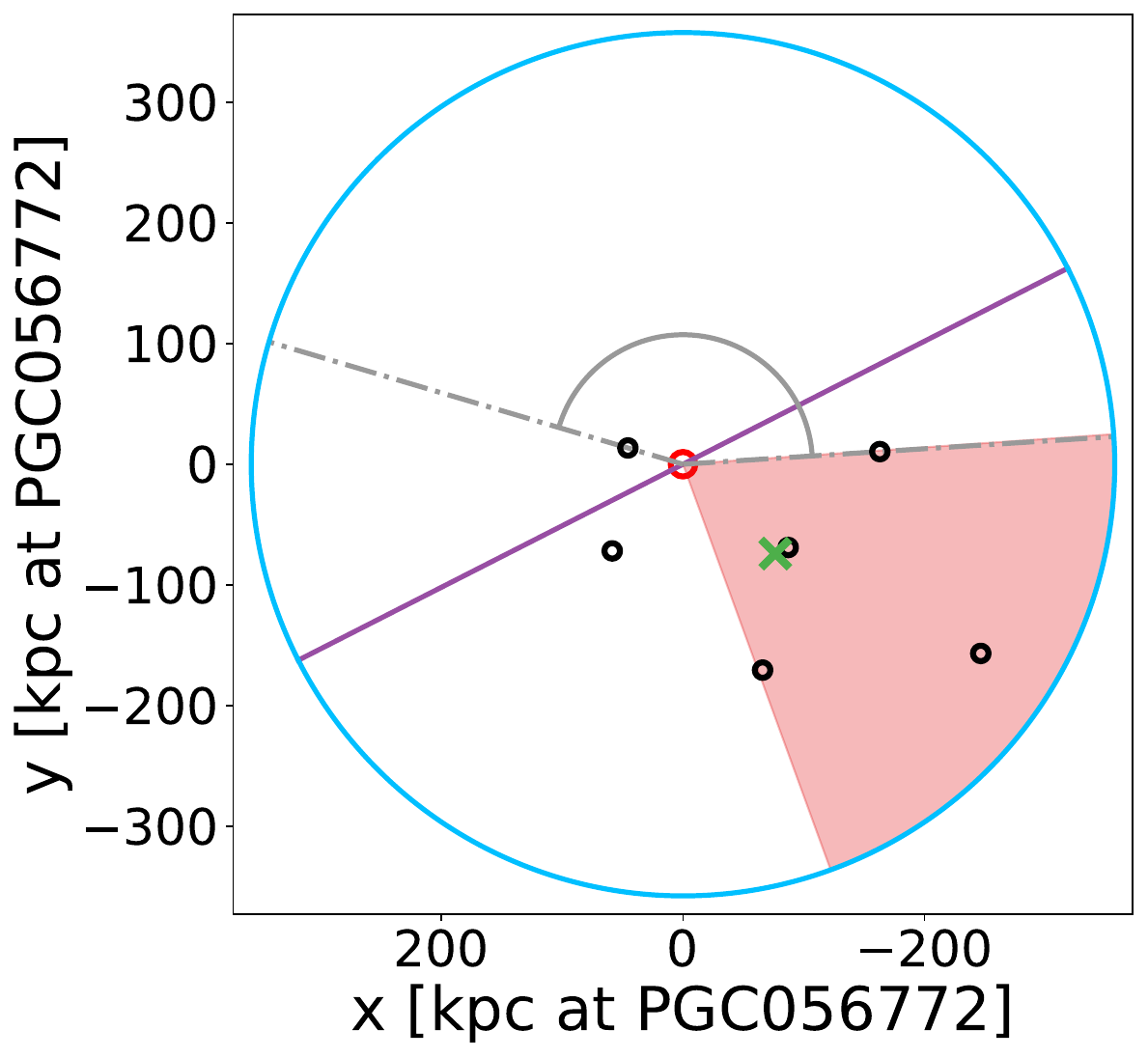}
    \includegraphics[width=6cm]{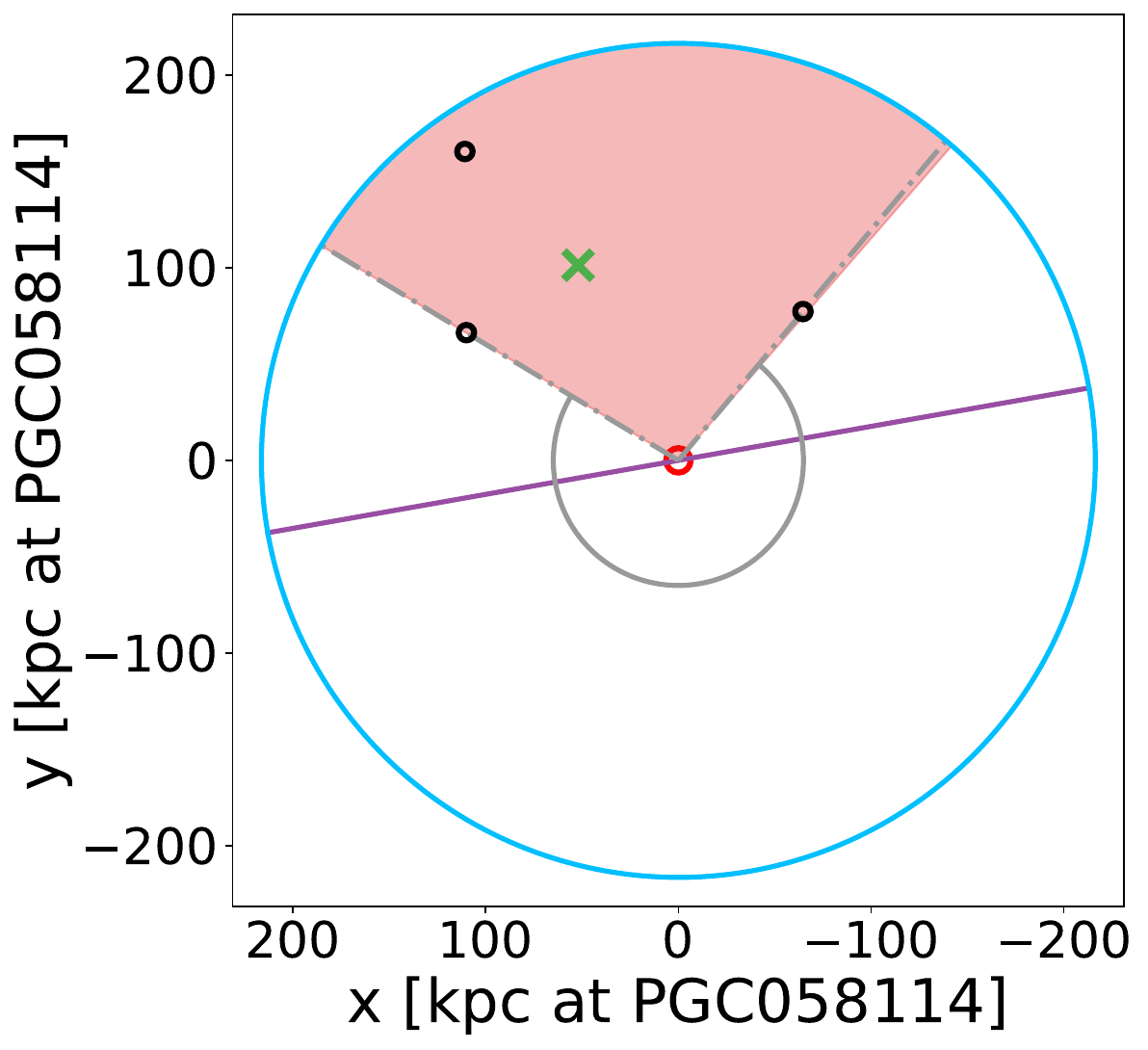}
    \includegraphics[width=6cm]{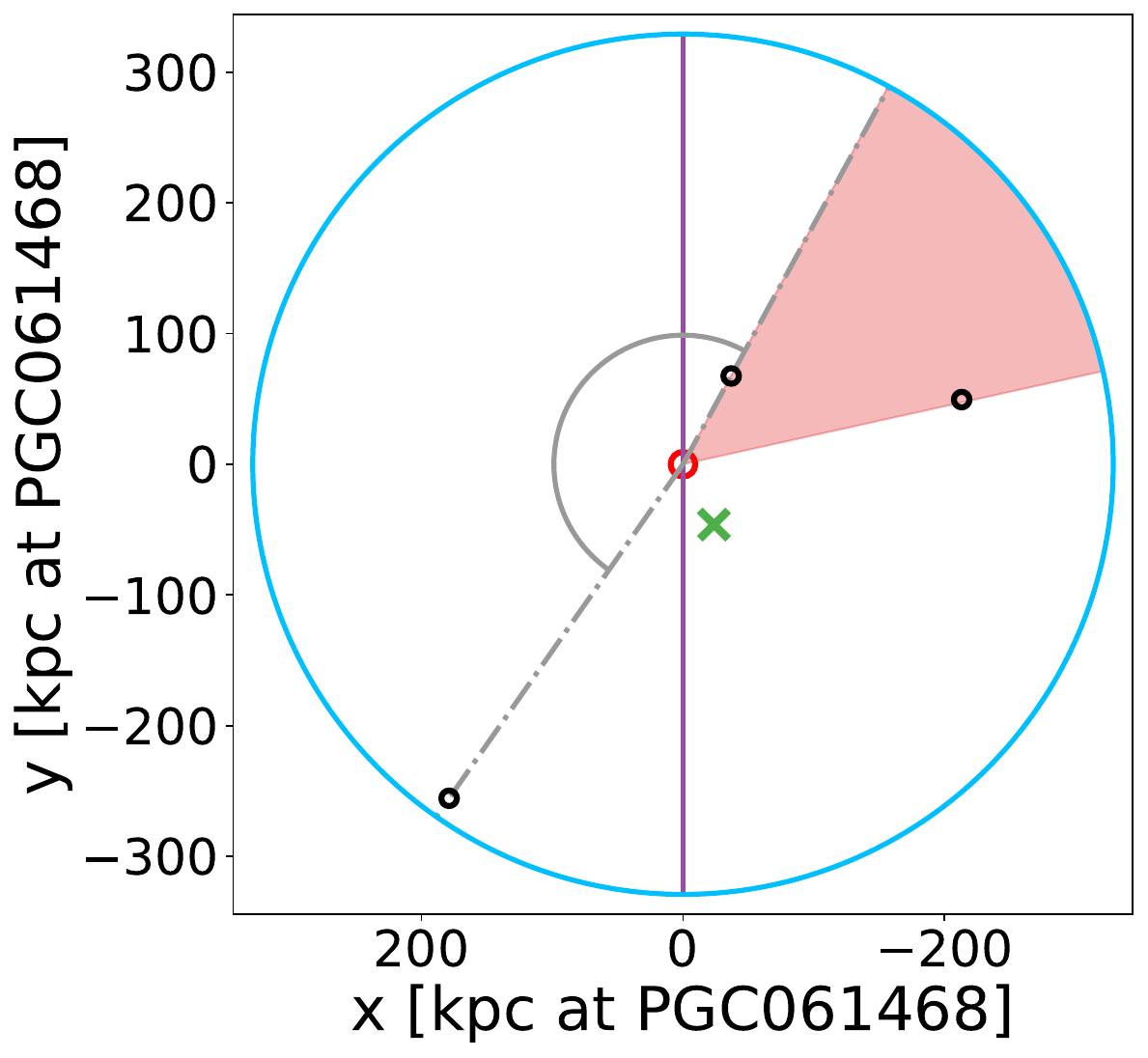}
    \includegraphics[width=6cm]{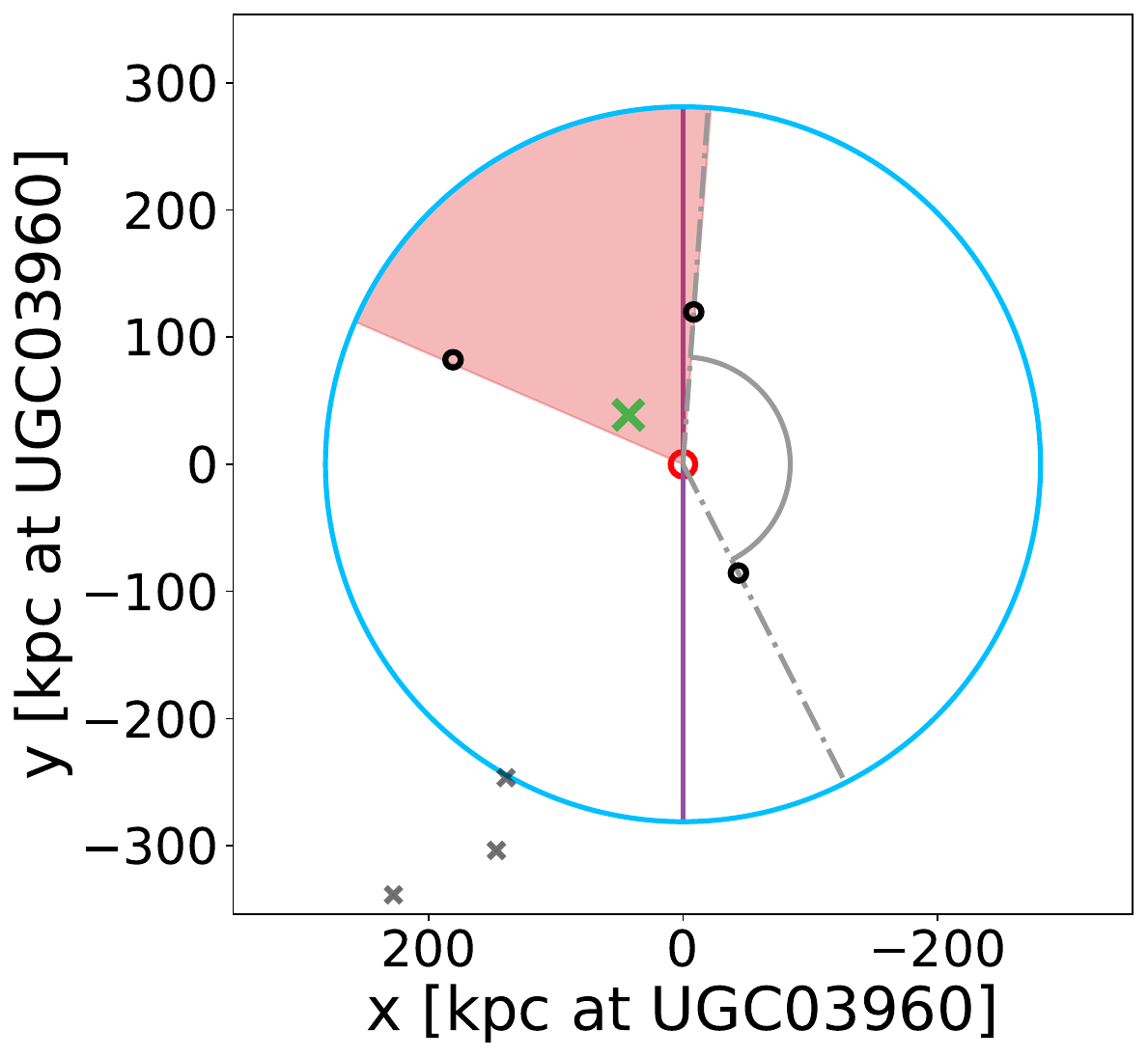}
    \includegraphics[width=6cm]{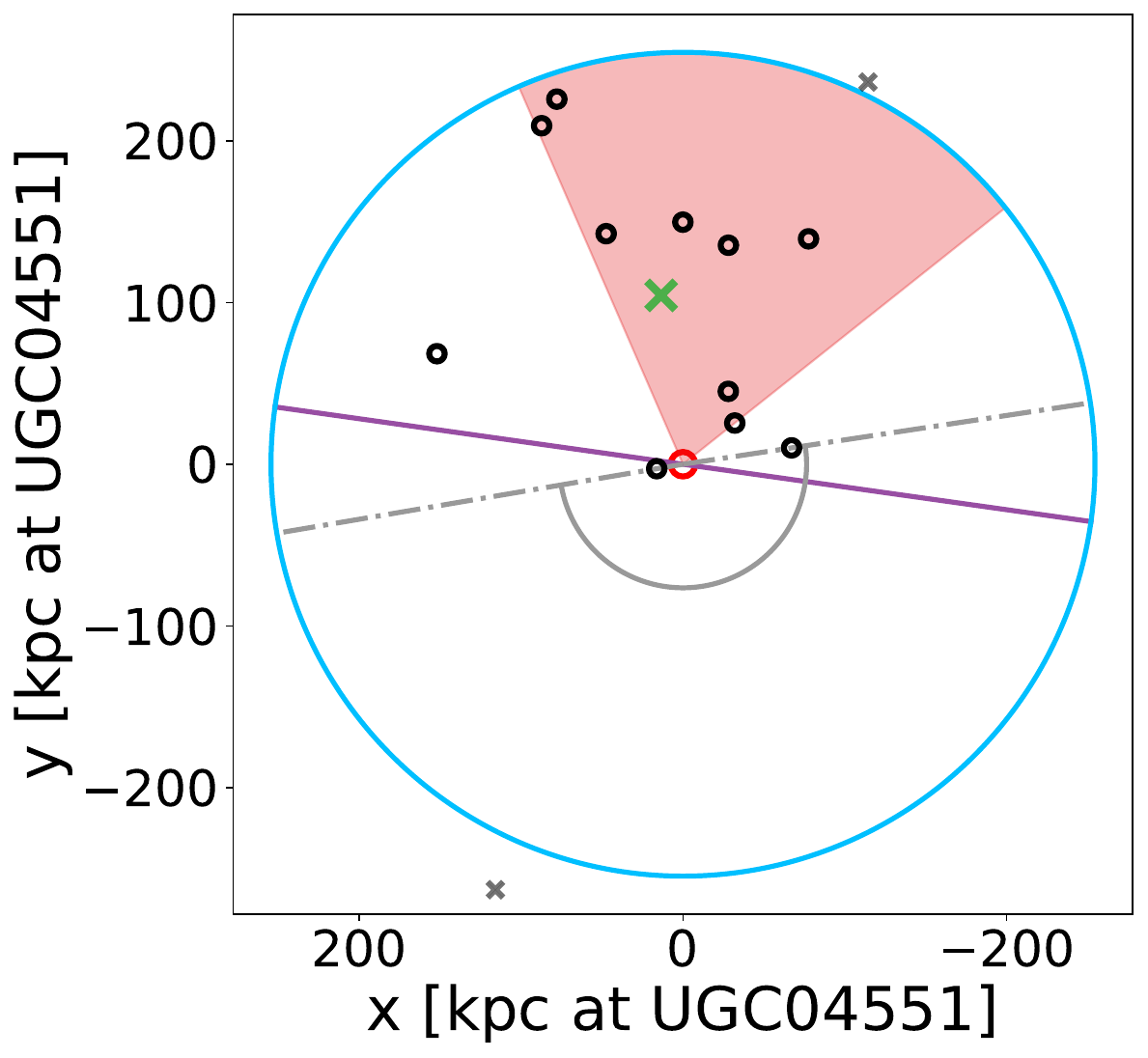}
    \includegraphics[width=6cm]{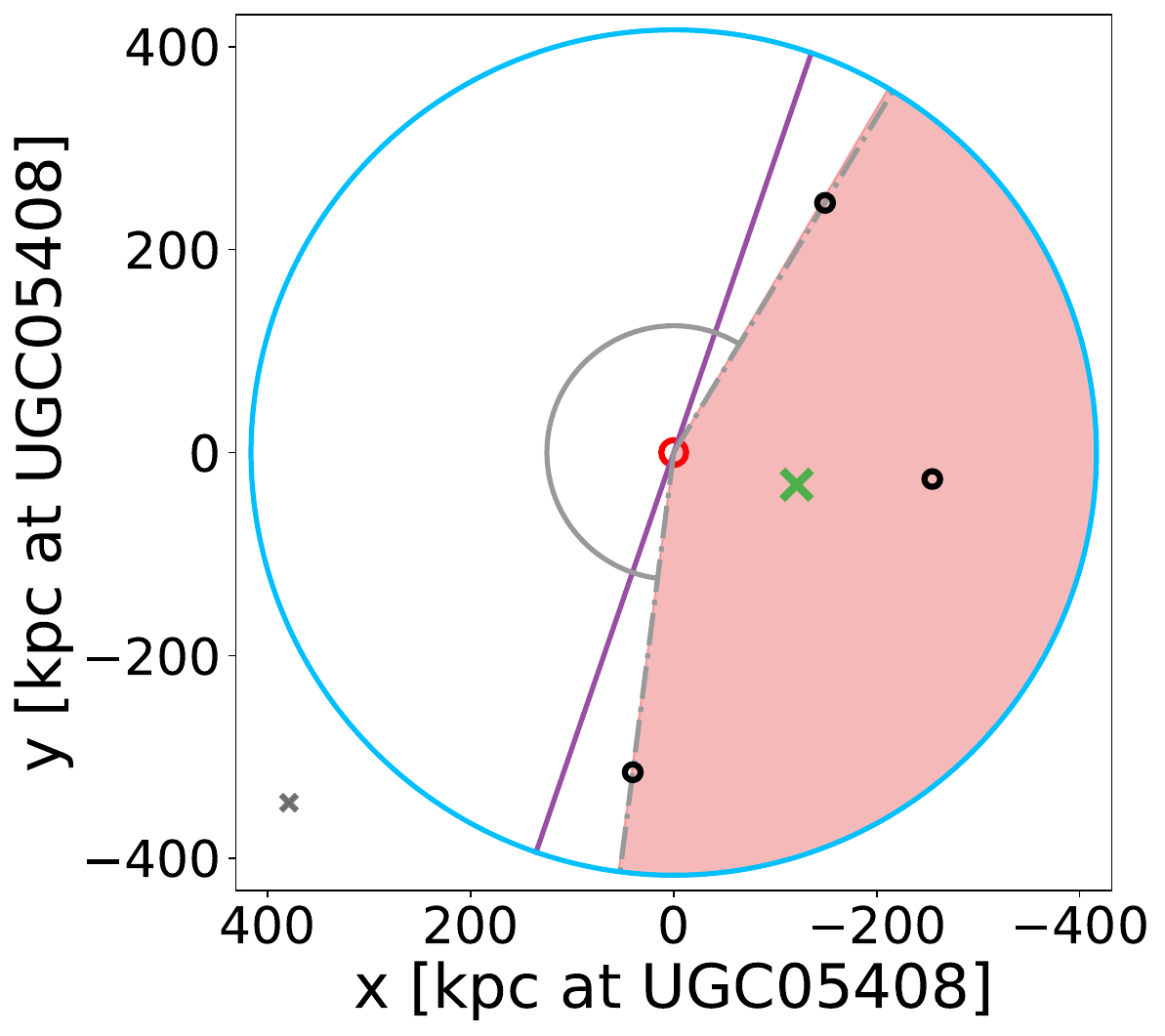}
    \includegraphics[width=6cm]{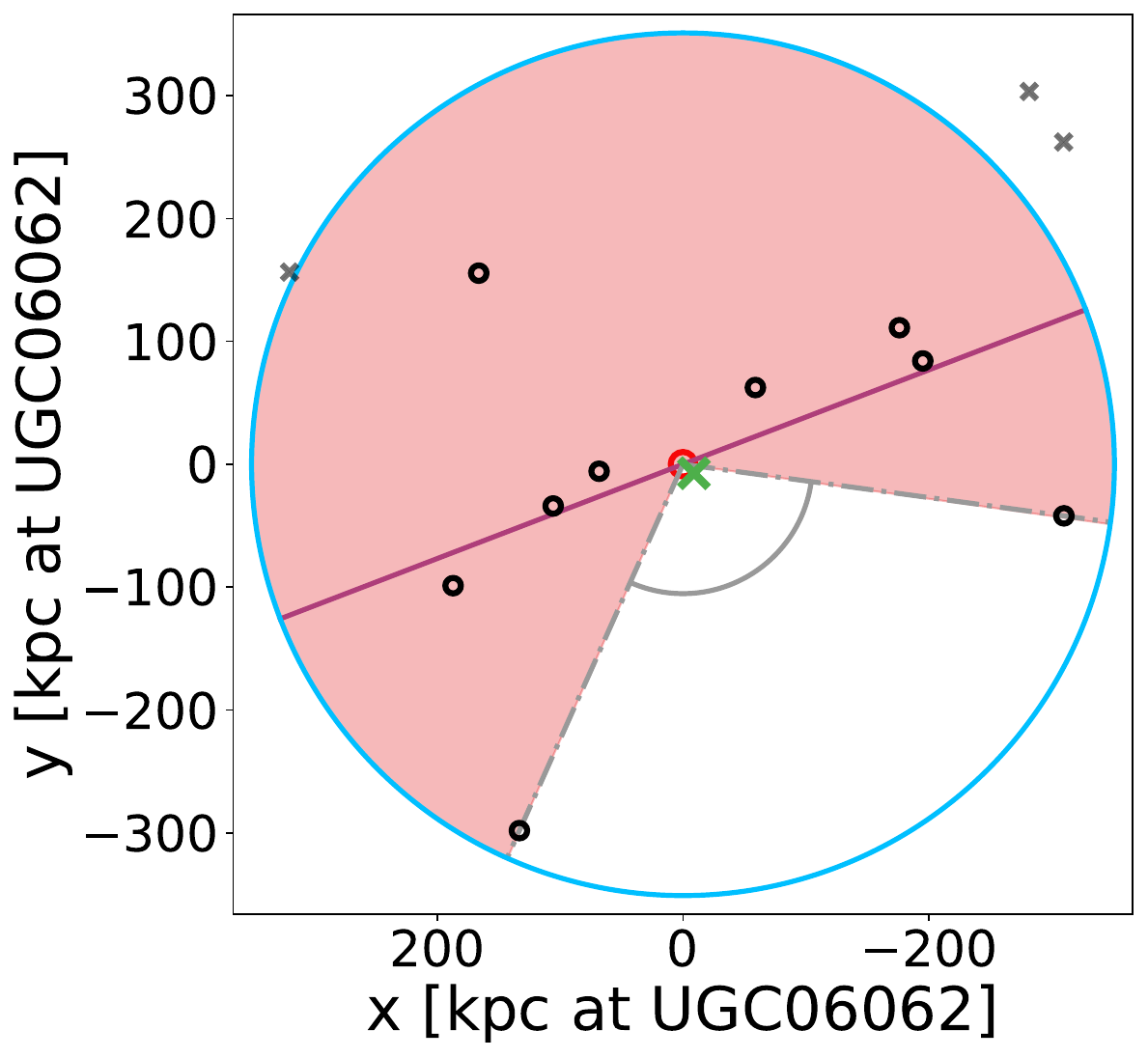}
    \includegraphics[width=6cm]{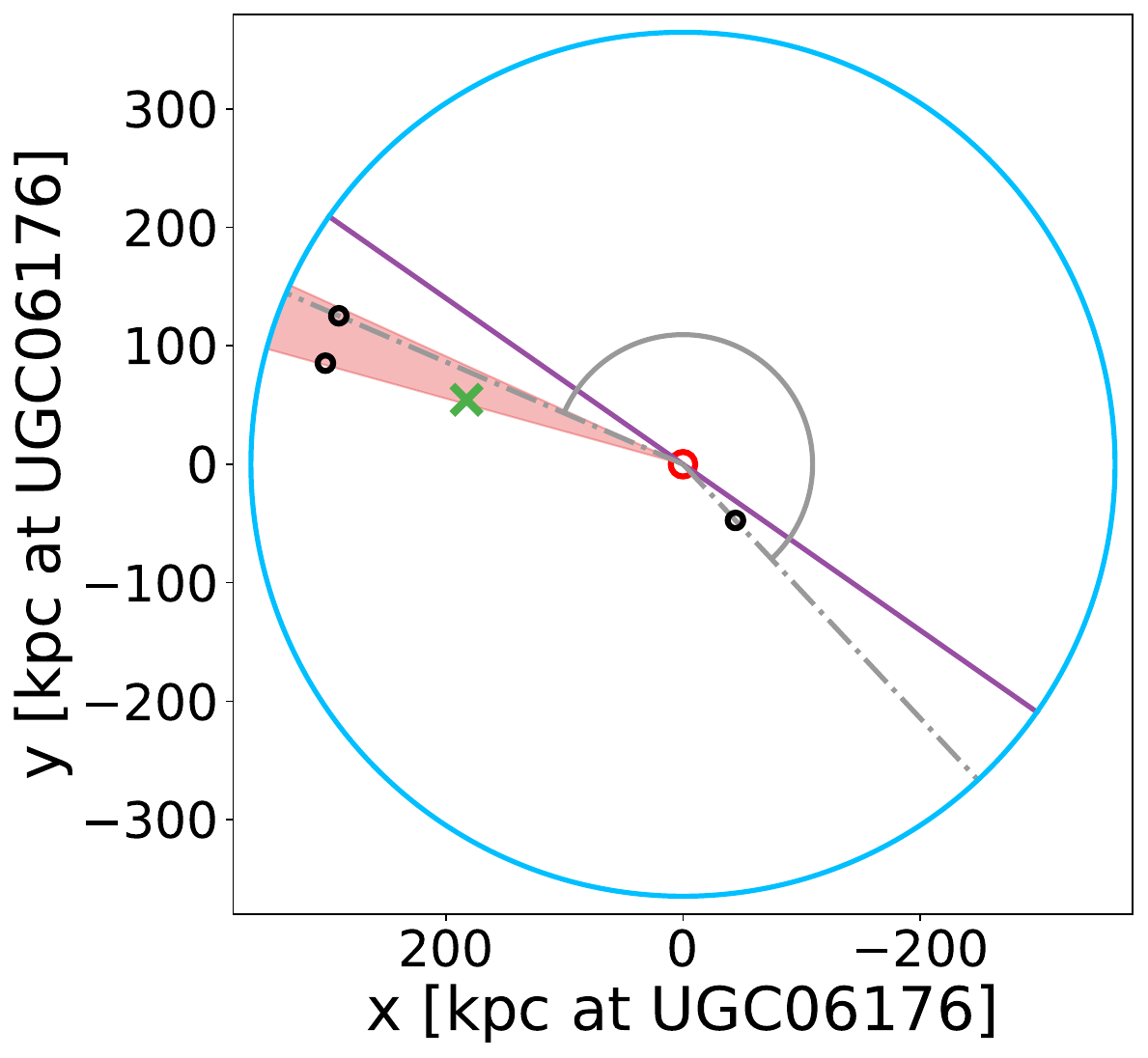}
    \includegraphics[width=6cm]{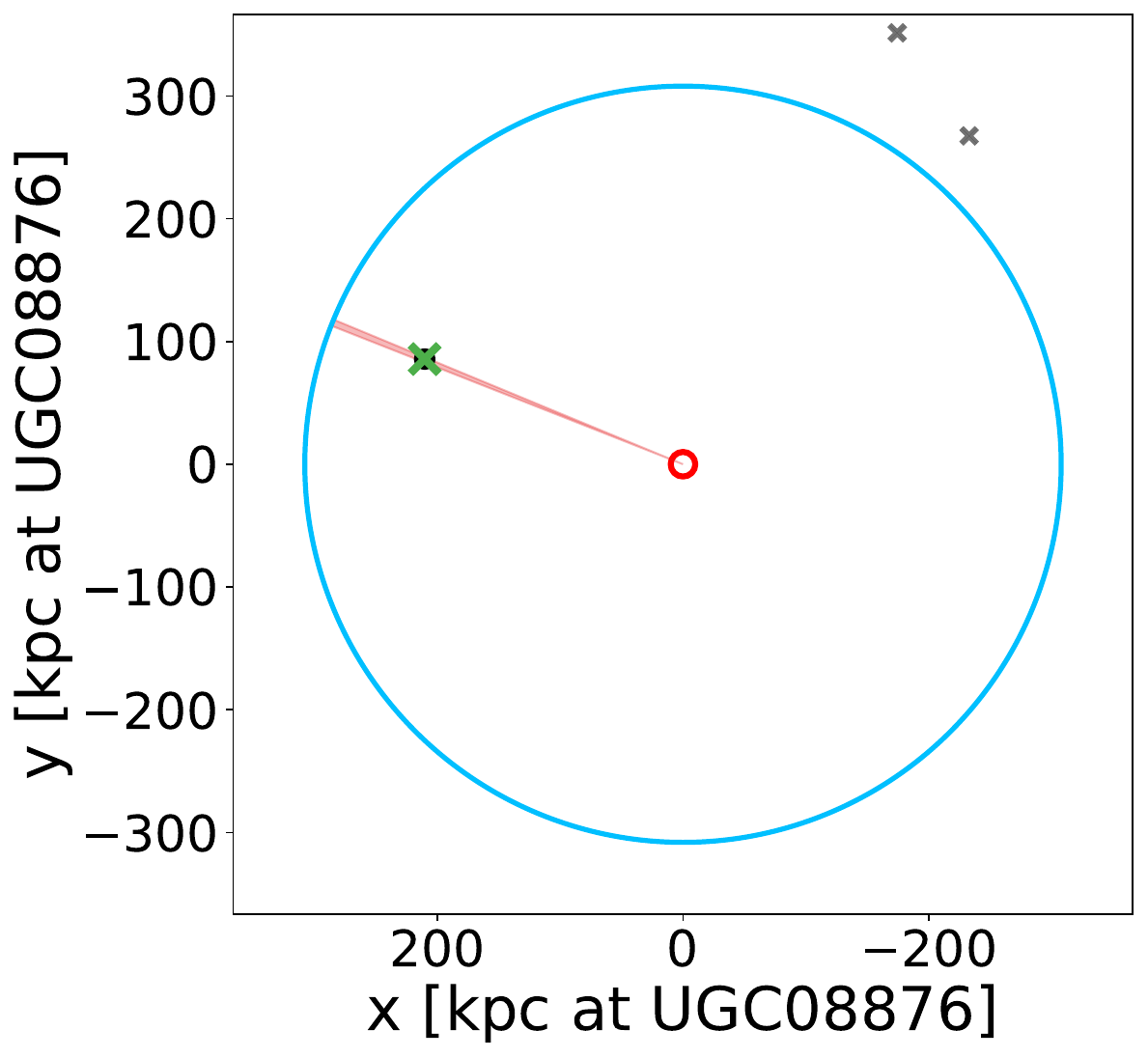}
    \includegraphics[width=6cm]{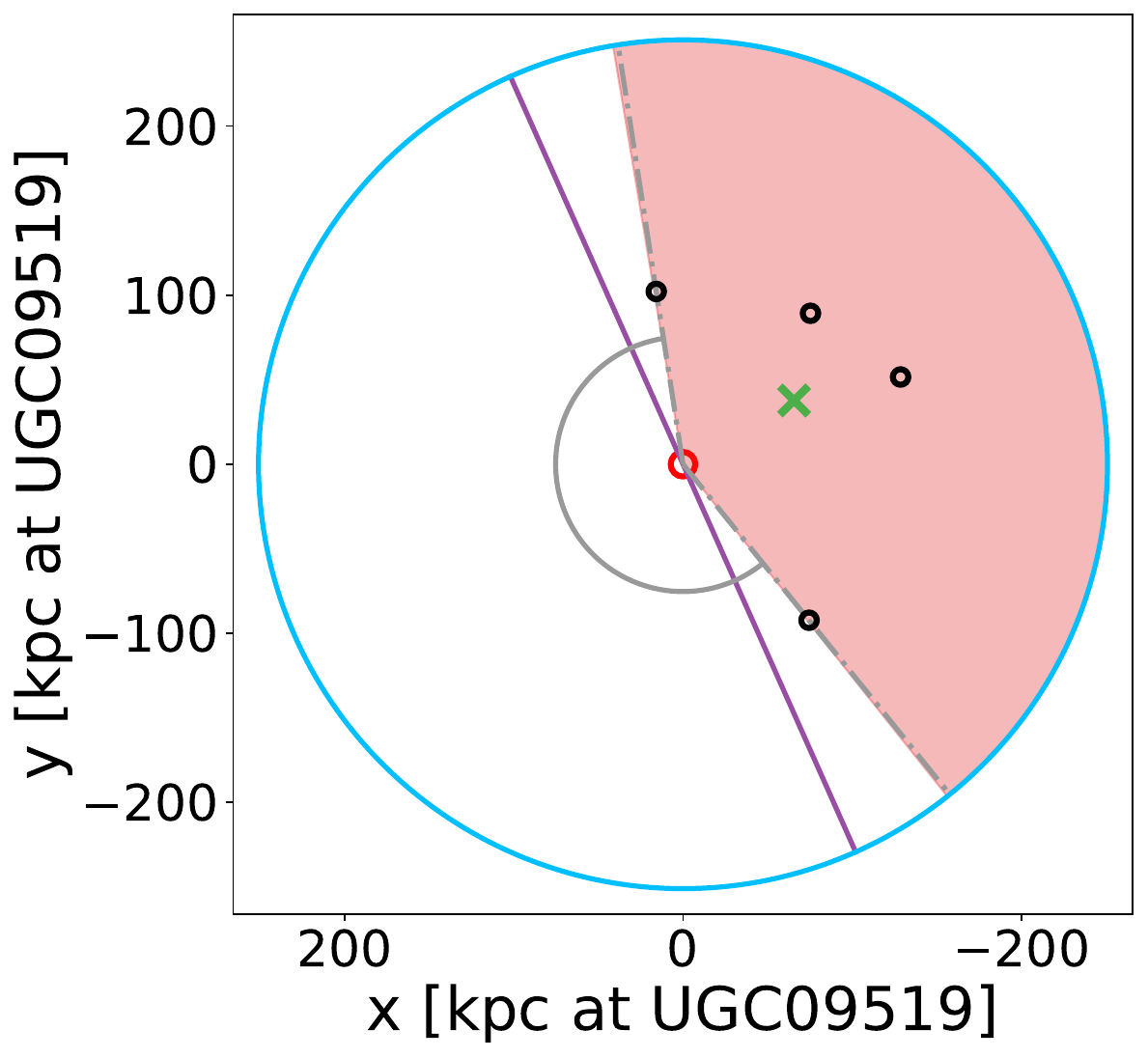}
\caption{Continued.}
    \label{fig:matlas_satellite_dists4}
\end{figure*}

\begin{figure*}[ht]
    \centering
    \includegraphics[width=6cm]{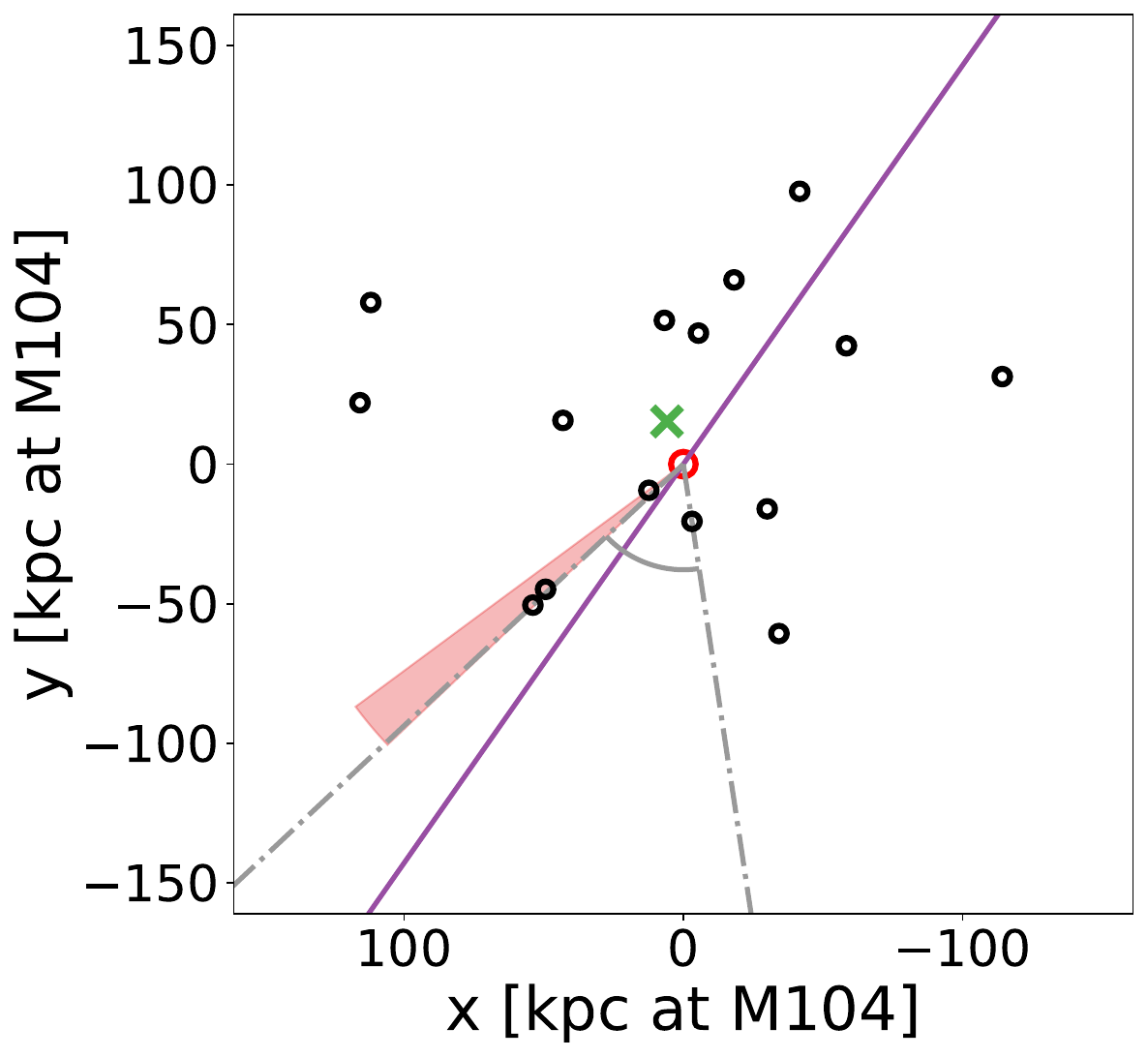}
    \includegraphics[width=6cm]{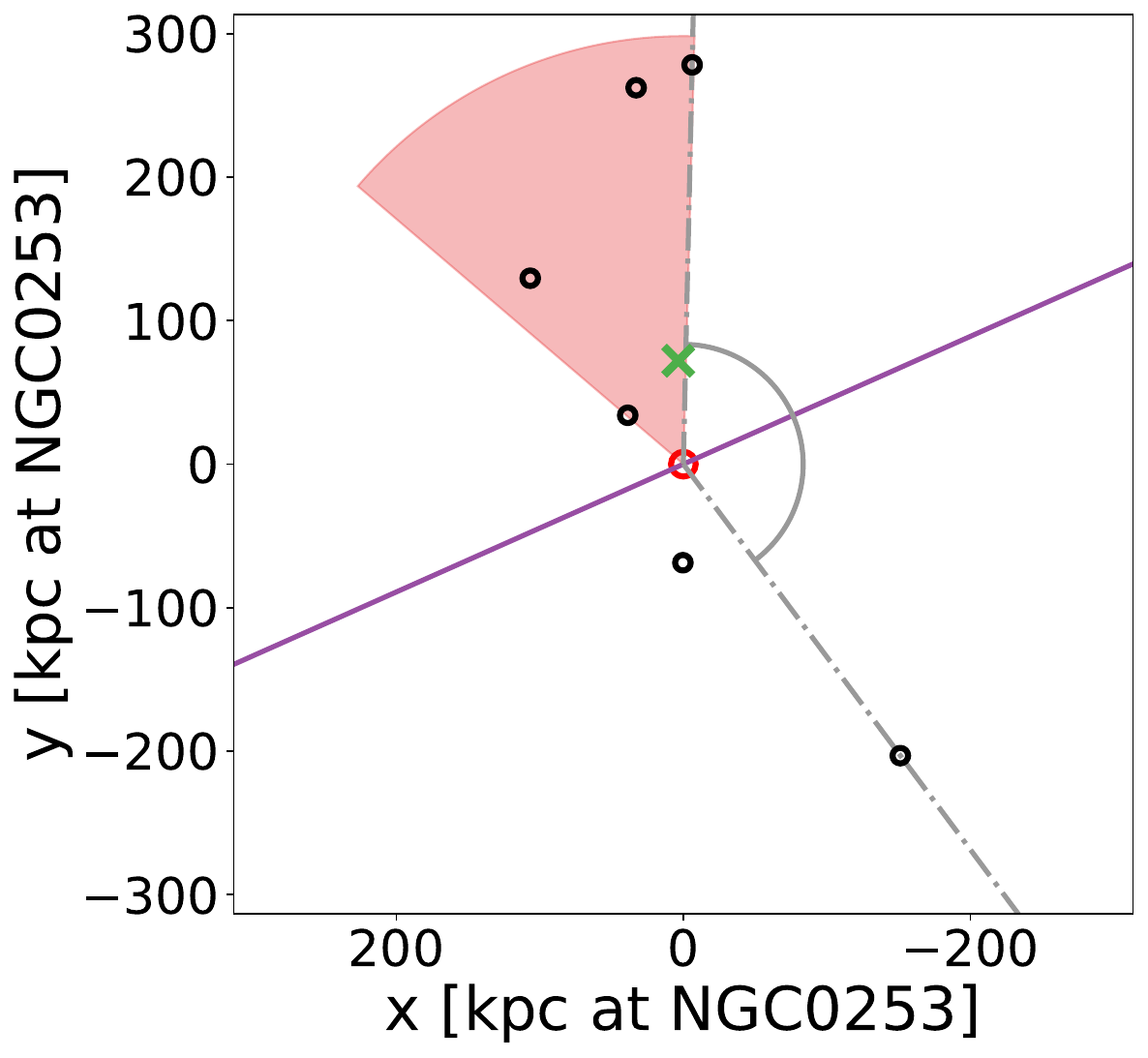}
    \includegraphics[width=6cm]{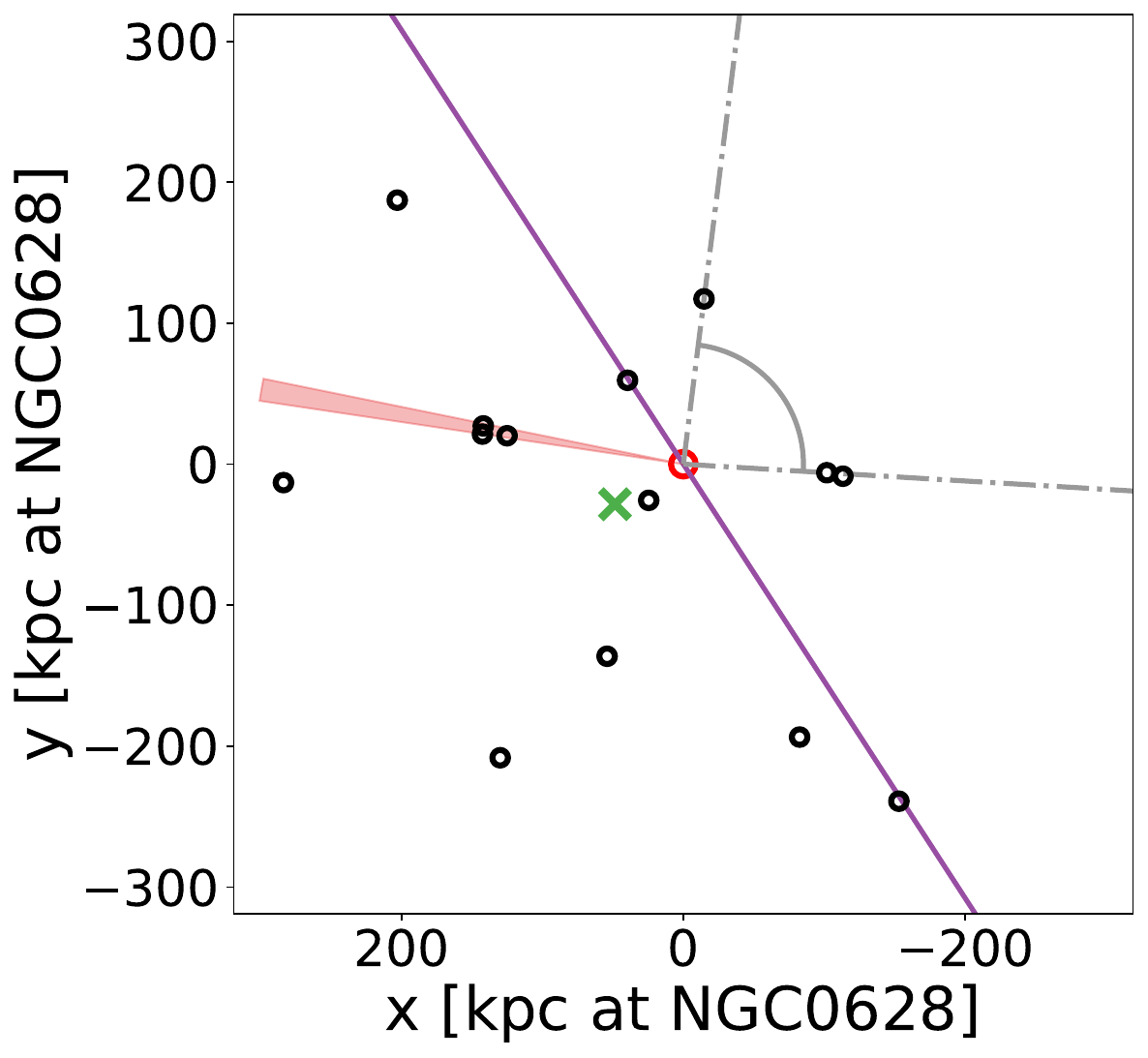}
    \includegraphics[width=6cm]{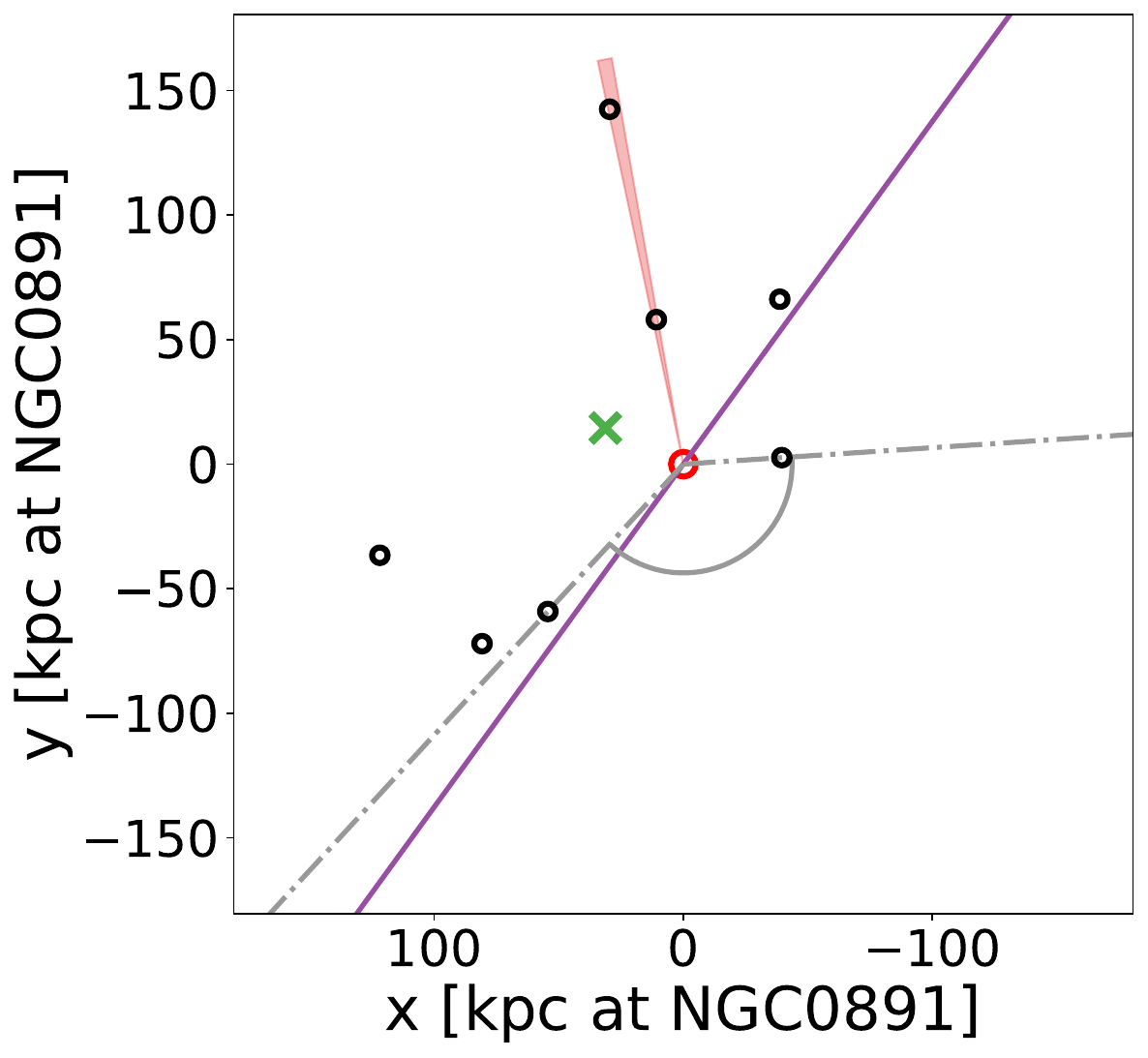}
    \includegraphics[width=6cm]{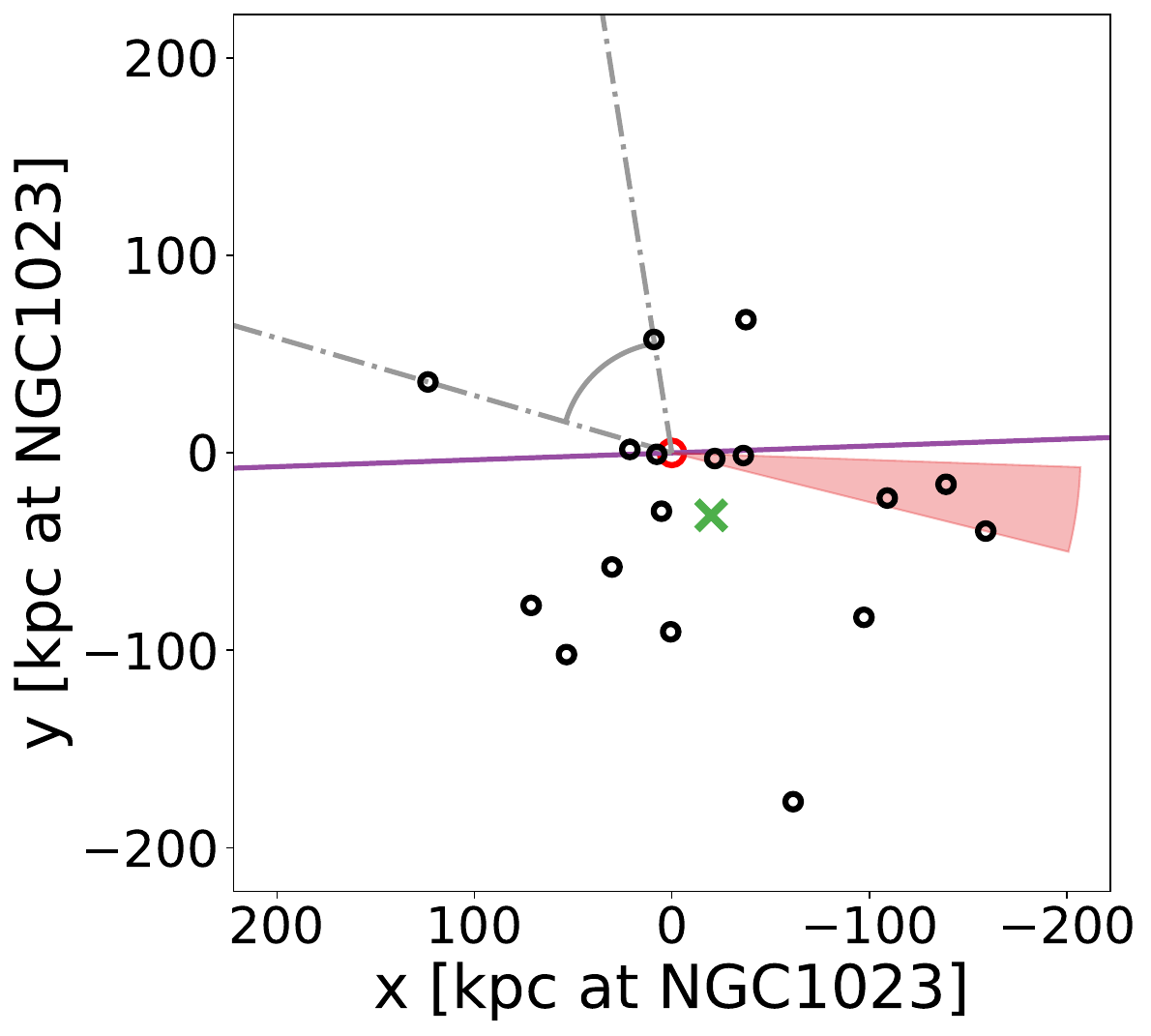}
    \includegraphics[width=6cm]{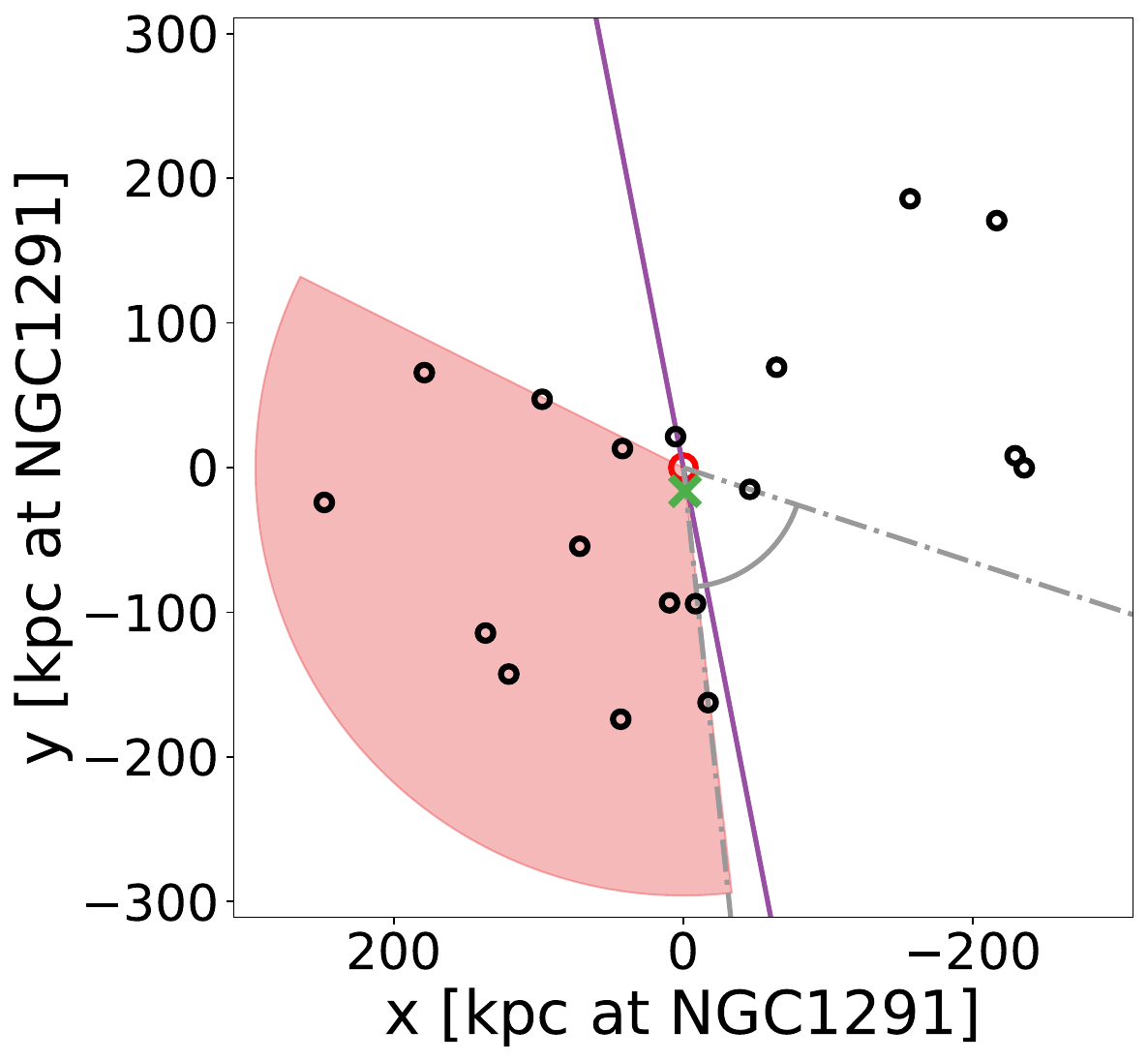}
    \includegraphics[width=6cm]{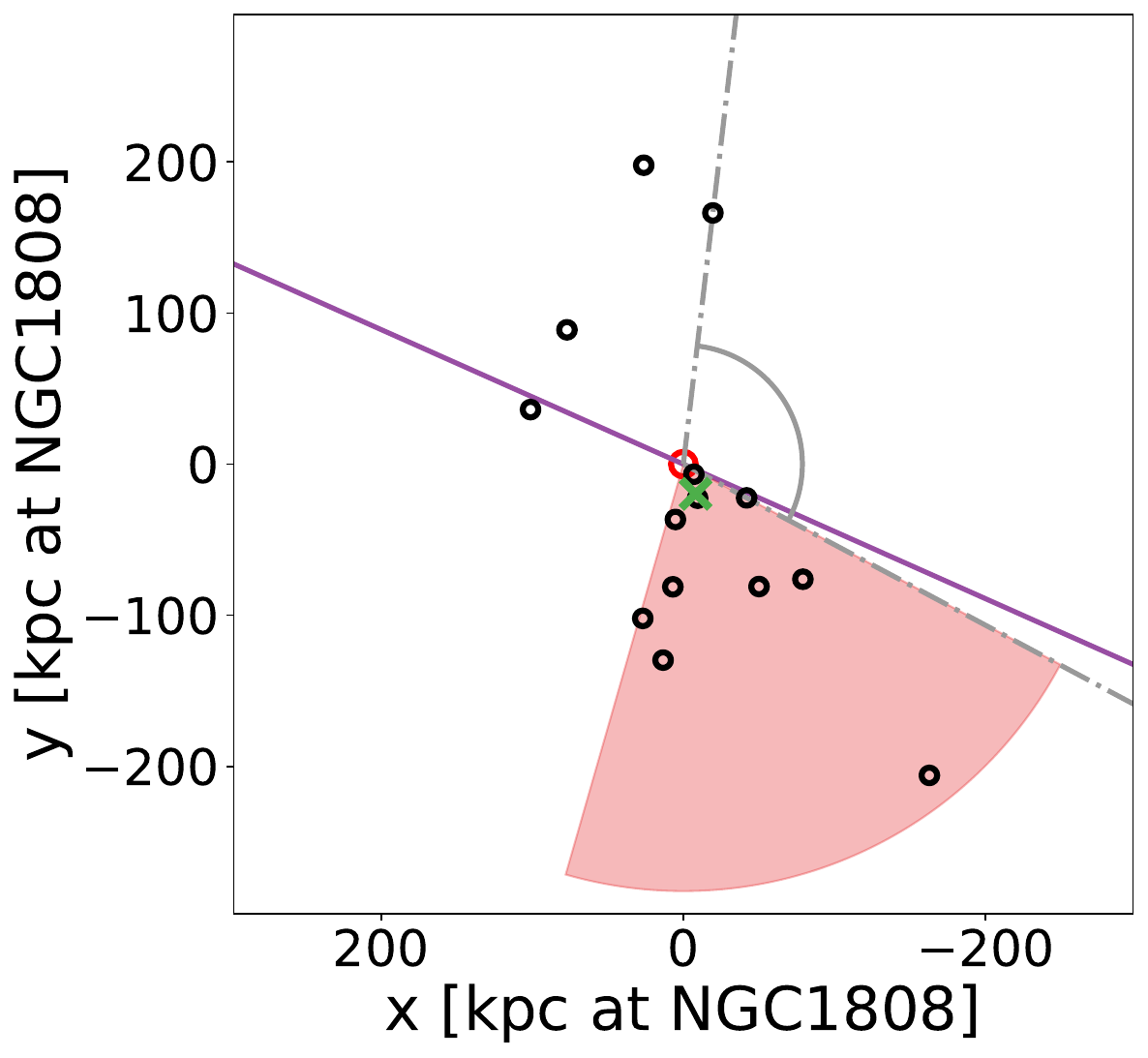}
    \includegraphics[width=6cm]{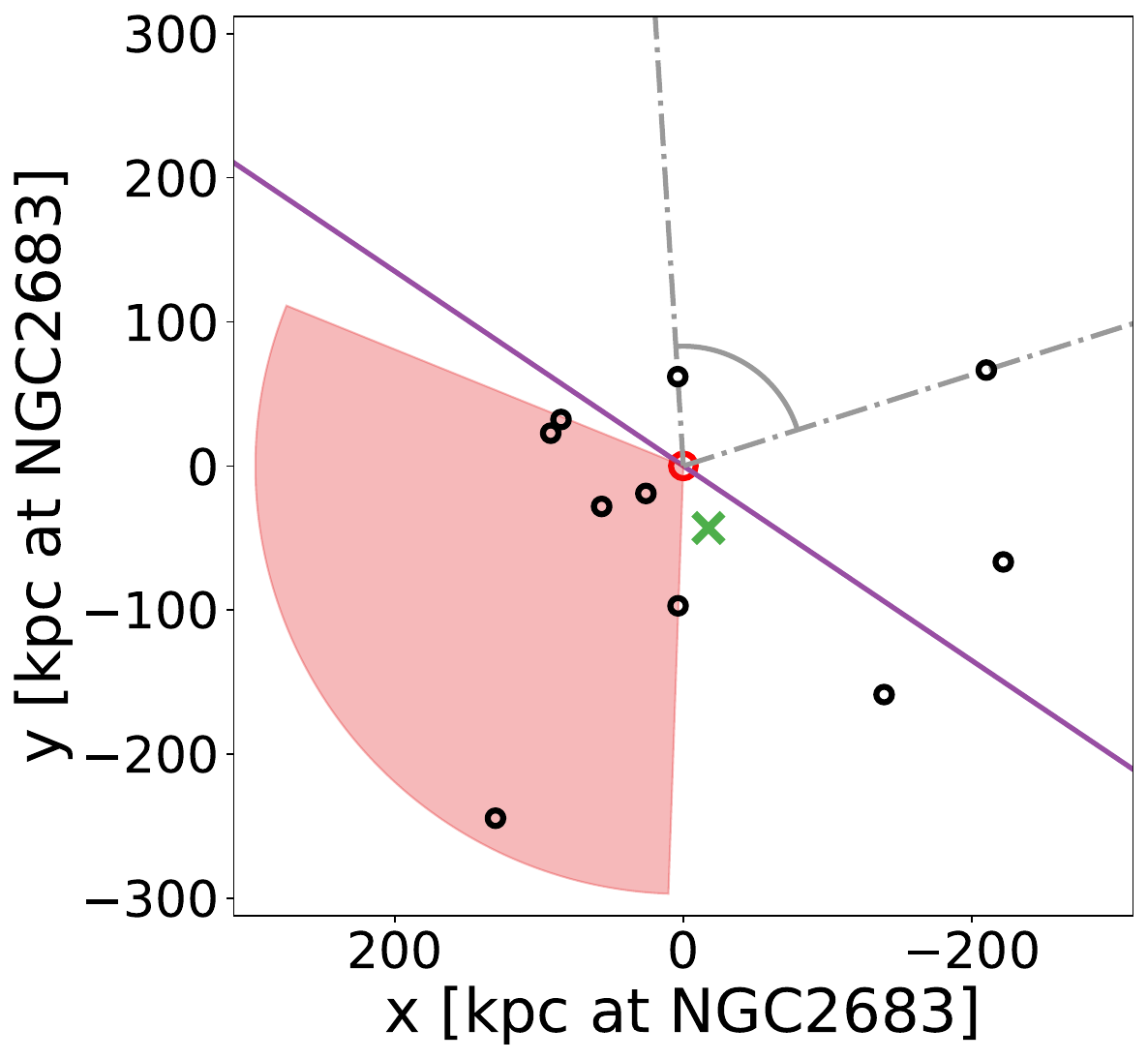}
    \includegraphics[width=6cm]{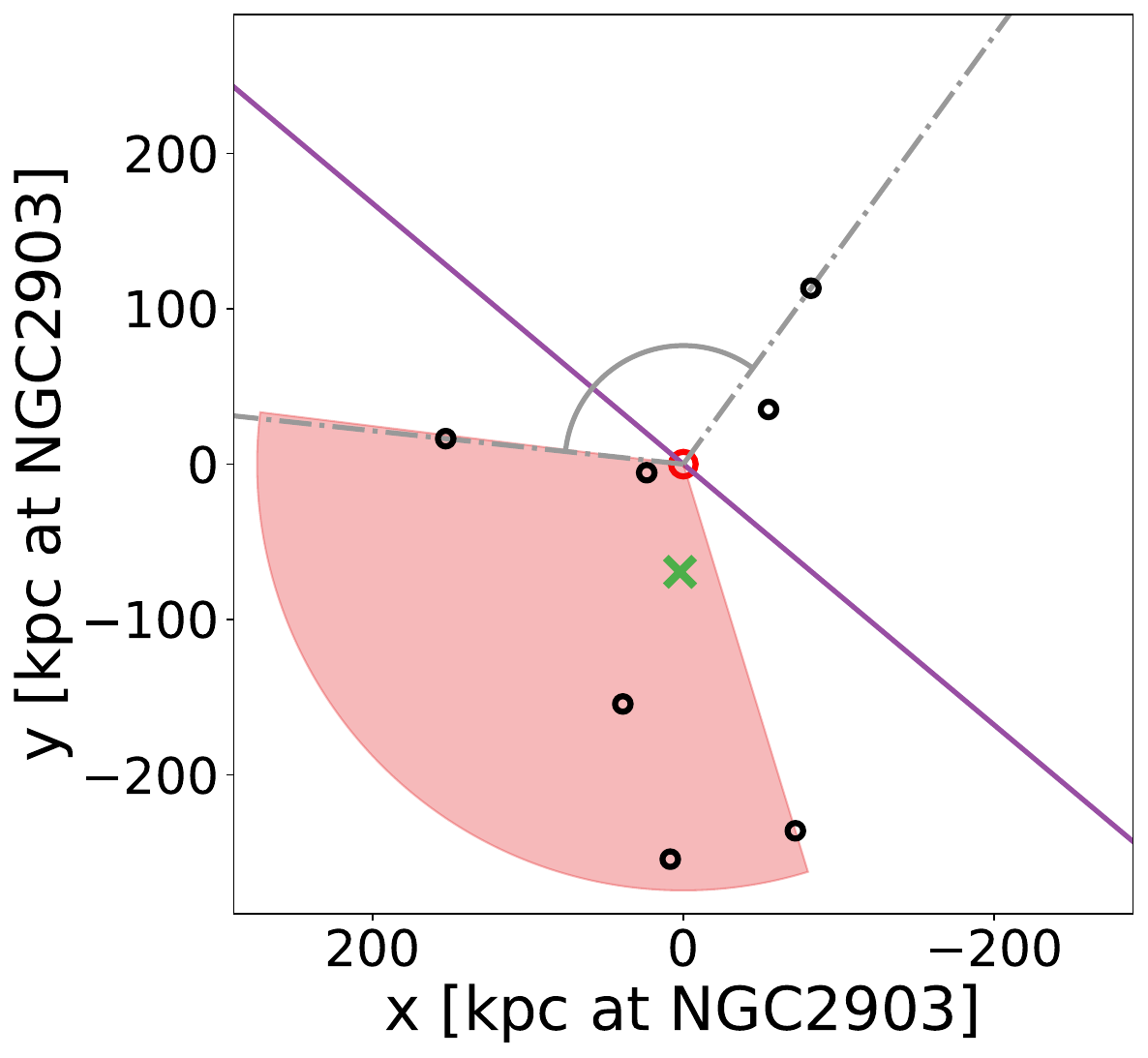}
    \includegraphics[width=6cm]{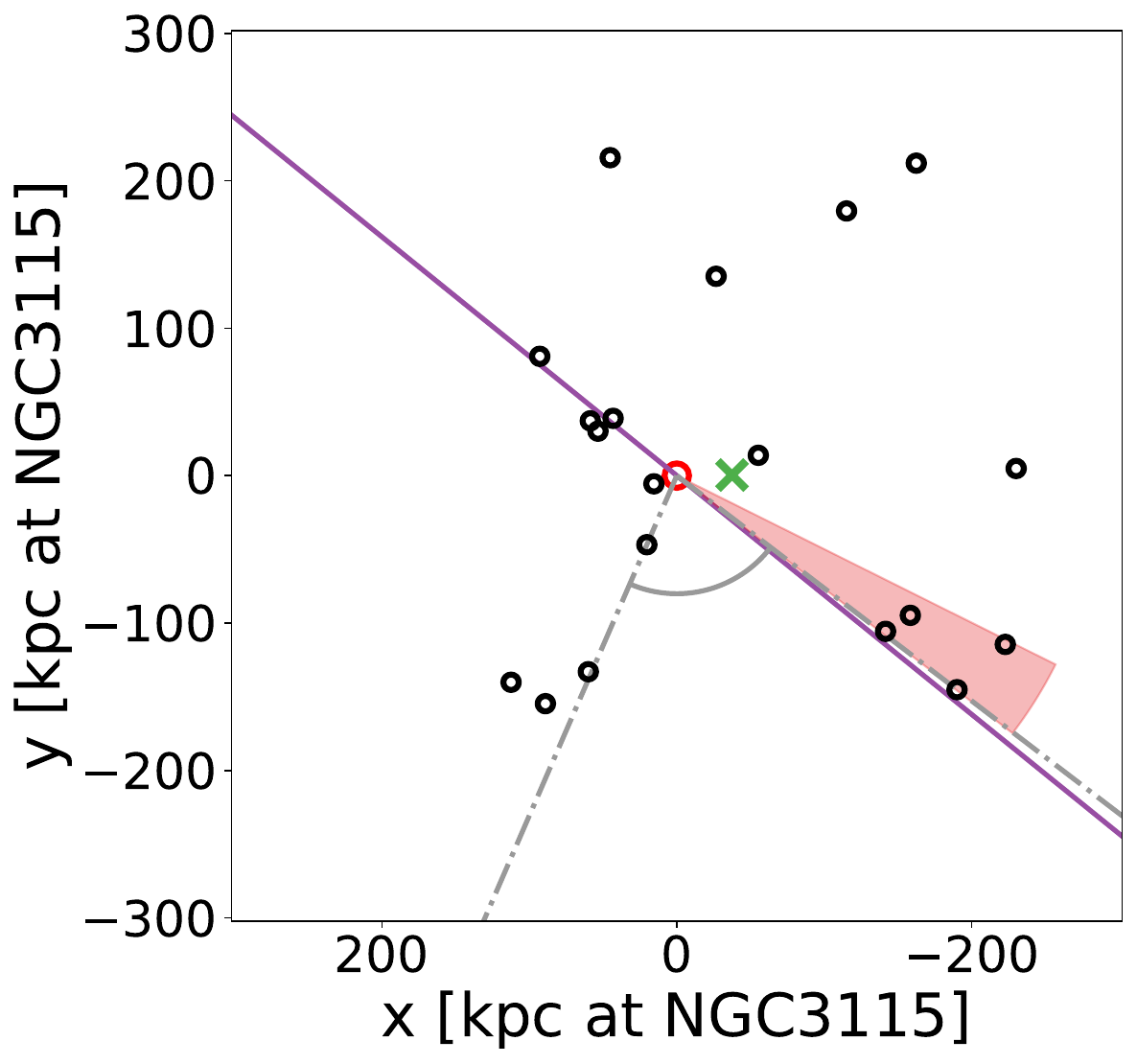}
    \includegraphics[width=6cm]{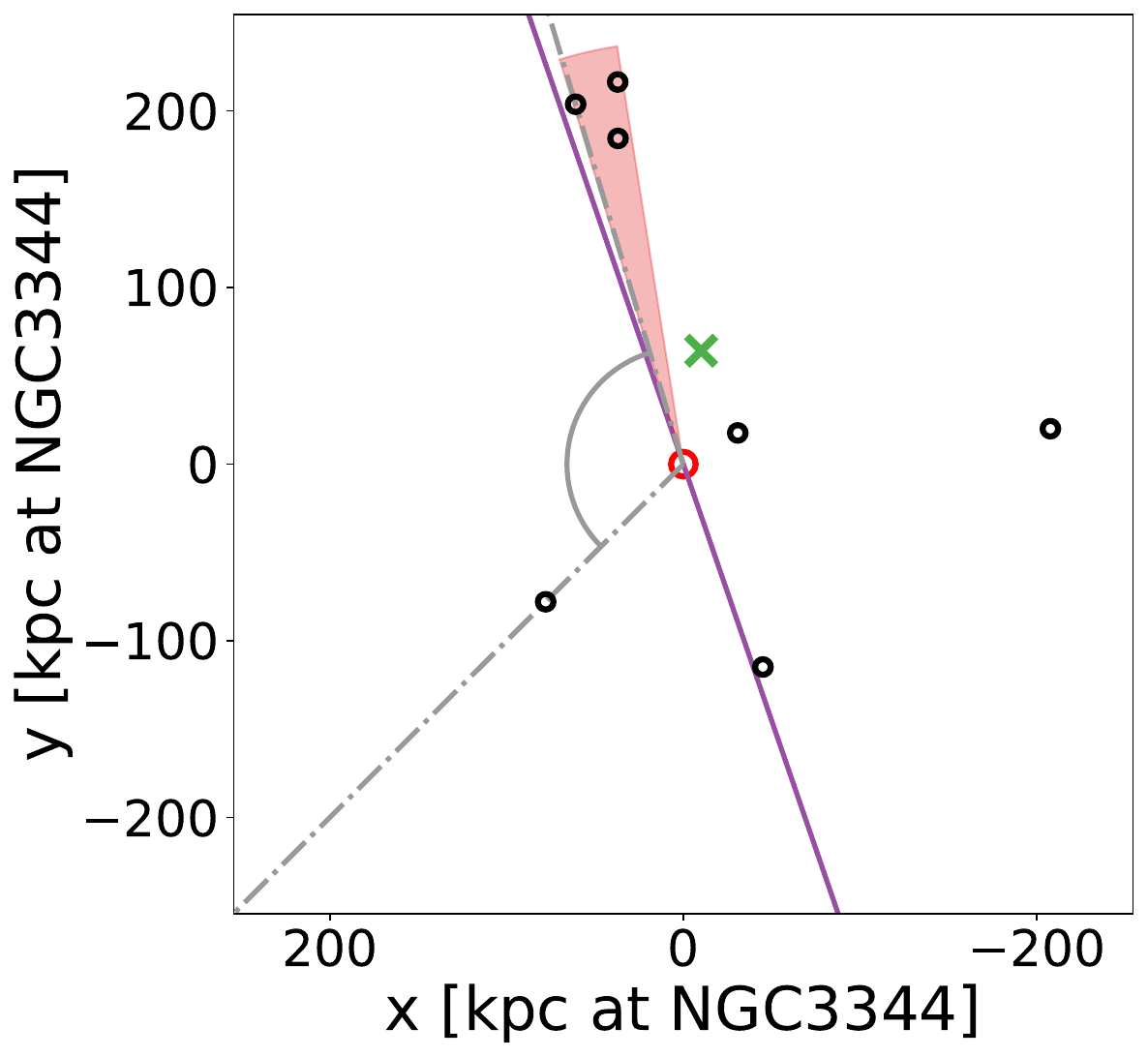}
    \includegraphics[width=6cm]{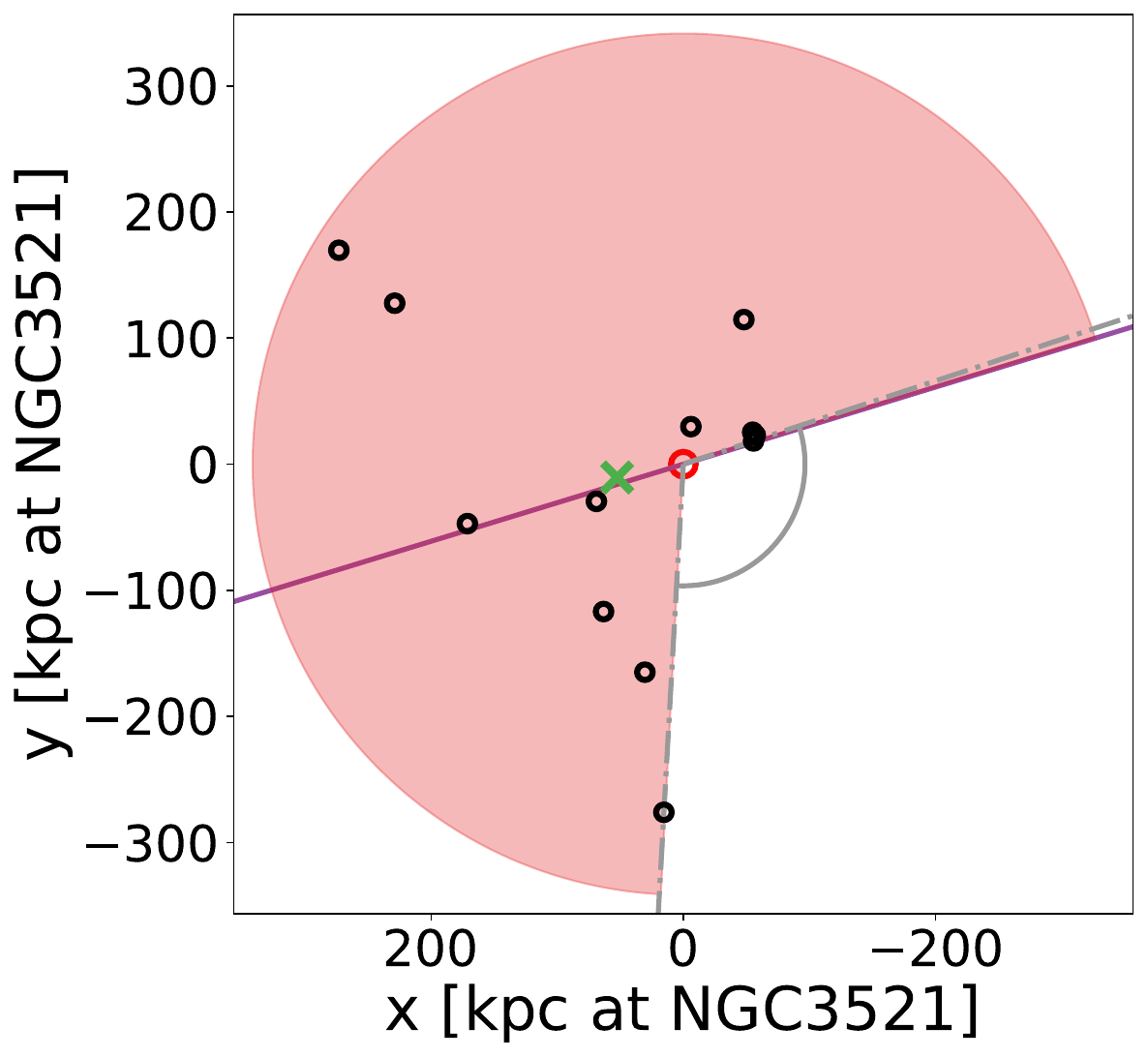}
    \caption{Same as Fig. \ref{fig:matlas_satellite_dists1} for the satellite distributions around host galaxies in the ELVES survey.
}
    \label{fig:elves_satellite_dists1}
\end{figure*}

\begin{figure*}[ht]
    \centering
    \ContinuedFloat
    \includegraphics[width=6cm]{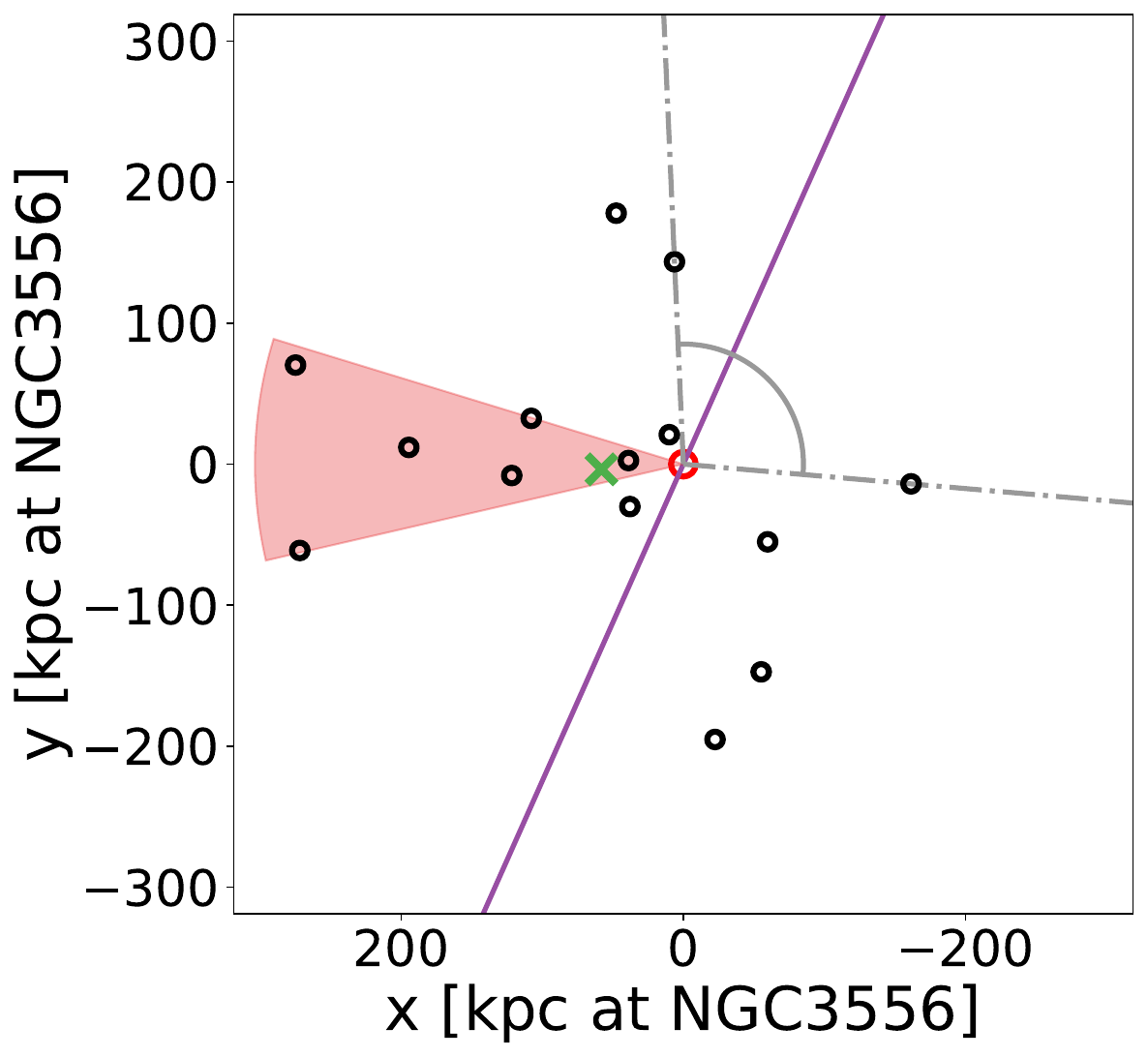}
    \includegraphics[width=6cm]{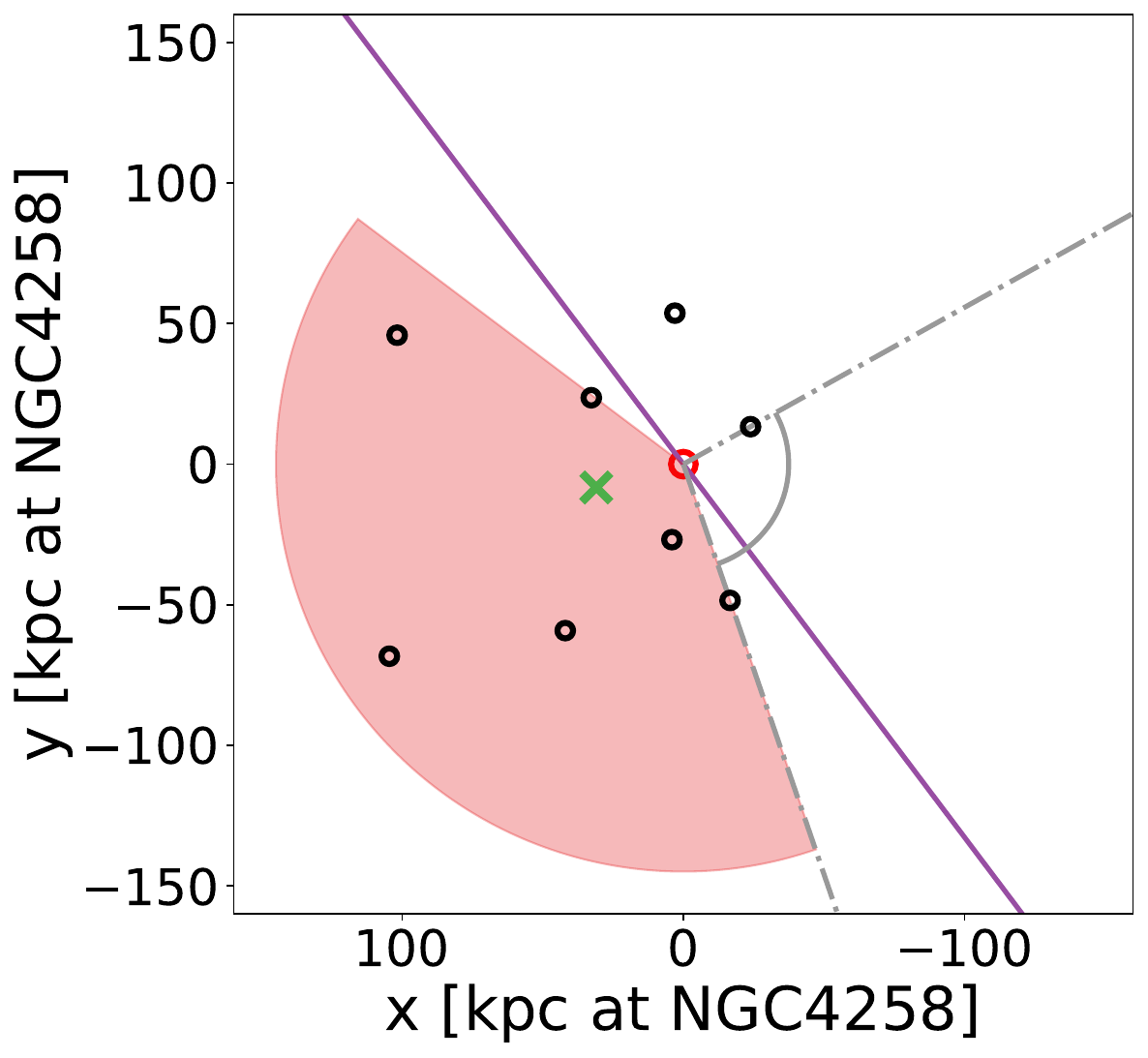}
    \includegraphics[width=6cm]{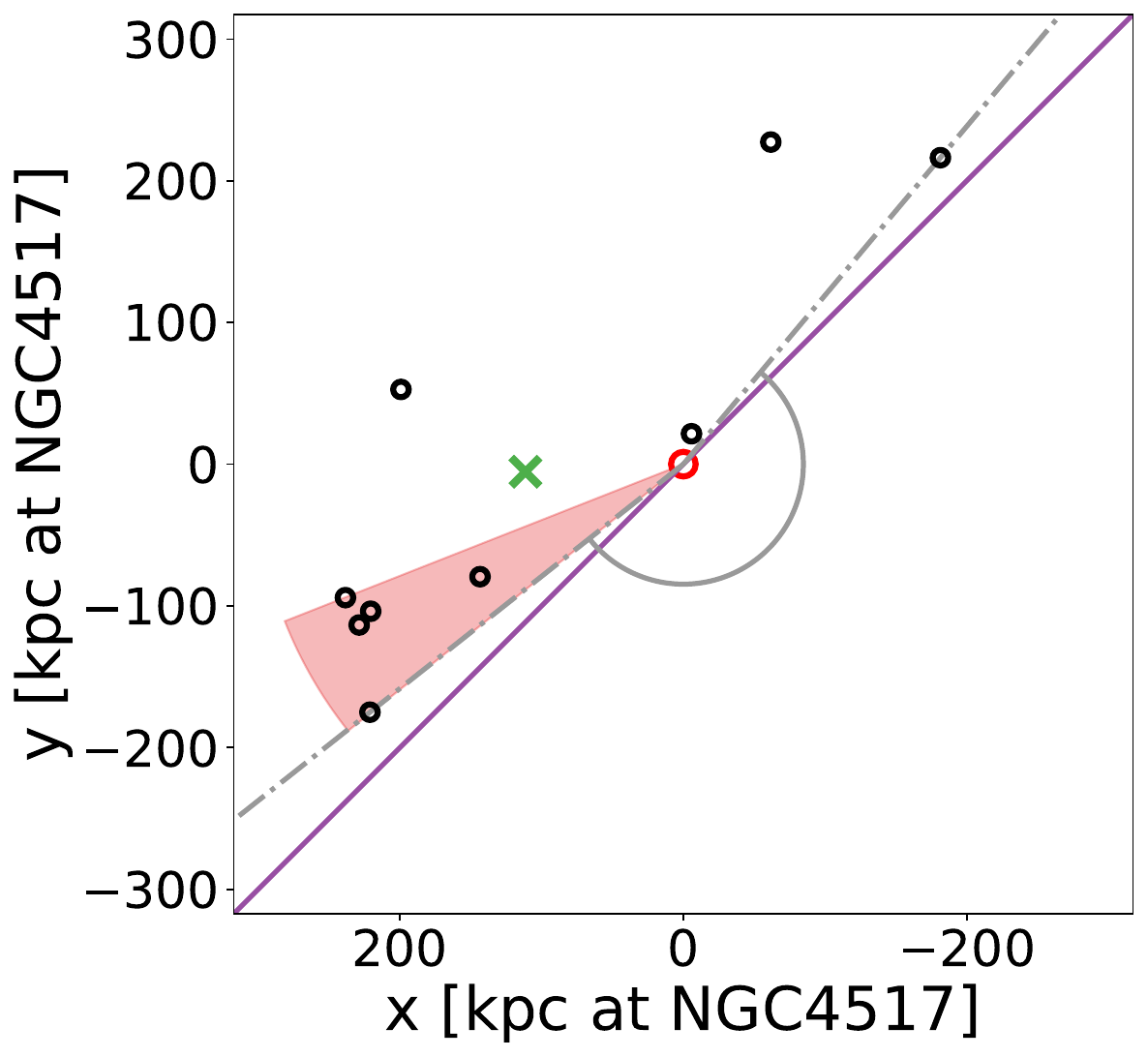}
    \includegraphics[width=6cm]{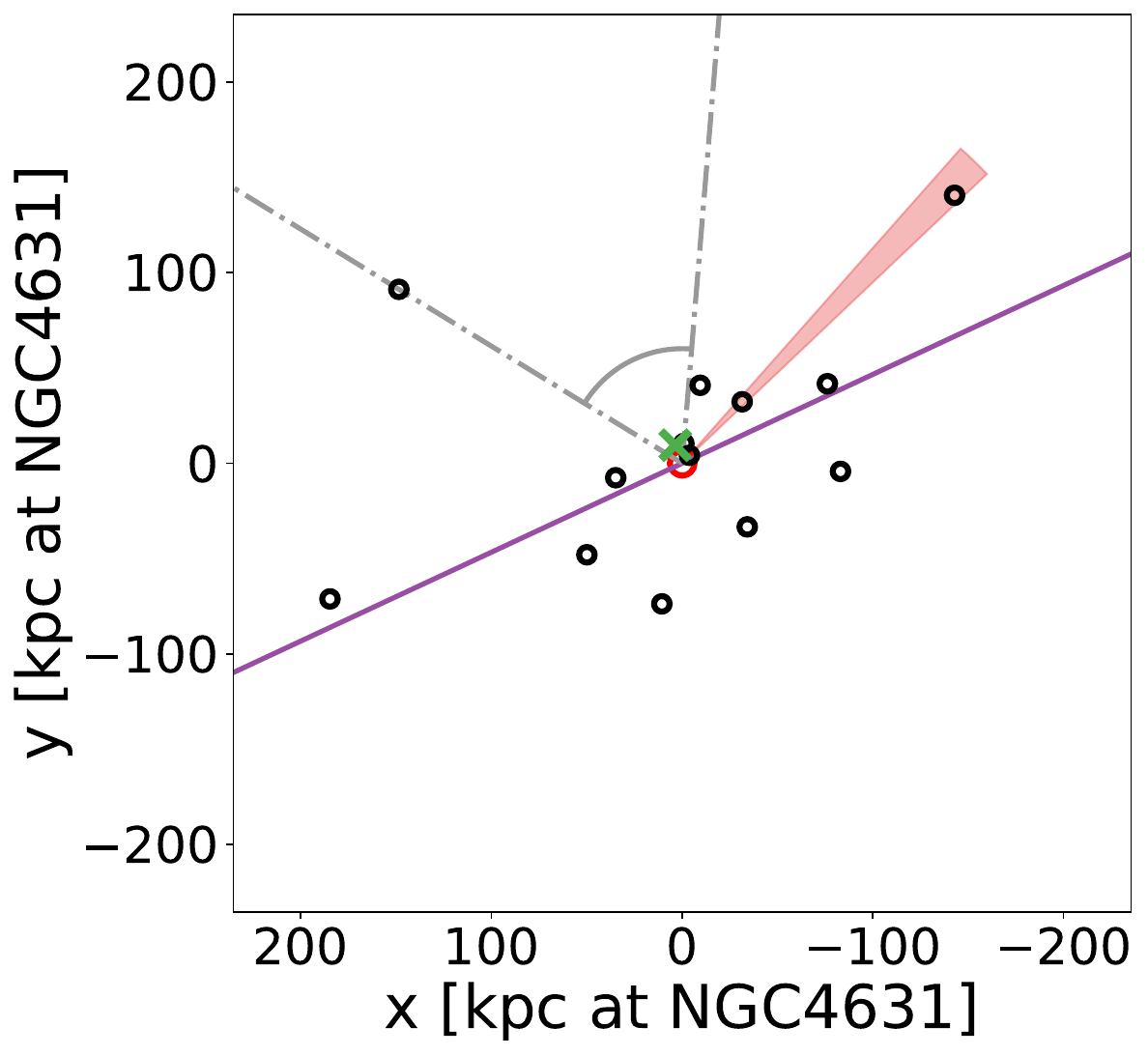}
    \includegraphics[width=6cm]{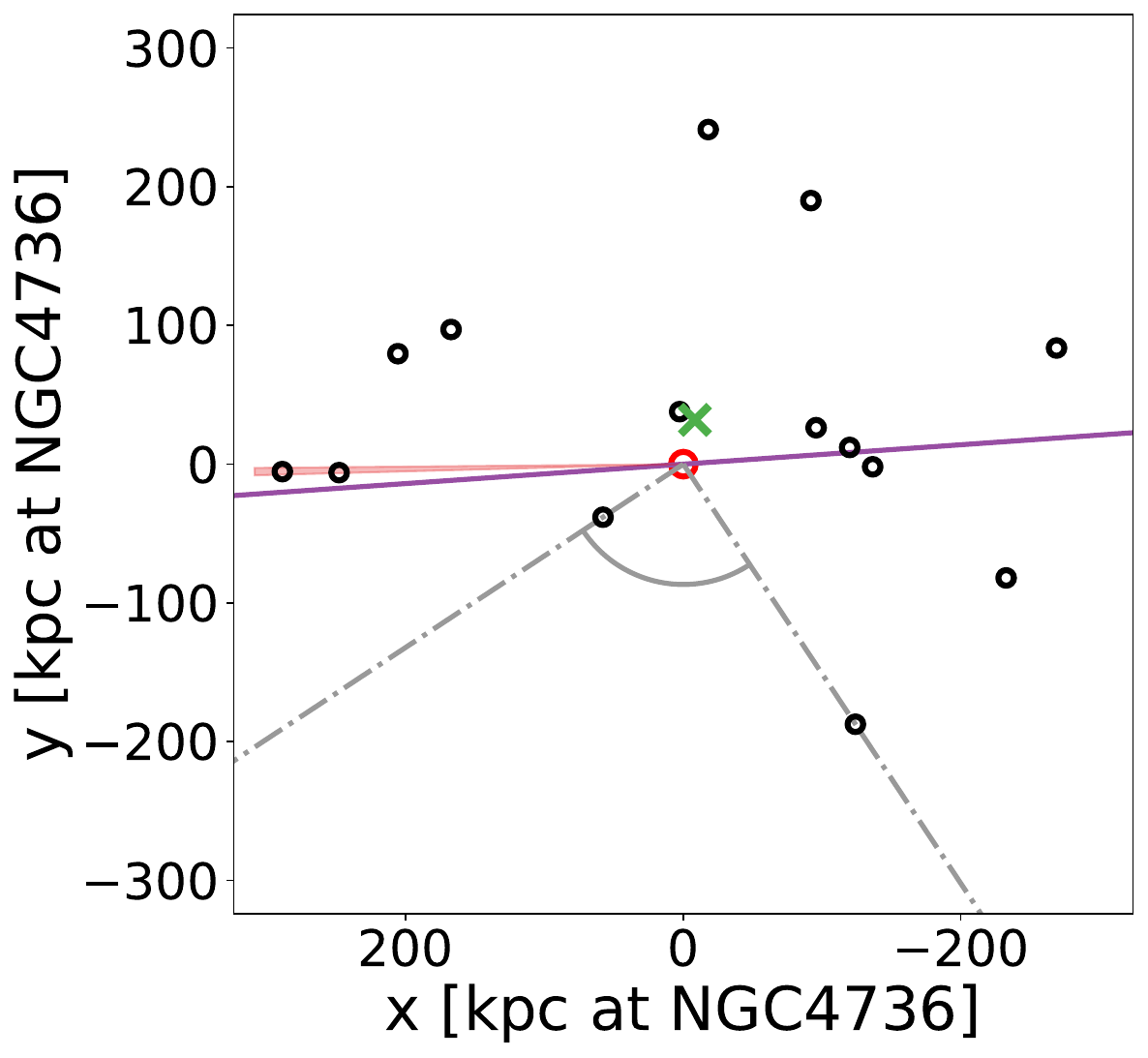}
    \includegraphics[width=6cm]{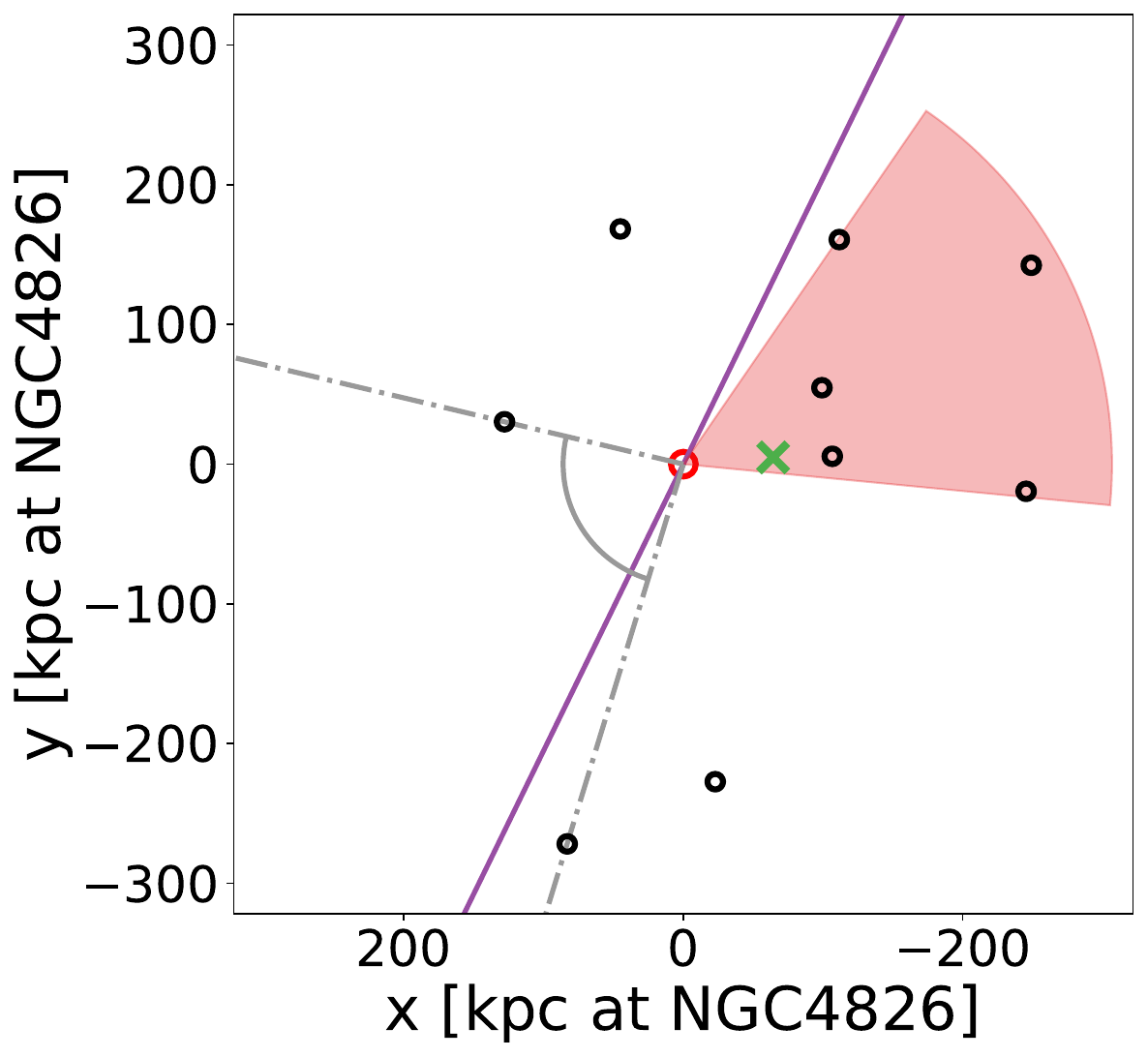}
    \includegraphics[width=6cm]{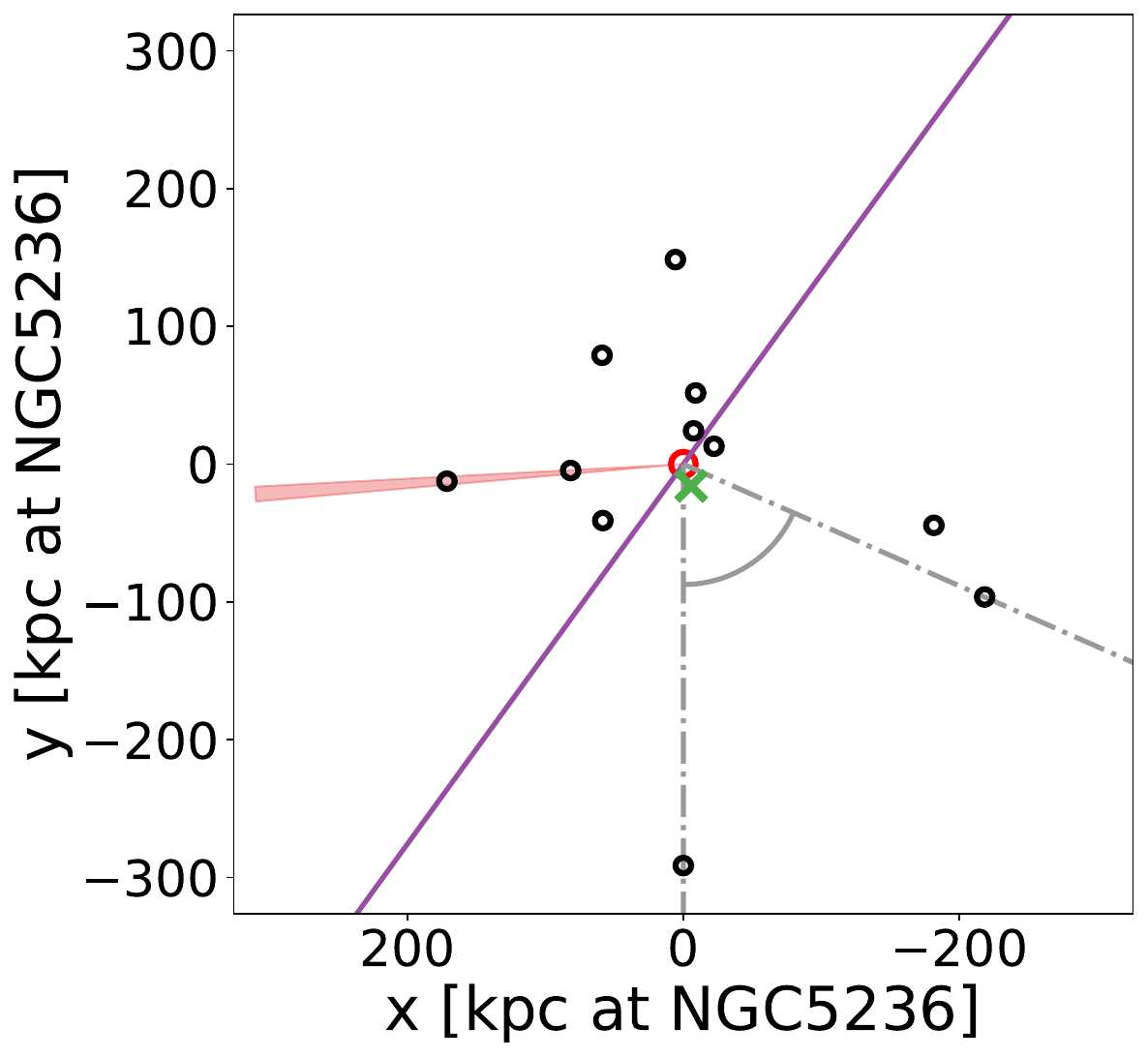}
    \includegraphics[width=6cm]{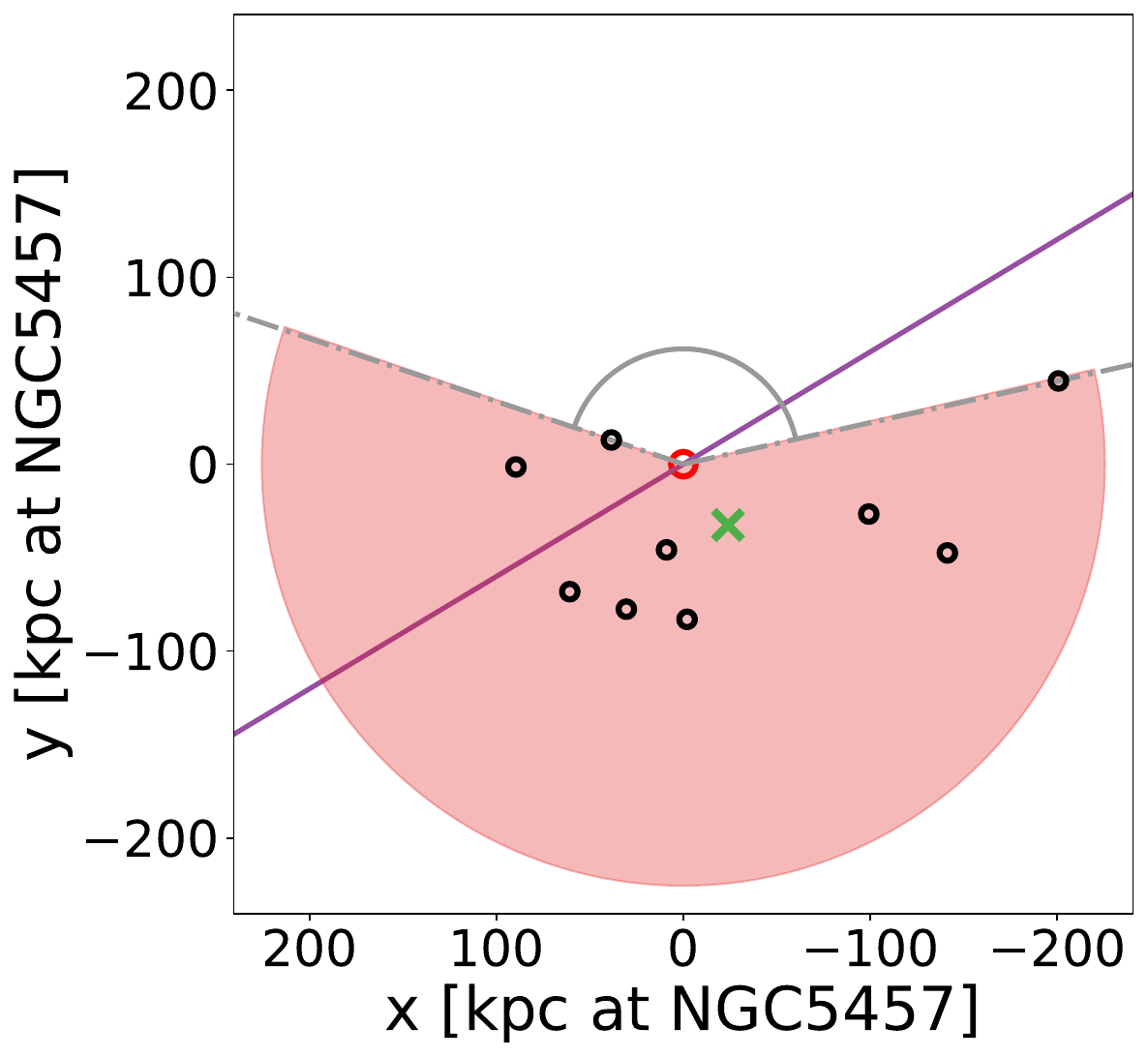}
    \includegraphics[width=6cm]{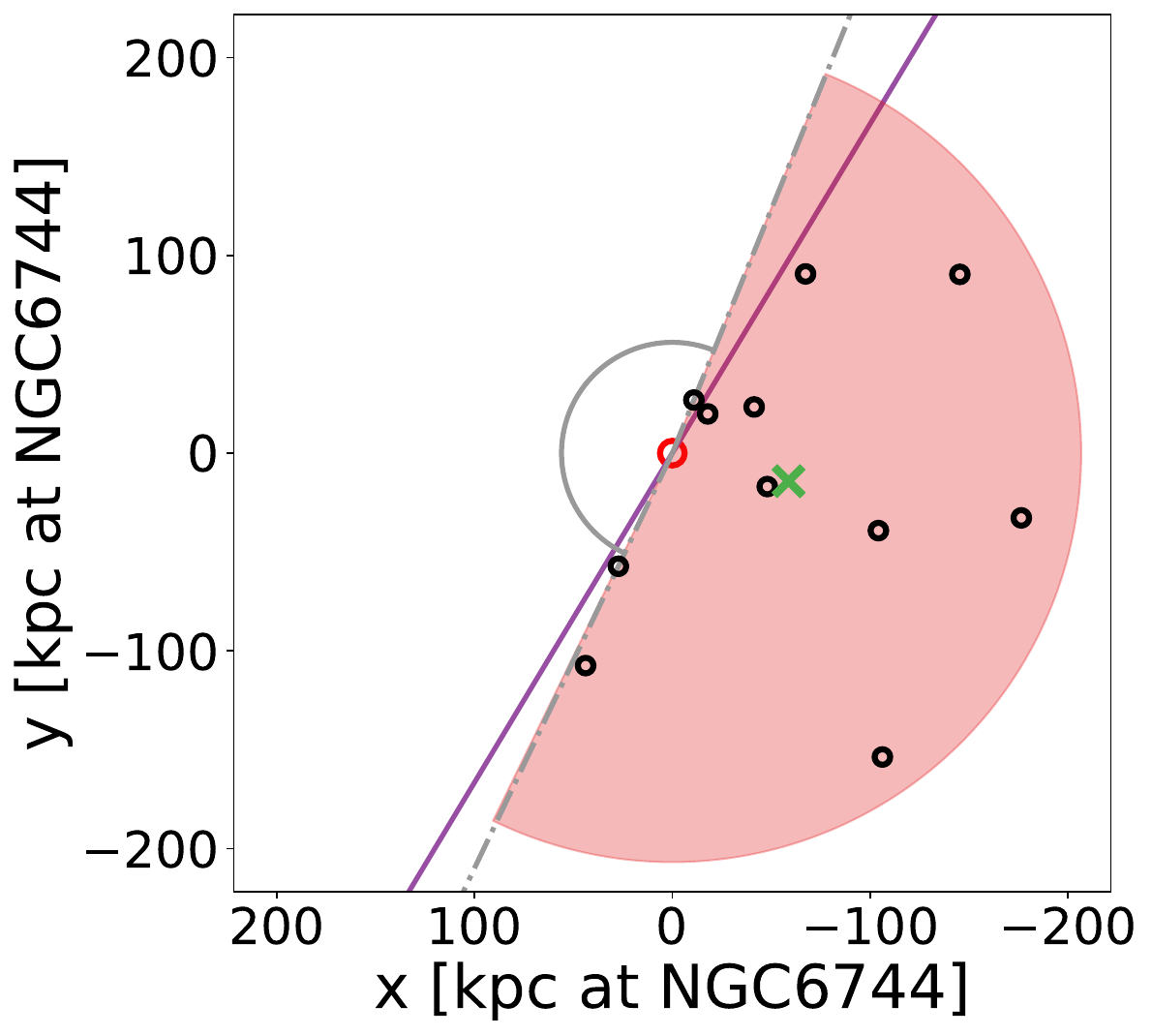}
    
    \caption{Continued.}
    \label{fig:elves_satellite_dists2}
\end{figure*}

\end{appendix}

\end{document}